\documentclass[runningheads,draft]{llncs}
\usepackage[T1]{fontenc}
\usepackage{graphicx}
\usepackage{xcolor}
\usepackage[final]{hyperref} %
\usepackage{amsmath}
\usepackage{amssymb}
\usepackage{cleveref}
\usepackage{macro}
\usepackage{color}

\NewDocumentCommand{\qedhere}{}{%
}

\usepackage[final]{microtype}

\usepackage[
style=lncs,
sortcites=true,
maxcitenames=2,
]{biblatex}

\IfFileExists{~/.texmf/bibtex/bib/library.bib}{
  \addbibresource{library.bib}
}{
  \IfFileExists{main_biber.bib}{\addbibresource{main_biber.bib}}{}
}
\defrule{a/f/var}{\arn{f/var}}{
  \japq{\blue{\Delta_\MU}, \blue{x} : \blue{A_\MF}}{\blue{x}}{\blue{A_\MF}}
}{
}

\defrule{a/f/lam}{\arn{f/lam}}{
  \japq{\Delta}{\ftlam{x}{\blue{M}}}{\FTlolly[\MF]{A_\MF}{B_\MF}}
}{
  \japq{\Delta, \blue{x} : \blue{A_\MF}}{\blue{M}}{\blue{B_\MF}}
}

\defrule{a/f/app}{\arn{f/app}}{
  \japq{\blue{\Delta_\MU}, \Delta_1, \Delta_2}{\ftapp{\blue{M}}{\blue{N}}}{\blue{B_\MF}}
}{
  \japq{\blue{\Delta_\MU}, \Delta_1}{\blue{M}}{\FTlolly[\MF]{\blue{A_\MF}}{B_\MF}}
  &
  \japq{\blue{\Delta_\MU}, \Delta_2}{\blue{N}}{\blue{A_\MF}}
}

\defrule{a/f/triv}{\arn{f/triv}}{
  \japq{\blue{\Delta_\MU}}{\fttriv}{\FTu[\MF]}
}{
  \mathstrut
}

\defrule{a/f/pair}{\arn{f/pair}}{
  \japq{\blue{\Delta_\MU}, \Delta_1, \Delta_2}{\ftpair{\blue{M_1}}{\blue{M_2}}}{\FTot[\MF]{A_1}{A_2}}
}{
  \japq{\blue{\Delta_\MU}, \Delta_1}{\blue{M_1}}{\blue{A_1}}
  &
  \japq{\blue{\Delta_\MU}, \Delta_2}{\blue{M_2}}{\blue{A_2}}
}

\defrule{a/f/susp}{\arn{f/susp}}{
  \japq{\Delta}{\ftsusp{P}}{\FTus[m][\MF]{A_m}}
}{
  \japq{\Delta}{P}{A_m}
  &
  \Delta \geq \MF
}

\defrule{a/f/force}{\arn{f/force}}{
  \japq{\Delta}{\ftforce{\blue{M}}}{\blue{A_\ML}}
}{
  \japq{\Delta}{\blue{M}}{\FTus[\ML][\MU]{\blue{A_\ML}}}
}

\defrule{a/f/down}{\arn{f/down}}{
  \japq{\Delta}{\ftdown{\blue{M}}}{\FTds[\ML][\MU]{A_\MU}}
}{
  \japq{\Delta}{\blue{M}}{\blue{A_\MU}}
}

\defrule{a/f/match}{\arn{f/match}}{
  \japq{\blue{\Delta_\MU}, \Delta_1, \Delta_2}{\ftm{P}{p}}{\blue{B_\MF}}
}{
  \japq{\blue{\Delta_\MU}, \Delta_1}{P}{A}
  &
  \japp{\blue{\Delta_\MU}, \Delta_2}{\blue{p}}{A}{\blue{B_\MF}}
}

\defrule{a/q/var}{\arn{q/var}}{
  \japq{\blue{\Delta_\MU}, \red{x} : \red{U_\MQ}}{\red{x}}{\red{U_\MQ}}
}{
}

\defrule{a/q/lam}{\arn{q/lam}}{
  \japq{{\Delta}}{\qtlam{x}{\red{C}}}{\QTlolly[\MQ]{U_\MQ}{S_\MQ}}
}{
  \japq{{\Delta}, \red{x} : \red{U_\MQ}}{\red{C}}{\red{S_\MQ}}
}

\defrule{a/q/app}{\arn{q/app}}{
  \japq{\blue{\Delta_\MU}, \Delta_1, \Delta_2}{\qtapp{\red{C_1}}{\red{C_2}}}{\red{U}}
}{
  \japq{\blue{\Delta_\MU}, \Delta_1}{\red{C_1}}{\QTlolly[\MQ]{S}{U}}
  &
  \japq{\blue{\Delta_\MU}, \Delta_2}{\red{C_2}}{\red{S}}
}

\defrule{a/q/triv}{\arn{q/triv}}{
  \japq{\blue{\Delta_\MU}}{\qttriv}{\QTu}
}{
  \mathstrut
}

\defrule{a/q/letu}{\arn{q/letu}}{
  \japq{\blue{\Delta_\MU}, \Delta_1, \Delta_2}{\qtletu{\red{C_1}}{\red{C_2}}}{\red{A_\MQ}}
}{
  \japq{\blue{\Delta_\MU}, \Delta_1}{\red{C_1}}{\red{\QTu_\MQ}}
  &
  \MQ \geq m
  &
  \japq{\blue{\Delta_\MU}, \Delta_2}{\red{C_2}}{\red{A_\MQ}}
}

\defrule{a/q/pair}{\arn{q/pair}}{
  \japq{\blue{\Delta_\MU}, \Delta_1, \Delta_2}{\qtpair{C_1}{C_2}}{\QTot[\MQ]{S_1}{S_2}}
}{
  \japq{\blue{\Delta_\MU}, \Delta_1}{\red{C_1}}{\red{S_1}}
  &
  \japq{\blue{\Delta_\MU}, \Delta_2}{\red{C_2}}{\red{S_2}}
}

\defrule{a/q/letp}{\arn{q/letp}}{
  \japq{\blue{\Delta_\MU}, \Delta_1, \Delta_2}{\qtletp{x_1}{x_2}{A_1}{A_2}{\red{C_1}}{\red{C_2}}}{\red{B_\MQ}}
}{
  \japq{\blue{\Delta_\MU}, \Delta_1}{\red{C_1}}{\QTot[\MQ]{A_1}{A_2}}
  &
  \japq{\blue{\Delta_\MU}, \Delta_2, \red{x_1} : \red{A_1}, \red{x_2} : \red{A_2}}{\red{C_2}}{\red{B_\MQ}}
}

\defrule{a/q/force}{\arn{q/force}}{
  \japq{\Delta}{\qtforce{\blue{M}}}{\red{A_\MQ}}
}{
  \japq{\Delta}{\blue{M}}{\FTus[\MQ][\ML]{\red{A_\MQ}}}
}

\defrule{a/q/gate}{\arn{q/gate}}{
  \japq{\blue{\Delta_\MU}}{\red{g}}{\QTlolly[\MQ]{S}{U}}
}{
  (\red{g} : \QTlolly[\MQ]{S}{U}) \in \Sigma_G
}

\defrule{a/q/down}{\arn{q/down}}{
  \japq{\Delta}{\qtdown{\blue{M}}}{\QTds[\MQ][\MF]{\blue{A_\MF}}}
}{
  \japq{\Delta}{\blue{M}}{\blue{A_\MF}}
}

\defrule{a/q/letd}{\arn{q/letd}}{
  \japq{\blue{\Delta_\MU}, \Delta_1, \Delta_2}{\qtletd{x}{\red{C_1}}{\red{C_2}}}{\red{B_\MQ}}
}{
  \japq{\blue{\Delta_\MU}, \Delta_1}{\red{C_1}}{\QTds[\MQ][\MF]{\blue{A_\MF}}}
  &
  \japq{\blue{\Delta_\MU}, \Delta_2, \blue{x} : \blue{A_\MF}}{\red{C_2}}{\red{B_\MQ}}
}

\defrule{a/q/match}{\arn{q/match}}{
  \japq{\blue{\Delta_\MU}, \Delta_1, \Delta_2}{\qtm{C}{q}}{\red{B_\MQ}}
}{
  \japq{\blue{\Delta_\MU}, \Delta_1}{\red{C}}{\red{S}}
  &
  \japp{\blue{\Delta_\MU}, \Delta_2}{\red{q}}{\red{S}}{\red{B_\MQ}}
}

\defrule{a/p/fpu}{\arn{p/f/u}}{
  \japp{\Delta}{\fpu{\blue{M}}}{\blue{\FTu_k}}{\blue{A_\MF}}
}{
  \japq{\Delta}{\blue{M}}{\blue{A_\MF}}
  &
  k \geq \MF
}

\defrule{a/p/fpp}{\arn{p/f/p}}{
  \japp{\Delta}{\blue{\fpp{x_1}{x_2}{M}}}{\blue{\FTot[\MF]{A_1}{A_2}}}{\blue{B_\MF}}
}{
  \japq{\Delta, \blue{x_1} : \blue{A_1}, \blue{x_2} : \blue{A_2}}{\blue{M}}{\blue{B_{\MF}}}
}

\defrule{a/p/fpd}{\arn{p/f/d}}{
  \japp{\Delta}{\blue{\fpd{x}{M}}}{\blue{\FTds[\ML][\MU]{A_\MU}}}{\blue{B_\ML}}
}{
  \japq{\Delta, \blue{x} : \blue{A_\MU}}{\blue{M}}{\blue{B_{\ML}}}
}

\defrule{a/p/qpu}{\arn{p/q/u}}{
  \japp{\Delta}{\qpu{P}}{\red{\QTu_\MQ}}{B_k}
}{
  \japq{\Delta}{P}{B_k}
  &
  \MQ \geq k
}

\defrule{a/p/qpp}{\arn{p/q/p}}{
  \japp{\Delta}{\qpp{x_1}{x_2}{P}}{\QTot[\MQ]{S_1}{S_2}}{A}
}{
  \japq{\Delta, \red{x_1} : \red{S_1}, \red{x_2} : \red{S_2}}{P}{A}
}

\defrule{x/f/var}{\xrn{f/var}}{
  \jxpq{\Delta, \blue{x} : \blue{A_\MF}}{\blue{x}}{\blue{A_\MF}}
}{
  \mathstrut
}

\defrule{x/f/lam}{\xrn{f/lam}}{
  \jxpq{\Delta}{\ftlam{x}{\blue{M}}}{\FTlolly[\MF]{A_\MF}{B_\MF}}
}{
  \jxpq{\Delta, \blue{x} : \blue{A_\MF}}{\blue{M}}{\blue{B_\MF}}
}

\defrule{x/f/app}{\xrn{f/app}}{
  \jxpq{\Delta}{\ftapp{\blue{M}}{\blue{N}}}{\blue{B_\MF}}
}{
  \jxpq{\Delta}{\blue{M}}{\FTlolly[\MF]{\blue{A_\MF}}{B_\MF}}
  &
  \jxpq{\Delta}{\blue{N}}{\blue{A_\MF}}
}

\defrule{x/f/triv}{\xrn{f/triv}}{
  \jxpq{\Delta}{\fttriv}{\FTu[\MF]}
}{
  \mathstrut
}

\defrule{x/f/pair}{\xrn{f/pair}}{
  \jxpq{\Delta}{\ftpair{\blue{M_1}}{\blue{M_2}}}{\FTot[\MF]{A_1}{A_2}}
}{
  \jxpq{\Delta}{\blue{M_1}}{\blue{A_1}}
  &
  \jxpq{\Delta}{\blue{M_2}}{\blue{A_2}}
}

\defrule{x/f/susp}{\xrn{f/susp}}{
  \jxpq{\Delta}{\ftsusp{P}}{\FTus[m][\MF]{A_m}}
}{
  \jxpq{\Delta}{P}{A_m}
}

\defrule{x/f/force}{\xrn{f/force}}{
  \jxpq{\Delta}{\ftforce{\blue{M}}}{\blue{A_\ML}}
}{
  \jxpq{\Delta}{\blue{M}}{\FTus[\ML][\MU]{\blue{A_\ML}}}
}

\defrule{x/f/down}{\xrn{f/down}}{
  \jxpq{\Delta}{\ftdown{\blue{M}}}{\FTds[\ML][\MU]{A_\MU}}
}{
  \jxpq{\Delta}{\blue{M}}{\blue{A_\MU}}
}

\defrule{x/f/match}{\xrn{f/match}}{
  \jxpq{\Delta}{\ftm{P}{p}}{\blue{B_\MF}}
}{
  \jxpq{\Delta}{P}{A}
  &
  \jxpp{\Delta}{\blue{p}}{A}{\blue{B_\MF}}
}

\defrule{x/q/var}{\xrn{q/var}}{
  \jxpq{\Delta, \red{x} : \red{A_\MQ}}{\red{x}}{\red{A_\MQ}}
}{
}

\defrule{x/q/lam}{\xrn{q/lam}}{
  \jxpq{\Delta}{\qtlam{x}{\red{C}}}{\QTlolly[\MQ]{S_\MQ}{U_\MQ}}
}{
  \jxpq{\Delta, \qtp{x}{S_\MQ}}{\red{C}}{\red{U_\MQ}}
}

\defrule{x/q/app}{\xrn{q/app}}{
  \jxpq{\Delta}{\qtapp{\red{C_1}}{\red{C_2}}}{\red{U}}
}{
  \jxpq{\Delta}{\red{C_1}}{\QTlolly[\MQ]{S}{U}}
  &
  \jxpq{\Delta}{\red{C_2}}{\red{S}}
}

\defrule{x/q/triv}{\xrn{q/triv}}{
  \jxpq{\Delta}{\qttriv}{\QTu}
}{
  \mathstrut
}

\defrule{x/q/pair}{\xrn{q/pair}}{
  \jxpq{\Delta}{\qtpair{C_1}{C_2}}{\QTot[\MQ]{S_1}{S_2}}
}{
  \jxpq{\Delta}{\red{C_1}}{\red{S_1}}
  &
  \jxpq{\Delta}{\red{C_2}}{\red{S_2}}
}

\defrule{x/q/force}{\xrn{q/force}}{
  \jxpq{\red{\Delta}}{\qtforce{\blue{M}}}{\red{A_\MQ}}
}{
  \jxpq{\red{\Delta}}{\blue{M}}{\FTus[\MQ][\ML]{\red{A_\MQ}}}
}

\defrule{x/q/gate}{\xrn{q/gate}}{
  \jxpq{\Delta}{\red{g}}{\QTlolly[\MQ]{S}{U}}
}{
  (\red{g} : \QTlolly[\MQ]{S}{U}) \in \Sigma_G
}

\defrule{x/q/down}{\xrn{q/down}}{
  \jxpq{\Delta}{\qtdown{\blue{M}}}{\QTds[\MQ][\MF]{\blue{A_\MF}}}
}{
  \jxpq{\Delta}{\blue{M}}{\blue{A_\MF}}
}

\defrule{x/q/match}{\xrn{q/match}}{
  \jxpq{\Delta}{\qtm{C}{q}}{\red{B_\MQ}}
}{
  \jxpq{\Delta}{\red{C}}{\red{S}}
  &
  \jxpp{\Delta}{\red{q}}{\red{S}}{\red{B_\MQ}}
}

\defrule{x/p/fpu}{\xrn{p/f/u}}{
  \jxpq{\Delta}{\fpu{M}}{\fpat{\FTu_k}{A_\MF}}
}{
  \jxpq{\Delta}{\blue{M}}{\blue{A_\MF}}
}

\defrule{x/p/fpp}{\xrn{p/f/p}}{
  \jxpp{\Delta}{\fpp{x_1}{x_2}{M}}{\blue{\FTot[\MF]{A_1}{A_2}}}{\blue{B_\MF}}
}{
  \jxpq{\Delta, \ftp{x_1}{A_1}, \ftp{x_2}{A_2}}{\blue{M}}{\blue{B_{\MF}}}
}

\defrule{x/p/fpd}{\xrn{p/f/d}}{
  \jxpp{\Delta}{\fpd{x}{M}}{\blue{\FTds[\ML][\MU]{A_\MU}}}{\blue{B_\ML}}
}{
  \jxpq{\Delta, \ftp{x}{A_\MU}}{\blue{M}}{\blue{B_{\ML}}}
}

\defrule{x/p/qpu}{\xrn{p/q/u}}{
  \jxpp{\Delta}{\qpu{P}}{\red{\QTu_\MQ}}{B}
}{
  \jxpq{\Delta}{P}{B}
}

\defrule{x/p/qpp}{\xrn{p/q/p}}{
  \jxpp{\Delta}{\qpp{x_1}{x_2}{P}}{\QTot[\MQ]{S_1}{S_2}}{B}
}{
  \jxpq{\Delta, \qtp{x_1}{S_1}, \qtp{x_2}{S_2}}{P}{B}
}

\defrule{x/cs/empty}{\xrn{fcs/empty}}{
  \jssub{\cdot}{\Psi}{\cdot}
}{
}

\defrule{x/cs/tm}{\xrn{fcs/tm}}{
  \jssub{(\sigma, V)}{\Psi}{\Delta, \ptp{x}{A}}
}{
  \jssub{\sigma}{\Psi}{\Delta}
  &
  \jxpq{\ctp{\Psi}}{V}{A}
  &
  \jnorm[\cneu{\Psi}]{V}
}

\defrule{neu/var}{\arn{neu/var}}{
  \jneu[\Pi]{x}
}{
  \jneu{x} \in \Pi
}

\defrule{neu/gate}{\arn{neu/gate}}{
  \jneu[\Pi]{\red{g}}
}{
}

\defrule{neu/app/neu}{\arn{neu/app/neu}}{
  \jneu[\Pi]{\qtapp{g}{R}}
}{
  \jneu[\Pi]{R}
}

\defrule{neu/app/can}{\arn{neu/app/can}}{
  \jneu[\Pi]{\qtapp{g}{K}}
}{
  \jcan[\Pi]{\red{K}}
}

\defrule{neu/force}{\arn{neu/force}}{
  \jneu[\Pi]{\ptforce{R}}
}{
  \jneu[\Pi]{R}
}

\defrule{can/triv}{\arn{can/triv}}{
  \jcan[\Pi]{\pttriv}
}{
}

\defrule{can/pair}{\arn{can/pair}}{
  \jcan[\Pi]{\ptpair{V_1}{V_2}}
}{
  \jnorm[\Pi]{V_1}
  &
  \jnorm[\Pi]{V_2}
}

\defrule{can/lam/c}{\arn{can/lam/c}}{
  \jcan[\Pi]{\qtlam{x}{V}}
}{
  \jnorm[\Pi, \jneu{\red{x}}]{\red{V}}
}

\defrule{can/lam/f}{\arn{can/lam/f}}{
  \jcan[\Pi]{\ftlam{x}{M}}
}{
}

\defrule{can/down}{\arn{can/down}}{
  \jcan[\Pi]{\ftdown{V}}
}{
  \jnorm[\Pi]{\blue{V}}
}

\defrule{can/susp}{\arn{can/susp}}{
  \jcan[\Pi]{\ftsusp{P}}
}{
}

\defrule{norm/match}{\arn{norm/match}}{
  \jnorm[\Pi]{\ptm{\red{R}}{\qpat{\pi}{P}}}
}{
  \jneu[\Pi]{R}
  &
  \jnorm[\Pi, \jneu{\pi}]{P}
}

\defrule{norm/p/q/u}{\arn{norm/p/q/u}}{
  \jnormp[\Pi]{\qpu{P}}
}{
  \jnorm[\Pi]{P}
}

\defrule{norm/p/q/p}{\arn{norm/p/q/p}}{
  \jnormp[\Pi]{\qpp{x}{y}{P}}
}{
  \jnorm[\Pi, \jneu{x}, \jneu{y}]{P}
}

\defrule{norm/letu}{\arn{norm/letu}}{
  \jnorm[\Pi]{\ptletu{R}{V}}
}{
  \jneu[\Pi]{R}
  &
  \jnorm[\Pi]{V}
}

\defrule{norm/letp}{\arn{norm/letp}}{
  \jnorm[\Pi]{\ptletp{x}{y}{R}{V}}
}{
  \jneu[\Pi]{R}
  &
  \jnorm[\Pi,
  \jneu{x},
  \jneu{y}]{V}
}

\defrule{norm/letd}{\arn{norm/letd}}{
  \jnorm[\Pi]{\ptletd{x}{R}{V}}
}{
  \jneu[\Pi]{R}
  &
  \jnorm[\Pi,
  \jneu{x}]{V}
}

\defrule{norm/can}{\arn{norm/can}}{
  \jnorm[\Pi]{K}
}{
  \jcan[\Pi]{K}
}

\defrule{norm/neu}{\arn{norm/neu}}{
  \jnorm[\Pi]{\red{R}}
}{
  \jneu[\Pi]{R}
}

\defrule{fstep/app/beta}{\arn{\(\afs\)/app/\(\beta\)}}{
  \fstep{\Pi}{
    \ftapp{(\ftlam{x}{M})}{\blue{V}}
  }{
    \subst{\blue{V}}{\blue{x}}{\blue{M}}
  }
}{
  \jnorm[\Pi]{\blue{V}}
}

\defrule{fstep/app/1}{\arn{\(\afs\)/app/1}}{
  \fstep{\Pi}{\ftapp{M}{N}}{\ftapp{M'}{N}}
}{
  \fstep{\Pi}{\blue{M}}{\blue{M'}}
}

\defrule{fstep/app/2}{\arn{\(\afs\)/app/2}}{
  \fstep{\Pi}{\ftapp{V}{N}}{\ftapp{V}{N'}}
}{
  \jnorm[\Pi]{\blue{V}}
  &
  \fstep{\Pi}{\blue{N}}{\blue{N'}}
}

\defrule{fstep/app/cc}{\arn{\(\afs\)/app/cc}}{
  \fstep{\Pi}{
    \ftapp{(\ftm{\red{R}}{\qpat{\pi}{V}})}{\blue{W}}
  }{
    \ftm{\red{R}}{\qpat{\pi}{\ftapp{V}{W}}}
  }
}{
  \jnorm[\Pi]{\ftm{\red{R}}{\qpat{\pi}{V}}}
  &
  \jnorm[\Pi]{\blue{W}}
}

\defrule{fstep/pair/1}{\arn{\(\afs\)/pair/1}}{
  \fstep{\Pi}{\ftpair{M}{N}}{\ftpair{M'}{N}}
}{
  \fstep{\Pi}{\blue{M}}{\blue{M'}}
}

\defrule{fstep/pair/2}{\arn{\(\afs\)/pair/2}}{
  \fstep{\Pi}{\ftpair{V}{N}}{\ftpair{V}{N'}}
}{
  \jnorm[\Pi]{\blue{V}}
  &
  \fstep{\Pi}{\blue{N}}{\blue{N'}}
}

\defrule{fstep/force}{\arn{\(\afs\)/force}}{
  \fstep{\Pi}{
    \ftforce{(\ftsusp{M})}
  }{
    \blue{M}
  }
}{
}

\defrule{fstep/force/1}{\arn{\(\afs\)/force/1}}{
  \fstep{\Pi}{
    \ftforce{\blue{M}}
  }{
    \ftforce{\blue{N}}
  }
}{
  \fstep{\Pi}{\blue{M}}{\blue{N}}
}

\defrule{fstep/force/cc}{\arn{\(\afs\)/force/cc}}{
  \fstep{\Pi}{
    (\ftforce{(\ftm{\red{R}}{\qpat{\pi}{V}})})
  }{
    (\ftm{\red{R}}{\qpat{\pi}{\ftforce{V}}})
  }
}{
  \jnorm[\Pi]{\ftm{\red{R}}{\qpat{\pi}{V}}}
}

\defrule{fstep/down}{\arn{\(\afs\)/down}}{
  \fstep{\Pi}{
    \ftdown{M}
  }{
    \ftdown{N}
  }
}{
  \fstep{\Pi}{\blue{M}}{\blue{N}}
}

\defrule{fstep/m/f}{\arn{\(\afs\)/m/f}}{
  \fstep{\Pi}{
    \ftm{\blue{M}}{p}
  }{
    \ftm{\blue{N}}{p}
  }
}{
  \fstep{\Pi}{\blue{M}}{\blue{N}}
}

\defrule{fstep/m/k}{\arn{\(\afs\)/m/k}}{
  \fstep{\Pi}{
    \ftm{K}{\blue{p}}
  }{
    \blue{M}
  }
}{
  \jcan[\Pi]{K}
  &
  \ecan{K}{\blue{p}}{\blue{M}}
}

\defrule{fstep/m/f/cc}{\arn{\(\afs\)/m/f/cc}}{
  \fstep{\Pi}{
    \ftm{(\ftm{\red{R}}{\qpat{\pi}{\blue{V}}})}{p}
  }{
    \ftm{\red{R}}{\qpat{\pi}{\ftm{V}{p}}}
  }
}{
  \jnorm[\Pi]{\ftm{\red{R}}{\qpat{\pi}{V}}}
}

\defrule{fstep/m/q}{\arn{\(\afs\)/m/q}}{
  \fstep{\Pi}{
    \ftm{\red{C}}{p}
  }{
    \ftm{\red{D}}{p}
  }
}{
  \cstep{\Pi}{\red{C}}{\red{D}}
}

\defrule{fstep/m/q/cc}{\arn{\(\afs\)/m/q/cc}}{
  \fstep{\Pi}{
    \ftm{(\qtm{\red{R}}{\qpat{\pi}{\red{V}}})}{p}
  }{
    \ftm{\red{R}}{\qpat{\pi}{\ftm{\red{V}}{p}}}
  }
}{
  \jnorm[\Pi]{\ftm{\red{R}}{\qpat{\pi}{V}}}
}

\defrule{fstep/m/q/r}{\arn{\(\afs\)/m/q/r}}{
  \fstep{\Pi}{
    \ftm{\red{R}}{\qpat{\pi}{M}}
  }{
    \ftm{\red{R}}{\qpat{\pi}{N}}
  }
}{
  \jneu[\Pi]{R}
  &
  \fstep{\Pi, \jneu{\pi}}{\blue{M}}{\blue{N}}
}

\defrule{cstep/lam}{\arn{\(\acs\)/lam}}{
  \cstep{\Pi}{
    \qtlam{x}{C}
  }{
    \qtlam{x}{D}
  }
}{
  \cstep{\Pi, \jneu{\red{x}}}{\red{C}}{\red{D}}
}

\defrule{cstep/app/beta}{\arn{\(\acs\)/app/\(\beta\)}}{
  \cstep{\Pi}{
    \qtapp{(\qtlam{x}{V})}{\red{W}}
  }{
    \subst{\red{W}}{\red{x}}{\red{V}}
  }
}{
  \jnorm[\Pi]{\qtlam{x}{V}}
  &
  \jnorm[\Pi]{\red{W}}
}

\defrule{cstep/app/1}{\arn{\(\acs\)/app/1}}{
  \cstep{\Pi}{\qtapp{C}{D}}{\qtapp{C'}{D}}
}{
  \cstep{\Pi}{\red{C}}{\red{C'}}
}

\defrule{cstep/app/2}{\arn{\(\acs\)/app/2}}{
  \cstep{\Pi}{\qtapp{V}{D}}{\qtapp{V}{D'}}
}{
  \jnorm[\Pi]{\red{V}}
  &
  \cstep{\Pi}{\red{D}}{\red{D'}}
}

\defrule{cstep/app/cc/1}{\arn{\(\acs\)/app/cc/1}}{
  \cstep{\Pi}{
    (\qtapp{(\qtm{\red{R}}{\qpat{\pi}{V}})}{\red{W}})
  }{
    (\qtm{\red{R}}{\qpat{\pi}{\qtapp{V}{W}}})
  }
}{
  \jnorm[\Pi]{\qtm{\red{R}}{\qpat{\pi}{V}}}
  &
  \jnorm[\Pi]{\red{W}}
}

\defrule{cstep/app/cc/2}{\arn{\(\acs\)/app/cc/2}}{
  \cstep{\Pi}{
    (\qtapp{g}{(\qtm{\red{R}}{\qpat{\pi}{V}})})
  }{
    (\qtm{\red{R}}{\qpat{\pi}{\qtapp{g}{V}}})
  }
}{
  \jnorm[\Pi]{\qtm{\red{R}}{\qpat{\pi}{V}}}
}

\defrule{cstep/pair/1}{\arn{\(\acs\)/pair/1}}{
  \cstep{\Pi}{\qtpair{C}{D}}{\qtpair{C'}{D}}
}{
  \cstep{\Pi}{\red{C}}{\red{C'}}
}

\defrule{cstep/pair/2}{\arn{\(\acs\)/pair/2}}{
  \cstep{\Pi}{\qtpair{V}{D}}{\qtpair{V}{D'}}
}{
  \jnorm[\Pi]{\red{V}}
  &
  \cstep{\Pi}{\red{D}}{\red{D'}}
}

\defrule{cstep/force}{\arn{\(\acs\)/force}}{
  \cstep{\Pi}{
    \qtforce{(\ftsusp{\red{C}})}
  }{
    \red{C}
  }
}{
}

\defrule{cstep/force/1}{\arn{\(\acs\)/force/1}}{
  \cstep{\Pi}{
    \qtforce{\blue{M}}
  }{
    \qtforce{\blue{N}}
  }
}{
  \fstep{\Pi}{\blue{M}}{\blue{N}}
}

\defrule{cstep/force/cc}{\arn{\(\acs\)/force/cc}}{
  \cstep{\Pi}{
    \qtforce{(\ftm{\red{R}}{\qpat{\pi}{V}})}
  }{
    (\qtm{\red{R}}{\qpat{\pi}{\qtforce{V}}})
  }
}{
  \jnorm[\Pi]{\ftm{\red{R}}{\qpat{\pi}{V}}}
}

\defrule{cstep/m}{\arn{\(\acs\)/m}}{
  \cstep{\Pi}{
    \qtm{C}{p}
  }{
    \qtm{D}{p}
  }
}{
  \cstep{\Pi}{\red{C}}{\red{D}}
}

\defrule{cstep/m/r}{\arn{\(\acs\)/m/r}}{
  \cstep{\Pi}{
    (\qtm{R}{\pat{\pi}{D}})
  }{
    (\qtm{R}{\pat{\pi}{D'}})
  }
}{
  \jneu[\Pi]{R}
  &
  \cstep{\Pi, \jneu{\pi}}{\red{D}}{\red{D'}}
}

\defrule{cstep/m/k}{\arn{\(\acs\)/m/k}}{
  \cstep{\Pi}{
    \qtm{K}{p}
  }{
    \red{C}
  }
}{
  \jcan[\Pi]{\red{K}}
  &
  \ecan{\red{K}}{\red{p}}{\red{C}}
}

\defrule{cstep/m/cc}{\arn{\(\acs\)/m/cc}}{
  \cstep{\Pi}{
    \qtm{(\qtm{\red{R}}{\qpat{\pi}{\red{V}}})}{p}
  }{
    \qtm{\red{R}}{\qpat{\pi}{\qtm{\red{V}}{p}}}
  }
}{
  \jnorm[\Pi]{\qtm{\red{R}}{\qpat{\pi}{V}}}
}

\defrule{ecan/triv}{\arn{e/triv}}{
  \ecan{\pttriv}{(\ppu{P})}{P}
}{
}

\defrule{ecan/pair}{\arn{e/pair}}{
  \ecan{\ptpair{V_1}{V_2}}{(\ppp{x_1}{x_2}{P})}{\subst{V_1, V_2}{x_1, x_2}{P}}
}{
}

\defrule{ecan/down}{\arn{e/down}}{
  \ecan{\ftdown{V}}{(\fpd{x}{M})}{\subst{\blue{V}}{\blue{x}}{\blue{M}}}
}{
}

\defrule{steps/refl}{\arn{steps/refl}}{
  \steps{\Pi}{P}{P}
}{
}
\defrule{steps/step}{\arn{steps/step}}{
  \steps{\Pi}{P}{Q}
}{
  \step{\Pi}{P}{P'}
  &
  \steps{\Pi}{P'}{Q}
}

\defrule{rtc/refl}{\arn{rtc/refl}}{
  X \mathrel{\mathcal{R}^*} X
}{
}
\defrule{rtc/trans}{\arn{rtc/trans}}{
  X \mathrel{\mathcal{R}^*} Z
}{
  X \mathrel{\mathcal{R}} Y
  &
  Y \mathrel{\mathcal{R}^*} Z
}

\defrule{llr/triv}{\arn{lr/\(\PTu\)}}{
  \llr{\Psi}{\pttriv}{\PTu[m]}
}{
}

\defrule{llr/pair}{\arn{lr/\(\otimes\)}}{
  \llr{\Psi}{\ptpair{P_1}{P_2}}{\PTot{A_1}{A_2}}
}{
  \llr{\Psi}{P_1}{A_1}
  &
  \llr{\Psi}{P_2}{A_2}
}

\defrule{llr/fun}{\arn{lr/\(\multimap\)}}{
  \llr{\Psi}{P}{\PTlolly{A}{B}}
}{
  \jhalts{\cneu{\Psi}}{P}
  &
  \llr{\Phi}{\ptapp{(\apprs{\sigma}{P})}{Q}}{B}
  &
  (\forall \jnctx{\sigma}{\Phi}{\Psi})
  &
  (\forall \llr{\Phi}{Q}{A})
}

\defrule{llr/ds}{\arn{lr/\(\blue{\downarrow}\)}}{
  \llr{\Psi}{\ftdown{M}}{\FTds[\ML][\MU]{A}}
}{
  \llr{\Psi}{\blue{M}}{\blue{A}}
}

\defrule{llr/us}{\arn{lr/\(\blue{\uparrow}\)}}{
  \llr{\Psi}{\blue{M}}{\FTus[][]{A}}
}{
  \llr{\Psi}{\ptforce{\blue{M}}}{A}
}

\defrule{llr/neu}{\arn{lr/neu}}{
  \llr{\Psi}{\red{R}}{\red{A}}
}{
  \jxpq{\ctp{\Psi}}{\red{R}}{\red{A}}
  &
  \jneu[\cneu{\Psi}]{R}
}

\defrule{llr/m}{\arn{lr/m}}{
  \llr{\Psi}{\ptm{\red{R}}{\qpat{\pi}{P}}}{A}
}{
  \jxpq{\ctp{\Psi}}{\red{R}}{\red{S}}
  &
  \jneu[\cneu{\Psi}]{R}
  &
  \llr{\Psi, \qtp{\pi}{S}}{P}{A}
}

\defrule{llr/steps}{\arn{lr/steps}}{
  \llr{\Psi}{P}{A}
}{
  \jxpq{\ctp{\Psi}}{P}{A}
  &
  \step{\cneu{\Psi}}{P}{P'}
  &
  \llr{\Psi}{P'}{A}
}

\defrule{llr/rs/nil}{\arn{lr/rs/nil}}{
  \llr{\Psi}{\cdot}{\cdot}
}{
}
\defrule{llr/rs/tm}{\arn{lr/rs/tm}}{
  \llr{\Psi}{\sigma, V}{\Delta, x : A}
}{
  \llr{\Psi}{\sigma}{\Delta}
  &
  \llr{\Psi}{V}{A}
}

\defrule{ns/empty}{\arn{ns/empty}}{
  \jnctx{\cdot}{\Psi}{\cdot}
}{
}

\defrule{ns/neu}{\arn{ns/neu}}{
  \jnctx{(\sigma, \red{R})}{\Psi}{\Phi, \red{x} : \red{A}, \jneu{x}}
}{
  \jnctx{\sigma}{\Psi}{\Phi}
  &
  \jxpq{\ctp{\Psi}}{\red{R}}{\red{A}}
  &
  \jneu[\cneu{\Psi}]{\red{R}}
}

\NewDocumentCommand{\pqvar}{m}{Proto-Quipper-#1\xspace}
\NewDocumentCommand{\pqs}{}{\pqvar{S}}
\NewDocumentCommand{\pqa}{}{\textsc{\pqvar{A}}\xspace}
\NewDocumentCommand{\pqx}{}{\textsc{\pqvar{X}}\xspace}
\NewDocumentCommand{\pqm}{}{\pqvar{M}}
\NewDocumentCommand{\pqdyn}{}{\pqvar{Dyn}}

\usepackage{wrapfig}
\usepackage{tikz}
\usepackage{pgf-extrect}
\pgfdeclarelayer{backlayer}
\pgfdeclarelayer{edgelayer}
\pgfdeclarelayer{nodelayer}

\pgfsetlayers{backlayer,edgelayer,nodelayer,main}
\usetikzlibrary{arrows}
\usetikzlibrary{arrows.meta}
\usetikzlibrary{automata}
\usetikzlibrary{calc}
\usetikzlibrary{fit}
\usetikzlibrary{petri}
\usetikzlibrary{positioning}
\usetikzlibrary{shapes}
\tikzset{
  node distance=0.25cm and 1cm,
  medium box/.style={
    rectangle,
    draw=black,
    fill=white,
    shape=rectangle,
    minimum height=0.75cm,
    minimum width=0.75cm
  },
  none/.style={},
  string/.style = {rounded corners},
}

\usetikzlibrary{decorations.pathreplacing}
\tikzset{%
  show curve controls/.style={
    postaction={
      decoration={
        show path construction,
        curveto code={
          \draw [blue]
          (\tikzinputsegmentfirst) -- (\tikzinputsegmentsupporta)
          (\tikzinputsegmentlast) -- (\tikzinputsegmentsupportb);
          \fill [red, opacity=0.5]
          (\tikzinputsegmentsupporta) circle [radius=.5ex]
          (\tikzinputsegmentsupportb) circle [radius=.5ex];
        }
      },
      decorate
    }}}

\usepackage[createShortEnv, conf={restate, no link to proof}]{proof-at-the-end}

\begin{document}
\title{Deconstructed Proto-Quipper: A Rational Reconstruction}

\author{Ryan Kavanagh\inst{1}\orcidID{0000-0001-9497-4276} \and
Chuta Sano\inst{2}\orcidID{0000-0002-8179-2307} \and
Brigitte Pientka\inst{2}\orcidID{0000-0002-2549-4276}}
\authorrunning{R. Kavanagh et al.}
\institute{Département d'informatique, Université du Québec à Montréal,\\
  201 avenue du Président-Kennedy, Montréal, QC, H2X 3Y8, Canada\\
  \email{kavanagh.ryan@uqam.ca}\\
\and
School of Computer Science, McGill University,\\
3480 rue University, Montréal, QC, H3A 0E9, Canada\\
\email{chuta.sano@mail.mcgill.ca}\;
\email{brigitte.pientka@mcgill.ca}}
\maketitle              %
\begin{abstract}
  The Proto-Quipper family of programming languages aims to provide a formal foundation for the Quipper quantum programming language.
  Unfortunately, Proto-Quipper languages have complex operational
  semantics: they are inherently effectful, and they rely on set-theoretic operations and fresh name generation to manipulate quantum circuits.
  This makes them difficult to reason about using standard programming
  language techniques and, ultimately, to mechanize.
  We introduce \pqa, a rational reconstruction of Proto-Quipper languages for static circuit generation.
  It uses a linear \(\lambda\)-calculus to describe quantum circuits with normal forms that closely correspond to box-and-wire circuit diagrams.
  Adjoint-logical foundations integrate this circuit language with a linear/non-linear functional language and let us reconstruct Proto-Quipper's circuit programming abstractions using more primitive adjoint-logical operations.
  \pqa enjoys a simple call-by-value reduction semantics, and to illustrate its tractability as a foundation for Proto-Quipper languages, we show that it is normalizing.
  We show how to use standard logical relations to prove normalization of linear and substructural systems, thereby avoiding the inherent complexity of existing linear logical relations.

\end{abstract}
\section{Introduction}
\label{sec:introduction}

Quantum computing is becoming increasingly accessible, with capabilities far surpassing those offered by classical computing.
While physicists and theorists may be content reasoning about quantum computation in terms of physical properties or Hilbert spaces, programmers need higher-level abstractions to effectively use quantum computation.
The current state of the art remains low level: quantum programs are quantum circuits formed of quantum gates that manipulate qubits.
Quantum programming languages can be viewed as meta-programming languages with abstractions for generating, composing, and manipulating quantum circuits.

Quipper~\cite{green_2013:_quipp} is a functional programming language that provides abstractions not only for manipulating individual circuits but also for manipulating \textit{families of circuits} that are parametrized by values known at circuit generation time.
Programmers can thus uniformly implement parametrized quantum algorithms like the quantum Fourier transform (parametrized by the number of qubits to be used) or Shor's prime factorization algorithm (parametrized by the integer to be factored).
Unfortunately, Quipper is an embedded domain-specific language in Haskell, making it difficult to reason about.
Mismatches between Haskell's and Quipper's type systems also make it unsafe: well-typed programs can have run-time errors.

To put Quipper on a sound footing, a series of essentialized \textit{Proto-Quipper} fragments~\cite{ross_2015:_algeb_logic_method_quant_comput,fu_2023:_proto_quipp_with_dynam_liftin,rios_selinger_2018:_categ_model_quant,fu_2025:_proto_quipp_with_rever_contr,fu_2020:_tutor_introd_to,colledan_dal_2023:_dynam_liftin_effec,lee_2021:_concr_categ_model} have been formally specified and studied.
Unfortunately, these Proto-Quipper languages remain challenging to reason about using standard programming languages techniques.
Though they provide functional-programming features, their computational model is inherently effectful: program state is given by a circuit prefix, and term evaluation extends this prefix.
Their operational semantics rely on fresh name generation to ensure that circuit wires are uniquely identified, circuits are modelled set-theoretically, and their circuit operations are operationally implemented using set-theoretic operations.
Moreover, safety properties are shown not for programs themselves but for pairs of programs and circuit prefixes that are endowed with their own type system.

We introduce \pqa, a rational reconstruction of \pqm~\cite{rios_selinger_2018:_categ_model_quant,rios_2021:_categ_sound_quant} in terms of familiar programming language concepts.
We focus on \pqm because it provides a core set of features that are extended by other fragments.
Our design of \pqa relies on three key observations.
First, circuits are morphisms in symmetric monoidal categories (SMCs), and the linear \(\lambda\)-calculus is the internal language of SMCs~\cite{abramsky_tzevelekos_2011:_introd_categ_categ_logic,selinger_2011:_survey_graph_languag_monoid_categ}.
This means that we can use it as a syntax for describing compositions of morphisms in SMCs, \ie, compositions of circuits.
Second, name binding and \(\alpha\)-conversion have been used to model fresh names with higher-order abstract syntax (HOAS)~\cite{pfenning_elliott_1988:_higher_order_abstr_syntax}, and we use the binding structure of the linear \(\lambda\)-calculus to easily model fresh names.
Finally, the interaction between quantum and classical systems has long been recognized as an instance of Benton's~\cite{benton_1995:_mixed_linear_non_linear_logic} linear/non-linear (LNL) model~\cite{rennela_staton_2020:_class_contr_quant,rios_selinger_2018:_categ_model_quant,selinger_valiron_2008:_linear_linear_model}.
Adjoint logic~\cite{reed_2009:_judgm_decon_modal_logic,pruiksma_2018:_adjoin_logic} generalizes LNL logic to combine logics with different structural properties, and its computational interpretations similarly combine languages with different structural properties.
We use this adjoint-logical foundation to soundly integrate a linear \(\lambda\)-calculus \emph{qua} circuit syntax language with a linear/non-linear \(\lambda\)-calculus \emph{qua} circuit meta-programming language into a single language: \pqa. %
These three observations allow us to endow \pqa with a \textit{pure} call-by-value reduction semantics, obviating the need to track circuit state, and a \emph{single} type system that closely resembles simple, well-known type systems for functional languages.

Concretely, we make the following contributions:
\begin{enumerate}
\item \textbf{We introduce \pqa}, a rational reconstruction of \pqm with adjoint-logical foundations.
  By eliminating much of the complexity inherent in prior work, we make it easier to reason about and mechanize Proto-Quipper languages using standard techniques.

\item We present a \textbf{simple logical relations proof technique for proving that linear and substructural systems are normalizing}.
  Logical relations for linear systems usually involve splitting and merging substitutions, which are complex to reason about and mechanize.
  We show that it is sufficient to define a structural approximation of our language, and to then use standard logical relations techniques to deduce that our substructural system is normalizing.
  We use this technique to show that \textbf{\pqa is normalizing}, thereby guaranteeing that every generated circuit reduces to a normal form that closely corresponds to its graphical representation.
\item Quantum gate constants in Proto-Quipper languages are uninterpreted function symbols, and we must perform commuting conversions past irreducible gate applications to reduce circuits to normal forms.
  These applications persist in normal forms, making them richer than usual, and logical relation techniques must be adapted to handle them.
  Consequently, we give an \textbf{analysis of commuting conversions and normalization in the presence of uninterpreted function symbols} that can be adapted to other functional languages with uninterpreted function symbols.
\end{enumerate}

To make our contributions accessible, we provide a gentle overview of quantum programming with Proto-Quipper in \cref{sec:gentle-intr-quant}.
We give an overview of \pqa's syntax and semantics in \cref{sec:overview-pqa} before defining its type system and operational semantics in \cref{sec:statics-dynamics}.
We show how to reconstruct \pqm's circuit manipulation abstractions using our more primitive operations in \cref{sec:recov-prot-abstr}.
Finally, we prove normalization in \cref{sec:normalization} before considering related and future work in \cref{sec:related-work,sec:conclusion}.

\section{A Gentle Introduction to Quantum Programming with Proto-Quipper Languages}
\label{sec:gentle-intr-quant}

Quantum computers integrate classical processing units with quantum coprocessors.
The classical processor prepares programs for the quantum coprocessor to execute.
Quantum coprocessors are often modelled as \textit{quantum random access memory} (QRAM) machines, where individual \textit{quantum bits} or \textit{qubits} can be addressed.
Qubits capture probability distributions on classical bits 0 and 1.
The basic operation of quantum processors is to apply \textit{quantum gates} to particular qubits to modify their probability distribution.
After executing a quantum program (a sequence of gate applications), the classical processor accesses its results by measuring individual qubits.
Measuring effectively samples from a qubit's distribution to transform it into a classical bit.

Quantum programs are given by quantum circuits, notated by box-and-wire diagrams whose wires represent addressable qubits and whose boxes represent gates acting on qubits.
We only consider \pqm's \textit{static circuit generation}, where a classical processor generates the entirety of a quantum circuit before execution.
We leave for future work \pqdyn's significantly more complex \textit{circuit generation with dynamic lifting}, where a system executes a circuit prefix, and iteratively extends it based on measurements of executed~prefixes.

Readers unfamiliar with quantum circuit programming are welcome to replace qubit and quantum gate by bit and logic gate throughout this paper.
Indeed, \pqm provides a \textit{circuit generation and description language} for any class of circuits that can be modelled as morphisms in symmetric monoidal categories (intuitively, as acyclic box-and-wire diagrams whose wires can cross~\cite{selinger_2011:_survey_graph_languag_monoid_categ}).

\begin{wrapfigure}{l}{0.5\textwidth}\label{page:cnot-h-figure}
    \begin{tikzpicture}
    \begin{pgfonlayer}{nodelayer}
      \node [style=medium box] (00) {\(\ms{CNOT}\)};
      \node [style=medium box, below=of 00] (10) {\(\ms{H}\)};
      \node [style=medium box, right=of 10] (11) {\(\ms{CNOT}\)};
    \end{pgfonlayer}
    \begin{pgfonlayer}{backlayer}
      \node [fit={(00) (10) (11)}, inner sep=0.5em] (rect) {};
      \coordinate (RE) at ($(rect.east)+(1,0)$);
      \coordinate (LE) at ($(rect.west)+(-1,0)$);
    \end{pgfonlayer}
    \begin{pgfonlayer}{edgelayer}
      \draw [string] (00.wnw -| LE) -- (00.wnw) node [above, near start] {\(x_1\)};
      \draw [string] (00.wsw -| LE) -- (00.wsw) node [above, near start, yshift=-0.2em] {\(x_2\)};
      \draw [string] (10.west -| LE) -- (10.west) node [above, near start] {\(x_3\)};

      \draw [string] (00.ese) to[out=0, in=180, edge node={node [above, xshift=0.3em] {\(y_2\)}}] (11.wnw);
      \draw [string] (10.east) to[out=0, in=180, edge node={node [above] {\(y_3\)}}] (11.wsw);

      \draw [string] (00.ene) -- (00.ene -| RE) node [above, pos=0.9] {\(y_1\)};
      \draw [string] (11.ene) -- (11.ene -| RE) node [above, near end] {\(z_2\)};
      \draw [string] (11.ese) -- (11.ese -| RE) node [above, near end, yshift=-0.2em] {\(z_3\)};
    \end{pgfonlayer}
  \end{tikzpicture}
\caption{Example Circuit \(E\)}
\end{wrapfigure}
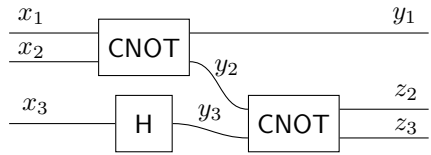
We use string-diagrammatic notation for symmetric monoidal categories \cite{selinger_2011:_survey_graph_languag_monoid_categ} to represent circuits, where diagrams are read from left to right.
For example, the quantum circuit on the left transforms three qubits \(x_1\), \(x_2\), \(x_3\).
It applies a \(\ms{CNOT}\) (controlled NOT) gate to qubits \(x_1\) and \(x_2\), and we use fresh names \(y_1\) and \(y_2\) to identify the resulting qubits to avoid any ambiguity.
It similarly applies a Hadamard gate \(\ms{H}\) to \(x_3\) to produce \(y_3\), and a \(\ms{CNOT}\) gate to qubits \(y_2\) and \(y_3\).
The names \(y_1\), \(z_2\) and \(z_3\) identify the circuit's results.

Proto-Quipper types wires using simple types \(S, U \Coloneqq \PTu \mid \Tq \mid \PTot{S}{U}\), where \(\PTot{S}{U}\) describes pairing. Circuits with inputs \(S\) and outputs \(U\) inhabit the datatype \(\Tcirc{S}{U}\) of boxed circuits.
For example, the \(\ms{CNOT}\) gate inhabits \(\Tcirc{\PTot{\Tq}{\Tq}}{\PTot{\Tq}{\Tq}}\).
Executing Proto-Quipper programs is inherently effectful: program state is given by a circuit prefix, and term evaluation extends this prefix.
Concretely, given a circuit prefix \(C\) with output wires \(x_1, \dotsc, x_n\), evaluating a term \(M\) can extend \(C\) by appending gates or circuits to wires \(x_1, \dotsc, x_n\) to produce a circuit \(C'\) and a value \(V\).
Functional programs extend circuits using the \(\tapply{M}{N}\) operator.
In its simplest form, it takes a tuple \(N : S\) of wire names, evaluates \(M : \Tcirc{S}{U}\) to get a \textit{bona fide} circuit \(C\), appends \(C\) to wires \(N\) in the current state, and produces a term \(\tapply{M}{N} : U\) identifying \(C\)'s output wires in the updated circuit state.
For example, evaluating the following term causes our example circuit \(E\) to be appended to wires \(x_1, x_2, x_3\), and it evaluates to the value \(\ptpair{y_1}{\ptpair{z_2}{z_3}}\) identifying \(E\)'s outputs: \(\ms{let}~\ptpair{y_1}{y_2} = \tapply{\ms{CNOT}}{\ptpair{x_1}{x_2}}~\ms{in}~\ms{let}~y_3 = \tapply{\ms{H}}{x_3}~\ms{in}~\ptpair{y_1}{\tapply{\ms{CNOT}}{\ptpair{y_2}{y_3}}}\).
The circuit boxing operator \(\tbox{S}{M}\) transforms duplicable linear functions \(M : {\Tbang{(S \multimap U)}}\) between simple types into boxed circuits of type \(\Tcirc{S}{U}\).
Functions \(M\) must be linear because physical constraints prohibit the duplication or discarding of qubits.
From a circuit-based perspective, it also ensures that input wires are not duplicated and used to satisfy multiple gate inputs.
Duplicability (captured by the bang modality \(!\)) ensures that the boxed circuit can be reused.

Proto-Quipper's effectful operational semantics is complex.
Though each language variant has its own particularities, they have many commonalities (\cf~\cites[10]{fu_2025:_proto_quipp_with_rever_contr}[175]{rios_selinger_2018:_categ_model_quant}[119\psqq]{ross_2015:_algeb_logic_method_quant_comput}).
They generally assume a circuit constant \(C\) for every possible circuit and model boxed circuits as triples \((\vec l, C, \vec k)\), where \(\vec l\) and \(\vec k\) are tuples of names labelling \(C\)'s input and output wires.
When boxing terms, the operational semantics relies on a meta-level function to generate fresh names to identify the circuit's wires.
Similarly, appending circuits with \(\ms{apply}\) involves a 5-ary function \(\mi{append}(C, \vec j, \vec k, D, \vec l) = (E, \vec m)\).
It first renames \(D\)'s output wires \(\vec l\) to fresh names \(\vec m\), producing a circuit \(D'\).
It connects \(C\)'s output wires \(\vec j\) to \(D'\)'s input wires \(\vec k\) to produce a circuit \(E\).
The reliance on meta-operations makes reasoning about Proto-Quipper's operational semantics complex, and fresh name generation makes mechanization particularly difficult.
Moreover, the absence of concrete syntax for circuits \(C\) makes it difficult for programmers to reason about the circuits produced by their programs, and it impedes future language extensions like meta-programming that rely on inspecting syntax.
Finally, safety properties like subject reduction are not proven for programs themselves, but rather for \textit{configurations} \((C, M)\) of circuits \(C\) being extended by evaluating \(M\).
Configurations have their own complex type system that types \(C\)'s input and output wires, tracks which of \(C\)'s output wires \(M\) is extending, and specifies \(M\)'s type in terms of the wires it extends.
In the next section, we explain how we can use the linear \(\lambda\)-calculus, binding and application instead of gate constants and complex meta-operations to describe circuit composition.

\section{Overview of \pqa}
\label{sec:overview-pqa}

We present an overview of \pqa's syntax, its structures for mediating between functional programs and circuits, and the challenges involved in designing an effect-free operational semantics for circuit generation.

\pqa is formed of two languages: a \blue{functional language}, and a \red{circuit-description language}.
We postpone the discussion of functional features until later.
Quantum circuits \(\red{C}\) are generated in part by the following grammar:
\[
  \red{C}, \red{D} \Coloneqq \red{x} \mid \red{g} \mid \qtlam{x}{\red{C}} \mid \qttriv \mid \qtpair{C}{D} \mid \qtapp{\red{C}}{\red{D}} \mid \qtm{C}{\qpu{D}} \mid \qtm{C}{\qpp{x}{y}{D}}
\]
Circuits are composed of variables \(x\), quantum gate constants \(\red{g}\), functions \(\qtlam{x}{\red{C}}\) and function applications \(\qtapp{C}{D}\), units and pairs \(\qttriv\) and \(\qtpair{C}{D}\), and unit and pair eliminators \(\qtm{C}{\qpu{D}}\) and \(\qtm{C}{\qpp{x}{y}{D}}\) that bind \(\red{x}\) and \(\red{y}\) in \(\red{D}\).
We write \(\qtm{C}{\qpu{D}}\) instead of \(\red{\ms{match}\ C\ \ms{with}\ \qpu{D}}\) because we will frequently encounter nested matches that can become exceedingly wide.
We use matches to bind fresh names to output wires.
For example, \(\qtm{\qtapp{\ms{CNOT}}{\qtpair{x}{y}}}{\qpp{u}{v}{\qtpair{\qtapp{\ms{H}}{u}}{v}}}\) applies \(\ms{CNOT}\) to qubits \(x\) and \(y\), binds the names \(\red{u}\) and \(\red{v}\) to its output wires, and then applies the quantum gate \(\red{\ms{H}}\) to \(\red{u}\).
Quantum gates are uninterpreted function constants of type \(\QTlolly{S}{U}\).
Programmers never write terms in this circuit language: it is solely a syntactic representation of circuits.
Instead, they use the functional language below and the abstractions of \cref{sec:recov-prot-abstr} to generate circuits.

The normal forms \(\red{V}\) of circuits closely match their graphical representation.
They are mutually inductively defined with neutral forms \(\red{R}\) and canonical forms~\(\red{K}\):
\begin{align*}
  \red{V} &\Coloneqq \red{R} \mid \red{K} \mid \qtm{\red{R}}{\qpu{V}} \mid \qtm{\red{R}}{\qpp{x}{y}{V}}\\
  \red{K} &\Coloneqq \qtlam{x}{V} \mid \qttriv \mid \qtpair{V_1}{V_2}\\
  \red{R} &\Coloneqq \red{g} \mid \qtapp{g}{K} \mid \qtapp{g}{R}
\end{align*}
Graphically, canonical forms \(\red{K}\) represent a single normalized circuit \(\qtlam{x}{V}\), a vertical stacking (or \textit{tensoring}) \(\qtpair{V_1}{V_2}\) of two normal circuits, or the tensor unit \(\qttriv\) (an empty diagram).
Neutral terms \(\red{R}\) are stuck terms caused by gates \(\red{g}\), \ie, uninterpreted function constants with no computational content.
Neutrals are either gates \(\red{g}\), a horizontal composition \(\qtapp{g}{K}\) of a gate after a vertical stack of normal circuits, or a horizontal composition \(\qtapp{g}{R}\) of a gate after another neutral circuit.
Normal circuits are formed by a stack of matches that sequence gate applications \(\red{R}\) and name output wires for later use, with a neutral or canonical circuit at the innermost level.
For example, the circuit \(E\) on \cpageref{page:cnot-h-figure} has as one of its normal forms \(\red{W} = \qtm{(\qtapp{\ms{CNOT}}{\qtpair{x_1}{x_2}})}{\qpp{y_1}{y_2}{\qtpair{y_1}{\qtapp{\ms{CNOT}}{\qtpair{y_2}{\qtapp{\ms{H}}{x_3}}}}}}\), assuming that its input wires are labelled \(\red{x_1}, \red{x_2}, \red{x_3}\).\footnote{Though each \pqa program reduces to a unique normal form, each graphical circuit may have multiple normal syntactic representations due to, \eg, reassociations of tensors or compositions that are graphically indistinguishable.}
Its closed canonical form is \(\qtlam{x}{\qtm{x}{\qpp{x_1}{x_{23}}{\qtm{x_{23}}{\qpp{x_2}{x_3}{W}}}}}\).
This function takes in a triple \(x\) of wires, and uses matches to name the individual wires \(\red{x_1}, \red{x_2}, \red{x_3}\) for use in \(\red{W}\).

To normalize circuits with gate constants, we must perform commuting conversions and reduce under binders, \eg, in the body of match-expressions.
Commuting conversions unlock additional redexes.
For example, the normal circuit \(\red{V} = \qtm{\qtapp{\ms{CNOT}}{\qtpair{x_1}{x_2}}}{\qpp{u}{v}{\qtpair{v}{u}}}\) applies \(\red{\ms{CNOT}}\) to two input wires, and then swaps the order of its output wires.
The non-normal term \(\qtm{V}{\qpp{y}{z}{\qtpair{z}{y}}}\) then composes \(\red{V}\) with another wire swap.
Circuit reduction \(\acs\) performs a commuting conversion:
\[
  \qtm{V}{\qpp{y}{z}{\qtpair{z}{y}}} \acs \qtm{\qtapp{\ms{CNOT}}{\qtpair{x_1}{x_2}}}{\qpp{u}{v}{\underline{(\qtm{\qtpair{v}{u}}{\qpp{y}{z}{\qtpair{y}{z}}})}}}
\]
By reducing the underlined subterm under the binder \(\qpp{u}{v}{{\ldots}}\), the term normalizes to:
\[
  \qtm{\qtapp{\ms{CNOT}}{\qtpair{x_1}{x_2}}}{\qpp{u}{v}{\qtpair{u}{v}}},
\]
\ie, an \(\eta\)-expansion of \(\qtapp{\ms{CNOT}}{\qtpair{x_1}{x_2}}\) without the redundant swaps.

\pqa's \blue{functional language} is a \(\lambda\)-calculus whose terms \(\blue{M}\) are generated in part by:
\[
  \blue{M}, \blue{N} \Coloneqq \blue{x} \mid \ftlam{x}{\blue{M}} \mid \fttriv \mid \ftpair{M}{N} \mid \ftapp{M}{N} \mid \ftm{\blue{M}}{\fpu{N}} \mid \ftm{\blue{M}}{\fpp{x}{y}{N}}.
\]
It is endowed with the usual call-by-value semantics that does not reduce under binders.
To manipulate circuits, we extend our grammars with new constructs:
\begin{align*}
  \blue{M}, \blue{N} &\Coloneqq \cdots \mid \ftsusp{\red{C}} \mid \ftm{\red{C}}{\qpu{N}} \mid \ftm{\red{C}}{\qpp{x}{y}{N}}\\
  \red{C} &\Coloneqq \cdots \mid \qtforce{M}
\end{align*}
The term \(\ftsusp{\red{C}}\) boxes a circuit term as a functional value, while \(\qtforce{M}\) unboxes a boxed circuit term.
To facilitate circuit composition from the functional layer, we allow pattern matching on circuit terms.
This lets the functional layer name wires from circuit terms for later use.
Though we provide explicit low-level multi-coloured constructs for manipulating the circuits from the functional layer, they are not used by programmers.
Instead, programmers use Proto-Quipper's \(\blue{\ms{box}}\) and \(\blue{\ms{apply}}\) abstractions that can be implemented in terms of the above elementary operations (see \cref{sec:recov-prot-abstr}), and a library of boxed circuits.

To safely manage boxed quantum resources like individual qubits, the functional language is endowed with an adjoint-logical or LNL-style type system that integrates linear and non-linear programming.
The linear layer ensures that open boxed circuit terms, like individual qubits, are not duplicated, while the non-linear layer allows for the duplication and reuse of closed boxed circuits.
Mediation between these two layers is given by the added constructs
\[
  \blue{M} \Coloneqq \cdots \mid \ftsusp{\blue{M}} \mid \ftforce{M} \mid \ftdown{M} \mid \ftm{\blue{M}}{\fpd{x}{N}}.
\]
Here, \(\ftsusp{\blue{M}}\) boxes a closed linear term as a value in the unrestricted layer, where it can freely be duplicated, while \(\ftforce{M}\) forces the evaluation of a boxed functional term in the linear layer.
Dually, the downshift operator \(\ftdown{M}\) forces the evaluation of an unrestricted term \(\blue{M}\) to a linear value that can be used via the corresponding match eliminator.
Composing downshifting with suspension recovers the usual bang \(!\) operator of linear functional programming languages.

Practically speaking, programmers implement boxed circuits using the blue language.
Given such a term \(\blue{M}\), we evaluate it to a \emph{purely red} circuit normal form by normalizing the term \(\qtforce{M}\).
Contrary to Proto-Quipper, evaluation and normalization in the functional and circuit languages are both given by pure call-by-value reduction semantics, with no notion of state.

\begin{example}
  \label{ex:overview:1}
  The following function \(\blue{M}\) composes two boxed circuits \(\blue{g}\) and \(\blue{f}\):
  \[
    \ftlam{g}{\ftlam{f}{\ftsusp{(\qtlam{x}{\qtapp{(\qtforce{g})}{(\qtapp{(\qtforce{f})}{x})}})}}}.
  \]
  Evaluating \(\blue{M}\) applied to boxed Hadamard and the Pauli Z gates produces a boxed circuit value:
  \(
  \ftapp{\ftapp{M}{(\ftsusp{\red{\ms{Z}}})}}{(\ftsusp{\red{\ms{H}}})} \afs^* \ftsusp{(\qtlam{x}{\qtapp{(\qtforce{\ftsusp{\red{\ms{Z}}}})}{(\qtapp{(\qtforce{\ftsusp{\red{\ms{H}}}})}{x})}})}
  \).
  Unboxing this boxed circuit value \(\blue{N}\) and normalizing it produces a normal circuit composing the two gates: \( \qtforce{\blue{N}} \acs^* \qtlam{x}{\qtapp{\ms{Z}}{(\qtapp{\ms{H}}{x})}} \).
\end{example}

\section{\pqa's Statics and Dynamics}
\label{sec:statics-dynamics}

\pqa is formed of \textit{two} different languages---a functional language and a circuit language---that each have different structural properties: its functional language integrates linear and non-linear programming, while the circuit language is linear.
To safely integrate these different languages and modes of use, \pqa builds on a computational interpretation of adjoint logic, a generalization of Benton's~\cite{benton_1995:_mixed_linear_non_linear_logic} LNL calculus.
This foundation enforces a clean separation between the two languages, and it ensures that linear resources are not misused by the non-linear functional programming features.
We maintain our colour scheme, using \blue{blue} for functional features, \red{red} for circuit features, and black to pun on features in both languages.
We present our types and the full syntax of our terms in \cref{sec:types-their-mode,sec:term-circ-lang}.
\Cref{sec:statics} presents our type system, while our reduction semantics is given \cref{sec:dynamics}.

\subsection{Types and their mode structure}
\label{sec:types-their-mode}

Resources in adjoint-logical languages are associated with \defin{modes} of use \(k\) that specify their structural properties, \ie, whether they can be discarded via weakening or duplicated via contraction.
We use the mode \(\MQ\) for all circuit terms, and modes \(\ML\) and \(\MU\) for functional terms.
The unrestricted mode \(\MU\) enjoys weakening and contraction, while the linear modes \(\ML\) and \(\MQ\) enjoy neither property.
A \defin{mode preorder} specifies permitted interdependencies between resources: a resource at mode \(k\) can only depend on resources at modes \(m \geq k\) with at least as many structural properties, a restriction known as the \defin{independence principle}.
Our mode preorder, generated by \(\MU > \ML\), \(\ML > \MQ\), and \(\MQ > \ML\), ensures that unrestricted terms cannot depend on linear functional or circuit variables.
This prevents, \eg, free linear variables from being embedded in duplicable unrestricted terms.
The cycle between \(\ML\) and \(\MQ\) permits linear quantum terms to be lifted to the linear functional layer, but also for linear functional programs to pattern match against quantum pairs of qubits to facilitate circuit programming.

We let \(m, k\) range over all modes, \(\MF\) over functional modes \(\MU\) and \(\ML\), and we write \(A_m\) for a type \(A\) at mode \(m\).
We lift the mode preorder to types and write \(A_m \geq A_k\) when \(m \geq k\).
Types are given by the following grammar:
\begin{align*}
  \blue{A_\MU}, \blue{B_\MU} &\Coloneqq \FTu[\MU] \mid \FTot[\MU]{A_\MU}{B_\MU} \mid \blue{A_\MU \Tlolly_\MU B_\MU} \mid \blue{\FTus[\ML][\MU]{A_\ML}}\\ %
  \blue{A_\ML}, \blue{B_\ML} &\Coloneqq \FTu[\ML] \mid \FTot[\ML]{A_\ML}{B_\ML} \mid \blue{A_\ML \Tlolly_\ML B_\ML} \mid \FTus[\MQ][\ML]{\red{A_\MQ}} \mid \blue{\FTds[\ML][\MU]{A_\MU}}\\
  \red{A_\MQ}, \red{B_\MQ} &\Coloneqq \red{S_\MQ} \mid \QTlolly[\MQ]{S_\MQ}{U_\MQ}\\
  \red{S_\MQ}, \red{U_\MQ} &\Coloneqq \QTu[\MQ] \mid \QTq \mid \QTot[\MQ]{S_\MQ}{U_\MQ}
\end{align*}
We can form unit types, tensors and linear function spaces at linear modes; the unrestricted functional mode has a unit type, cartesian products, and unrestricted function spaces.
\defin{Upshift types} \(\PTus[k][m]{A_k}\) have suspended terms from
level \(A_k\) as values; our type system ensures that they only contain free variables whose mode is at least \(m\).
The \defin{downshift type} \(\FTds[\ML][\MU]{A_\MU}\) allows unrestricted values to be used in the linear functional layer, and downshifting is left-adjoint to upshifting.
The circuit layer has a restricted function spaces: because we are modelling diagrams, \ie, morphisms in an arbitrary symmetric monoidal category, we cannot assume any monoidal closed structure.
Thus, our function spaces only contain functions between \defin{simple types} \(\red{S_\MQ}\), \ie, tuples of circuit wires.
Though a full adjoint-logical system allows for a downshift from \(\blue{A_\ML}\) to \(\red{A_\MQ}\), we do not include it.
This is because the circuit layer is given and the functional layer is an \textit{a posteriori} meta-language for generating circuits, but downshifting would require values from the functional layer to be embeddable in the circuit layer.

\subsection{Term and circuit syntax}
\label{sec:term-circ-lang}

As explained in \cref{sec:overview-pqa}, terms \(\blue{M}\), circuits \(\red{C}\) and programs \(P\) are generated by:
\begin{align*}
  \blue{M}, \blue{N} &\Coloneqq \blue{x} \mid \ftlam{x}{\blue{M}} \mid \fttriv \mid \ftpair{M}{N} \mid \ftsusp{\blue{M}} \mid \ftsusp{\red{C}} \mid \ftdown{M}\\
  &\grammid \ftapp{M}{N} \mid \ftforce{M} \mid \ftm{\blue{M}}{p_{\MF\MF}} \mid \ftm{\red{C}}{p_{\MQ\MF}}\\
  \blue{p}_{\MF\MF} &\Coloneqq \blue{\fpu{M}} \mid \blue{\fpp{x}{y}{M}} \mid \blue{\fpd{x}{M}}\\
  \blue{p}_{\MQ\MF} &\Coloneqq \blue{\qpu{M}} \mid \blue{\qpp{x}{y}{M}}\\
  \red{C}, \red{D} &\Coloneqq \red{x} \mid \qtlam{x}{\red{C}} \mid \qttriv \mid \qtpair{C}{D} \mid \red{g} \mid \qtapp{\red{C}}{\red{D}} \mid \qtforce{\blue{M}} \mid \qtm{C}{p_{\MQ\MQ}}\\
  \red{p_{\MQ\MQ}} &\Coloneqq \red{\qpu{C}} \mid \red{\qpp{x}{y}{C}}\\
  P, Q &\Coloneqq \blue{M} \mid \red{C}
\end{align*}
Compared to \cref{sec:overview-pqa}, we abstract over patterns with metavariable \(p\) to simplify our type system and operational semantics.
A pattern \(p_{mk}\) eliminates a canonical form from mode \(m\) into a body at mode \(k\), and it binds variables in the obvious way.
We also use the syntax \(\pat{\pi}{P}\) to range over patterns binding variables from \(\pi\) in \(P\).
The syntax makes the interactions between languages clear.
Though functional term \(\ftm{\red{C}}{p_{\MQ\MF}}\) pattern matches against circuit syntax, the red circuit variables bound by \(\blue{p_{\MQ\MF}}\) can ultimately only be used in another pattern match or in suspended circuit term \(\ftsusp{\red{C}}\).
Suspension \(\ftsusp{\red{C}}\) and forcing \(\qtforce{M}\) are the only other interlanguage operations.

\subsection{Statics}
\label{sec:statics}

The typing judgment \(\japq{\Delta}{P}{A}\) means that \(P\) has type \(A\) given typing assumptions \(\Delta = {x_1 : A_1}, \dotsc, {x_n : A_n}\).
It is mutually defined with the judgment \(\japp{\Delta}{p}{A}{B}\), which means that pattern \(p\) eliminates a term of type \(A\) to produce a term of type \(B\).
Adjoint-logical foundations ensure that assumptions are only used in ways that respect their mode's structural properties.
We write \(\Delta \geq k\) to mean that every assumption \(y : B_m\) in \(\Delta\) is at least at mode \(k\), \ie, that \(m \geq k\).
A typing judgment \(\japq{\Delta}{P}{A_k}\) is well formed only if it satisfies the independence principle, \ie, only if \(\Delta \geq k\).
Similarly, \(\japp{\Delta}{p}{A}{B_k}\) is well formed only if \(\Delta \geq k\) and \(A \geq B_k\).
We write \(\Delta_\MU\) to mean \(\Delta_\MU \geq \MU\), \ie, that every assumption is unrestricted.
The type system is parameterized by an implicit global signature \(\Sigma_G\) that specifies the type of each atomic gate \(\red{g}\).
We write \((\red{g} : \QTlolly[\MQ]{U}{S}) \in \Sigma_G\) to mean that gate \(\red{g}\) has type \(\QTlolly[\MQ]{U}{S}\).
Typing judgments are inductively defined by the following rules, where \(\MF\) still ranges over \(\MU\) and \(\ML\).
\begingroup\small
\allowdisplaybreaks
\begin{gatherrules}
  \judgmentbox{\japq{\Delta}{P}{A}}{Term or circuit \(P\) has type \(A\) assuming \(\Delta\)}\\
  \getrule*{a/f/var}
  \qquad
  \getrule*{a/f/lam}
  \\
  \getrule*{a/f/triv}
  \qquad
  \getrule*{a/f/pair}
  \\
  \getrule*{a/f/susp}
  \qquad
  \getrule*{a/f/down}
  \\
  \getrule*{a/f/app}
  \\
  \getrule*{a/f/force}
  \qquad
  \getrule*{a/f/match}
  \\
  \getrule*{a/q/var}
  \qquad
  \getrule*{a/q/lam}
  \\
  \getrule*{a/q/triv}
  \qquad
  \getrule*{a/q/pair}
  \\
  \getrule*{a/q/force}
  \qquad
  \getrule*{a/q/app}
  \\
  \getrule*{a/q/match}
  \quad
  \getrule*{a/q/gate}
\end{gatherrules}
\begin{gatherrules}
  \judgmentbox{\japp{\Delta}{p}{A}{B}}{Pattern \(p\) eliminates terms of type \(A\) to produce terms of type \(B\)}
  \\
  \getrule*{a/p/fpu}
  \quad
  \getrule*{a/p/fpp}
  \\
  \getrule*{a/p/fpd}
  \quad
  \getrule*{a/p/qpu}
  \\
  \getrule*{a/p/qpp}
\end{gatherrules}
\endgroup

The quantum variable rule \getrn{a/q/var} restricts quantum variables to simple types to ensure that we cannot define higher-order quantum circuits.
The variable, unit and gate rules ensure that any unused variables belong to the unrestricted mode that allows weakening.
Similarly, the pairing, function application and matching rules split the conclusion's assumptions \(\blue{\Delta_\MU}, \Delta_1, \Delta_2\) between the two premisses, while also using contraction to allow each premise to access the unrestricted resources \(\blue{\Delta_\MU}\).
When these rules type a term at mode \(\MU\), the independence principle ensures that \(\Delta_1\) and \(\Delta_2\) are empty, giving the usual pairing, application, and matching rules.
The suspension rule and unit pattern matching rules include a mode restriction to ensure the independence principle.
The \getrn{a/f/match} rule handles pattern matching on both circuits and functional terms, and ensures that the pattern body is a functional term.
An induction on typing derivations establishes:

\begin{proposition}[Independence principle]
  If \(\japq{\Delta}{P}{A_m}\), then \(\Delta \geq m\).
\end{proposition}

\subsection{Dynamics}
\label{sec:dynamics}

The functional and circuit languages have their own operational semantics.
The functional language is defined by a conventional call-by-value reduction semantics that appeals to the circuit reduction semantics to reduce pattern-matched circuits.
Similarly, the circuit language also uses a call-by-value reduction semantics, with several key extensions.
As described in \cref{sec:overview-pqa}, closed circuits include functions \(\qtlam{x}{C}\), and we must reduce under binders to reach a normal form that closely resembles graphical representations.
Moreover, the gate constants cause stuck neutral terms, and we must perform commuting conversions to unblock additional redexes during circuit normalization, both in the circuit and the functional language.
As is usual for functional languages, we only execute \textit{closed} programs.
However, our desire to reduce under circuit binders forces us to nuance this stance: we only execute \defin{functionally closed} programs, \ie, programs that have no free functional variables \(\blue{x}\) but that may have free circuit variables \(\red{y}\).

\subsubsection{Neutral contexts}

To reduce open circuit terms, we track in-scope free circuit variables using \defin{neutral variable contexts} \(\Pi\) generated by \( \Pi, \Nu \Coloneqq \cdot \grammid \Pi,\, \jneu{\red{x}} \).
Our dynamics is untyped.
However, several results involve well-typed terms, and \defin{typed neutral variable contexts} \(\Psi\) type neutral variables:
\(
  \Psi, \Phi \Coloneqq {\cdot} \grammid \Psi, \red{x} : \red{U}, \jneu{\red{x}}
\).
As in \cref{sec:statics}, circuit variables only range over simple types \(\red{U}\).
We do not include functional variables because we only reduce functionally closed programs.
We write \(\ctp{\Psi}\) for the typing context given by erasing all neutral variable judgments from \(\Psi\), and \(\cneu{\Psi}\) for the neutral variable context obtained by erasing all type assignments.

\subsubsection{Normal forms}

The judgment \(\jnorm{P}\) means that program \(P\) is in \defin{normal form}, and we range over normal forms with \(V\).
Because we reduce open circuits, we write \(\jnorm[\Pi]{V}\) to mean \(V\) is normal given neutral variables \(\Pi\).
Normal forms are mutually inductively defined with canonical forms and neutral forms, which we explain below.
We recast the grammar defining these term classes in \cref{sec:overview-pqa} in judgmental form to simplify the statement, proof and mechanization of theorems.
We write ``\(\jneu{\pi}\)'' for the empty neutral  context ``\(\cdot\)'' when \(\red{\pi}\) is \(\qttriv\), and for the context ``\(\jneu{x}, \jneu{y}\)'' when \(\red{\pi}\) is \(\qtpair{x}{y}\).
\begingroup\allowdisplaybreaks
\begin{gatherrules}
  \judgmentbox{\jnorm[\Pi]{V}}{Term or circuit \(V\) is normal}\\
  \getrule*{norm/can}
  \quad
  \getrule*{norm/neu}
  \\
  \getrule*{norm/match}
\end{gatherrules}
\begin{gatherrules}
  \judgmentbox{\jcan[\Pi]{K}}{Term or circuit \(K\) is a canonical form}\\
  \getrule*{can/triv}
  \quad
  \getrule*{can/pair}
  \\
  \getrule*{can/lam/f}
  \quad
  \getrule*{can/lam/c}
  \\
  \getrule*{can/susp}
  \quad
  \getrule*{can/down}
\end{gatherrules}
\begin{gatherrules}
  \judgmentbox{\jneu[\Pi]{R}}{Circuit \(R\) is neutral}\\
  \getrule*{neu/var}
  \qquad
  \getrule*{neu/gate}
  \\
  \getrule*{neu/app/neu}
  \qquad
  \getrule*{neu/app/can}
\end{gatherrules}
\endgroup

\defin{Canonical forms} \(\jcan[\Pi]{K}\) are normal forms that consist of a constructor with a potentially fully normalized body.
Functions are always normal in the functional layer, while circuit abstractions are only normal if their body is normal: the binder \(\qtlam{x}{({-})}\) serves only to name the input wires in the circuit body, so should not block normalization.
Canonical forms will play a special role when showing that match eliminators satisfy our logical normalization predicate.
As described in \cref{sec:overview-pqa}, canonical circuits correspond to a vertical stacking of normal circuits.

\defin{Neutral terms} \(\jneu[\Pi]{R}\) arise from gate applications and from reducing under binders.
Neutral variables appear as base cases when normalizing open circuits.
Similarly, gates applied to neutral terms or canonical forms are irreducible because gates are uninterpreted constants.
Compared to most systems with open term reduction, variables cannot appear in head position of neutral terms because circuit variables never have function type.

\defin{Neutral matches} are terms \(\ptm{\red{R}}{p}\) that match on neutral terms.
A neutral match with a normal body is again normal: the match serves only to name the output wires of the gate application, and because the gate application is neutral, the match cannot be eliminated.
Our semantics commutes elimination forms past neutral matches until they reach a canonical form that can be eliminated.
Thus, a gate applied to a normal match is never neutral because we commute the gate into the match body.
For simplicity, canonical forms can contain normal neutral matches: commuting constructors into neutral matches would require checking if the bound variable appears in the neutral term in the case of \getrn{can/lam/c}, and if both components in \getrn{can/pair} are neutral matches, then it would require an arbitrary choice on the commutation order.

Unrestricted terms cannot depend on linear quantum variables, so their normal forms are all canonical and closed.
Normal forms at linear types are elegantly characterized as neutral matches surrounding a canonical or neutral form.
In the case of linear function types, linearity ensures that any circuit variables bound by neutral matches or appearing in the context must be used; typing ensures that they can only occur in other neutral matches or in suspended circuits.
We remark that circuit normal forms \textit{never refer to the functional language}!
\begin{lemma}[Well-Typed Normal Forms]
  \label{lemma:apq:2}
  If \(\jnorm[\cneu{\Psi}]{V}\) is normal, then where \(\red{R}\) ranges over neutral terms such that \(\jneu[\cneu{\Psi}]{R}\),
  \begin{itemize}
  \item if \(\japq{\ctp{\Psi}}{\blue{V}}{\FTu[\MU]}\), then \(\Psi = \cdot\) and \(\blue{V}\) is generated by the grammar
    \(
    \blue{V_{\FTu[\MU]}} \Coloneqq \fttriv
    \)
  \item if \(\japq{\ctp{\Psi}}{\blue{V}}{\FTot[\MU]{A}{B}}\), then \(\Psi = \cdot\) and \(\blue{V}\) is generated by
    \(
    \blue{V_{\FTot[\MU]{A}{B}}} \Coloneqq \ftpair{V_A}{V_B}
    \)
  \item if \(\japq{\ctp{\Psi}}{\blue{V}}{\FTlolly[\MU]{A}{B}}\), then \(\Psi = \cdot\) and \(\blue{V}\) is generated by
    \(
    \blue{V}_{\FTlolly[\MU]{A}{B}} \Coloneqq \ftlam{x}{M}
    \)
  \item if \(\japq{\ctp{\Psi}}{\blue{V}}{\FTus[\ML][\MU]{\blue{A}}}\), then \(\Psi = \cdot\) and \(\blue{V}\) is generated by
    \(
    \blue{V_{\FTus[\ML][\MU]{A}}} \Coloneqq \ftsusp{\blue{M}}
    \)
  \item if \(\japq{\ctp{\Psi}}{\blue{V}}{\FTu[\ML]}\), then \(\blue{V}\) is generated by
    \(
    \blue{V_{\FTu[\ML]}} \Coloneqq \fttriv \mid \ftm{\red{R}}{\qpat{\pi}{V_{\FTu[\ML]}}}
    \)
  \item if \(\japq{\ctp{\Psi}}{\blue{V}}{\FTot[\ML]{A}{B}}\), then \(\blue{V}\) is generated by the grammar
    \(
    \blue{V_{\FTot[\ML]{A}{B}}} \Coloneqq \ftpair{V_A}{V_B} \mid \ftm{\red{R}}{\qpat{\pi}{V_{\FTot[\ML]{A}{B}}}}
    \)
  \item if \(\japq{\ctp{\Psi}}{\blue{V}}{\FTlolly[\ML]{A}{B}}\), then \(\blue{V}\) is generated by the grammar
    \(
    \blue{V}_{\FTlolly[\ML]{A}{B}} \Coloneqq \ftlam{x}{M} \mid \ftm{\red{R}}{\qpat{\pi}{V_{\FTlolly[\ML]{A}{B}}}}
    \)
  \item if \(\japq{\ctp{\Psi}}{\blue{V}}{\FTds[\ML][\MU]{\blue{A}}}\), then \(\Psi = \cdot\) and \(\blue{V}\) is generated by
    \(
    \blue{V_{\FTds[][]{\blue{A}}}} \Coloneqq \ftdown{\blue{V_A}}%
    \)
  \item if \(\japq{\ctp{\Psi}}{\blue{V}}{\FTus[\MQ][\ML]{\red{A}}}\), then \(\blue{V}\) is generated by
    \(
    \blue{V_{\FTus[\MQ][\ML]{\red{A}}}} \Coloneqq \ftsusp{\red{C}} \mid \ftm{\red{R}}{\qpat{\pi}{V_{\FTus[\MQ][\ML]{\red{A}}}}}
    \)
  \item if \(\japq{\ctp{\Psi}}{\red{V}}{\QTu[\MQ]}\), then \(\red{V}\) is generated by
    \(
      \red{V_{\QTu[\MQ]}} \Coloneqq \qttriv \mid \red{R} \mid \qtm{R}{\pat{\pi}{V_{\QTu[\MQ]}}}
    \)
  \item if \(\japq{\ctp{\Psi}}{\red{V}}{\QTq}\), then \(\red{V}\) is generated by
    \(
      \red{V}_{\QTq} \Coloneqq \red{R} \mid \qtm{R}{\pat{\pi}{V_{\QTq}}}
    \)
  \item if \(\japq{\ctp{\Psi}}{\red{V}}{\QTot[\MQ]{U}{S}}\), then \(\red{V}\) is generated by
    \(
      \red{V_{\QTot[\MQ]{U}{S}}} \Coloneqq \qtpair{V_U}{V_S} \mid \red{R} \mid \qtm{R}{\pat{\pi}{V_{\QTot[\MQ]{U}{S}}}}
    \)
  \item if \(\japq{\ctp{\Psi}}{\red{V}}{\QTlolly[\MQ]{U}{S}}\), then \(\red{V}\) is generated by
    \(
    \red{V}_{\QTlolly[\MQ]{U}{S}} \Coloneqq \qtlam{x}{V_S} \mid \red{g} \mid \qtm{R}{\pat{\pi}{V_{\QTlolly[\MQ]{U}{S}}}}
    \).
  \end{itemize}
\end{lemma}

\subsubsection{Reduction}
\label{sec:reduction}

The mutually defined judgments \(\fstep{\Pi}{\blue{M}}{\blue{N}}\) and \(\cstep{\Pi}{\red{C}}{\red{D}}\) respectively give a structural operational semantics to functional terms and circuits.
The circuit reduction semantics assumes a neutral variable context \(\Pi\) so that we can reduce open terms.
Though we only ever reduce functionally closed terms, functional terms can still depend on free circuit variables introduced by naming circuit output wires in terms \(\ftm{\red{R}}{\qpat{\pi}{M}}\).
Thus, they too depend on a neutral variable context.
The judgment \(\fstep{\Pi}{\blue{M}}{\blue{N}}\) is inductively defined by:

\begingroup\small\allowdisplaybreaks
\begin{gatherrules}
  \judgmentbox{\fstep{\Pi}{\blue{M}}{\blue{N}}}{Term \(M\) reduces to \(N\)}
  \\
  \getrule*{fstep/app/1}
  \quad
  \getrule*{fstep/app/2}
  \\
  \getrule*{fstep/app/beta}
  \\
  \getrule*{fstep/app/cc}
  \\
  \getrule*{fstep/force}
  \quad
  \getrule*{fstep/force/1}
  \\
  \getrule*{fstep/force/cc}
  \\
  \getrule*{fstep/pair/1}
  \quad
  \getrule*{fstep/pair/2}
  \\
  \getrule*{fstep/down}
  \quad
  \getrule*{fstep/m/f}
  \\
  \getrule*{fstep/m/k}
  \quad
  \getrule*{fstep/m/q}
  \\
  \getrule*{fstep/m/q/r}
  \\
  \getrule*{fstep/m/f/cc}
  \\
  \getrule*{fstep/m/q/cc}
\end{gatherrules}
\endgroup

\begingroup\small\allowdisplaybreaks
\begin{gatherrules}
  \judgmentbox{\ecan{K}{p}{P}}{Eliminating canonical form \(K\) with pattern \(p\) produces \(P\)}
  \\
  \getrule*{ecan/down}
  \qquad
  \getrule*{ecan/triv}
  \\
  \getrule*{ecan/pair}
\end{gatherrules}
\endgroup

Most rules are standard for a deterministic call-by-value functional language.
For example, \getrn{fstep/app/1}, \getrn{fstep/app/2}, and \getrn{fstep/app/beta} reduce subterms from left to right and then perform a call-by-value \(\beta\)-reduction.
Similarly, \getrn{fstep/m/f} and \getrn{fstep/m/q} serve to reduce a match's discriminee until it is normal.
If it is canonical, then \getrn{fstep/m/k} appeals to an auxiliary judgment \(\ecan{K}{p}{P}\) to eliminate it into the pattern body.
The definition of canonical forms ensures that \getrn{fstep/m/k} only ever substitutes normal forms for variables.
New is \getrn{fstep/m/q/r}, which reduces under a pattern binding a neutral circuit's output wires.
To do so, we must extend the context reducing the body with the neutral variables bound by the pattern.
Also new are the commuting conversions that commute an elimination form \(\ftapp{({-})}{\blue{W}}\), \(\ftforce{({-})}\) and \(\ftm{({-})}{p}\) into the body of a neutral match.
It is required because not all functional normal forms are canonical forms that can be eliminated in the usual way.
However, \cref{lemma:apq:2} ensures that the commuted elimination forms eventually reach a canonical form that can be eliminated.

The reduction rules for circuits are analogous.
We additionally allow reduction under function binders.
We must also commute a gate application past a neutral match.
These two new rules are below; the analogous rules are given in \cref{sec:full-reduct-semant}.
\begingroup\small
\begin{gatherrules}
  \judgmentbox{\cstep{\Pi}{\red{C}}{\red{D}}}{Circuit \(C\) reduces to \(D\)}
  \\
  \getrule*{cstep/lam}
  \\
  \getrule*{cstep/app/cc/2}
\end{gatherrules}
\endgroup

For convenience, we write \(\step{\Pi}{P}{P'}\) if \(\fstep{\Pi}{P}{P'}\) or \(\cstep{\Pi}{P}{P'}\).
We also write \(\ecanop{K}{p}\) for \(P\) when \(\ecan{K}{p}{P}\).

\Cref{prop:apq:3,prop:apq:2} jointly imply that every program reduces to at most one normal form.
We show in \cref{sec:normalization} that every well-typed program reduces to a normal form.

\begin{theoremEnd}{proposition}[Finality]
  \label{conj:apq:9}
  \label{prop:apq:3}
  Judgmentally normal terms do not reduce:
  if \({\jnorm[\Pi]{P}}\), then for no \(P'\) does \(\step{\Pi}{P}{P'}\).
\end{theoremEnd}

\begin{proofEnd}
  By induction on the derivation of \({\jnorm[\cneu{\Psi}]{P}}\).
\end{proofEnd}

\begin{theoremEnd}{proposition}[Determinism]
  \label{prop:apq:2}
  \leavevmode\\
  If \(\step{\Pi}{P}{Q}\) and \(\step{\Pi}{P}{Q'}\), then \(Q = Q'\).
\end{theoremEnd}

\begin{proofEnd}
  By induction on the derivation of \(\fstep{\Pi}{\blue{P}}{\blue{Q}}\) or \(\cstep{\Pi}{\red{P}}{\red{Q}}\).
  In each case, we use \cref{conj:apq:9} to observe that at most one rule is applicable, so the result is immediate.
\end{proofEnd}

\begin{theoremEnd}{proposition}[Subject Reduction]
  \label{conj:apq:4}
  \leavevmode\\
  If \(\japq{\ctp{\Psi}}{P}{A}\) and \(\step{\cneu{\Psi}}{P}{P'}\), then \(\japq{\ctp{\Psi}}{P'}{A}\).
\end{theoremEnd}

\section{Reconstructing Proto-Quipper's Abstractions}
\label{sec:recov-prot-abstr}

\defrule{rs/box}{(\ensuremath{\mi{RS/box}})}{
  \Gamma;\ Q \vdash \tbox S M : \Tcirc{S}{U}
}{
  \Gamma;\ Q \vdash M : {!}(S \multimap U)
}
\defrule{rs/apply}{(\ensuremath{\mi{RS/apply}})}{
  \Phi, \Gamma_1, \Gamma_2;\ Q_1, Q_2 \vdash \tapply{M}{N} : U
}{
  \Phi, \Gamma_1;\ Q_1 \vdash M : \Tcirc{S}{U}
  &
  \Phi, \Gamma_2;\ Q_2 \vdash N : S
}

Functional terms in \pqa may include circuit-language subterms.
However, Proto-Quipper users never write any circuit terms or directly use multilingual operators.
Instead, they restrict themselves to the functional fragment
\[
  \blue{M}, \blue{N} \Coloneqq \blue{x} \mid \ftlam{x}{\blue{M}} \mid \fttriv \mid \ftpair{M}{N} \mid \ftsusp{\blue{M}} \mid \ftdown{M} \mid \ftapp{M}{N} \mid \ftforce{M} \mid \ftm{\blue{M}}{p}
\]
and rely on a library of premade circuits, along with circuit manipulating operators \(\blue{\ms{box}}\) and \(\blue{\ms{apply}}\).
(We refer to \cref{sec:gentle-intr-quant} for an overview of circuit boxing and application.)
We show how to reconstruct these two abstractions, as defined by \textcite[173]{rios_selinger_2018:_categ_model_quant}, using \pqa's primitive multilingual operators.
They type boxing and unboxing using the rules:
\begin{gatherrules}
  \getrule*{rs/box}
  \\
  \getrule*{rs/apply}
\end{gatherrules}
Here, the contexts \(\Phi\) and \(\Gamma\) type parameter variables, while \(Q_i\) types wire names.
The \getrn{rs/box} rule takes a duplicable linear function between simple types and boxes it as a circuit.
Rule \getrn{rs/apply} appends a circuit \(M\) to an argument \(N\) giving circuit wires of shape \(S\), to provide circuit output wires of shape \(U\).

To reconstruct these abstractions, we must first encode Proto-Quipper types as \pqa types.
Proto-Quipper implicitly embeds its wire types into its functional language, where they can be manipulated by functional programs.
In contrast, \pqa explicitly embeds simple circuit types as functional types with an upshift.
Thus, a Proto-Quipper wire \(x : \Tq\) corresponds to a suspension \(\ftsusp{\red{x}} : \FTus[\MQ][\ML]{\QTq}\).
To ensure that programmers can access individual wires in tensors, we must distribute shifts over tensors.
If we were to encode a pair of wires \(\Tq \otimes \Tq\) directly as an upshift \(\FTus[\MQ][\ML]{(\QTot{\Tq}{\Tq})}\), then programmers would be unable to bind its components, as their fragment has no operators for eliminating suspensions.
By distributing the shift over the tensor so that a pair \(\ptpair{x}{y} : \Tq \otimes \Tq\) corresponds to a pair \(\ftpair{\ftsusp{\red{x}}}{\ftsusp{\red{y}}}\), programmers can access each component using the purely functional match eliminator.
The full encoding \(\enc{A}\) of types as linear types is defined by induction on~\(A\), where \(S\) ranges over simple types:
\begin{align*}
  \enc{\PTu} &= \FTus[\MQ][\ML]{\QTu}
  & \enc{\Tq} &= \FTus[\MQ][\ML]{\QTq}
  \\
  \enc{S \otimes U} &= \FTot[\ML]{\enc{S}}{\enc{U}}
  & \enc{\Tcirc{S}{U}} &= \FTus[\MQ][\ML]{(\QTlolly{S}{U})}
  \\
  \enc{A \multimap B} &= \FTlolly[\ML]{\enc{A}}{\enc{B}}
  & \enc{\Tbang{A}} &= \FTds[\ML][\MU]{\FTus[\ML][\MU]{\enc{A}}}
\end{align*}
Our encoding \(\enc{\Tcirc{S}{U}}\) of circuit types is defined as the direct embedding of the circuit function space \(\QTlolly{S}{U}\), and it \emph{does not} involve any encodings of the circuit input or output types.
This encoding ensures that boxed circuits \(\blue{M} : \enc{\Tcirc{S}{U}}\) can easily be converted into normal circuits simply by normalizing \(\qtforce{M}\).
The encoding of linear functions is direct, while the encoding of the bang comonad \(\Tbang{A}\) is given by the its usual adjoint decomposition (\cf~\cite[\S~2.1]{benton_1995:_mixed_linear_non_linear_logic}).
Proto-Quipper contexts \(\Phi, \Gamma; Q\) are encoded pointwise in the obvious manner.

Our encoding of \(\otimes\) recursively distributes the shift over tensors.
However, to apply a circuit to a tuple of wires, \eg, a term of type \(\FTot[\ML]{\FTus[\MQ][\ML]{\QTq}}{\FTus[\MQ][\ML]{\QTq}}\), we must undistribute the shift to get a term of type \(\FTus[\MQ][\ML]{(\QTot{\QTq}{\QTq})}\) that we can force back down to the circuit layer.
We do so with two closed helper functions:
\begin{align*}
  \ftlax{S,U} &: \FTlolly{
                          \FTot[\ML]{(\FTus[\MQ][\ML]{\red{S}})}{(\FTus[\MQ][\ML]{\red{U}})}
                          }{
                          \FTus[\MQ][\ML]{(\red{\QTot{S}{U}})}
                          }
  \\
  \ftlax{S,U} &= \ftlam{x}{\ftm{x}{\fpp{a}{b}{\ftsusp{\qtpair{\qtforce{a}}{\qtforce{b}}}}}}
  \\
  \ftoplax{S,U} &: \FTlolly{
                            \FTus[\MQ][\ML]{(\red{\QTot{S}{U}})}
                            }{
                            \FTot[\ML]{(\FTus[\MQ][\ML]{\red{S}})}{(\FTus[\MQ][\ML]{\red{U}})}
                            }
  \\
  \ftoplax{S,U} &= \ftlam{x}{\ftm{(\qtforce{x})}{\qpp{a}{b}{\ftpair{\ftsusp{\red{a}}}{\ftsusp{\red{b}}}}}}
\end{align*}
It is the definition of \(\blue{\ms{oplax}}\) that motivates the inclusion of pattern matching on circuits from the functional language.
Checking that these implementations have their specified types is routine.
For each simple type \(S\), it is now straightforward to define the corresponding functions \( \ftlax{S} : \FTlolly{\enc{S}}{\FTus[\MQ][\ML]{\red{S}}}\) and \(\ftoplax{S} : \FTlolly{\FTus[\MQ][\ML]{\red{S}}}{\enc{S}}\).

We encode \(\ms{box}\) and \(\ms{apply}\) as the following closed higher-order functions, whose domains and codomains correspond to the encodings of the types appearing in \getrn{rs/box} and \getrn{rs/apply}.
The rules can then be encoded using \getrn{a/f/app}.
\begin{align*}
  \blue{\ms{BOX}}_{S,U} &: \FTlolly[\ML]{
                          (\FTlolly[\ML]{\enc{S}}{\enc{U}})
                          }{
                          \FTus[\MQ][\ML]{(\QTlolly[\MQ]{S}{U})}
                          }
  \\
  \blue{\ms{BOX}}_{S,U} &= \ftlam{f}{\ftsusp{(
                          \qtlam{s}{\qtforce{(
                          \ftapp{\ftlax{U}}
                          {(\ftapp{f}
                          {(\ftapp{\ftoplax{S}}
                          {(\ftsusp{\red{s}})}
                          )}
                          )}
                          )}
                          )})}}
  \\
  \ftbox_{S,U} &:
                 \FTlolly[\ML]{
                 \FTds[\ML][\MU]{\FTus[\ML][\MU]{(\FTlolly[\ML]{\enc{S}}{\enc{U}}})}
                 }{
                 \FTus[\MQ][\ML]{(\QTlolly{S}{U})}
                 }
  \\
  \ftbox_{S,U} &= \ftlam{x}{\ftm{\blue{x}}{\fpd{f}{\ftapp{\blue{\ms{BOX}}_{\black{S,U}}}{(\ftforce{x})}}}}
  \\
  \ftapply_{S,U} &: \FTlolly[\ML]{
                   \FTus[\MQ][\ML]{(\QTlolly{S}{U})}
                   }{
                   (\FTlolly[\ML]{\enc{S}}{\enc{U}})
                   }
  \\
  \ftapply_{S,U} &= \ftlam{f}{\ftlam{s}{
                   \ftapp{\ftoplax{U}}
                   {(\ftsusp{(
                   \qtapp{
                   (\qtforce{f})
                   }{
                   (\qtforce{(
                   \ftapp{\ftlax{S}}{s})})
                   }
                   )}
                   )}
                   }}
\end{align*}
The helper function \(\blue{\ms{BOX}}_{S,U}\) boxes a linear function \(\blue{f} : \FTlolly[\ML]{\enc{S}}{\enc{U}}\) as a suspended circuit \(\FTus[\MQ][\ML]{(\QTlolly[\MQ]{S}{U})}\).
The circuit is defined by distributing the shifts on \(\ftsusp{\red{s}} : \FTus[\MQ][\ML]{\red{S}}\) to get an argument for \(\blue{f}\), and then undistributing shifts on its image to get a value of type \(\FTus[\MQ][\ML]{\red{U}}\) that is forced back down to the circuit layer.
The function \(\ftbox{}\) simply eliminates the shifts that ensure the function being boxed is duplicable before passing it to \(\blue{\ms{BOX}}_{S,U}\).
The definition of \(\ftapply_{S,U}\) is analogous.

\section{Normalization}
\label{sec:normalization}

We show that every well-typed term reduces to a normal form.
This not only ensures that every circuit term reduces to a normal form that closely corresponds to its graphical representation, but also illustrates the tractability of \pqa as a foundation for Proto-Quipper languages.
A term normalizes or \defin{halts} under neutral context \(\Pi\), written \(\jhalts{\Pi}{P}\), if:
\defrule{halts}{\arn{halts}}{
  \jhalts{\Pi}{P}
}{
  \steps{\Pi}{P}{V}
  &
  \jnorm[\Pi]{V}
}
\begin{gatherrules}
  \judgmentbox{\jhalts{\Pi}{P}}{Term \(P\) halts under neutral context \(\Pi\)}
  \\
  \getrule*{halts}
  \quad
  \\
  \judgmentbox{\steps{\Pi}{P}{Q}}{\(P\) reflexively or transitively steps to \(Q\) under neutral context \(\Pi\)}
  \\
  \getrule*{steps/refl}
  \quad
  \getrule*{steps/step}
\end{gatherrules}

We show that our system is normalizing using a Kripke logical relation.
Logical relations are a standard proof technique for showing that a rewriting system for a statically typed language is normalizing.
It involves defining a type-indexed family of predicates by induction on the structure of types, such that a term satisfies the logical predicate for type \(A\) only if it reduces to a normal form of type \(A\).
The fundamental theorem of logical relation implies that every well-typed term satisfies its corresponding logical predicate, \ie, that every well-typed term normalizes.
However, its statement is more general: it says that for every substitution \(\sigma\) of terms that satisfy the predicate (including the identity substitution), if a program \(P\) is well-typed, then \(\apprs{\sigma}{P}\) satisfies the logical predicate.
The fundamental theorem is proven by induction on typing derivations, but because its statement involves a universal quantification on substitutions, in order to appeal to the induction hypotheses, we must first split substitutions in inductive cases like \getrn{a/f/pair} where the context in the conclusion is a merge of the contexts of the premisses.
This creates significant complexity.

A simple observation avoids this complexity altogether: normalization is a property of a rewriting system, and our rewriting systems are untyped.
More importantly, they are agnostic on whether they reduce terms typed by a substructural or a structural type system.
Since logical relations for structural type systems are well understood and do not involve splitting and merging substitutions, we prove the fundamental theorem of logical relations for a structural (unrestricted) approximation of our substructural system.
We define our structural approximation so that every substructurally well-typed term is structurally well typed.
It follows that every substructurally well-typed term is also normalizing.

This approximate system, \pqx, has the same syntax and operational semantics as \pqa.
Its type system is given by erasing all mode restrictions on the rules of \cref{sec:statics}, including the independence principle, and by preserving contexts instead of splitting them.
We give a representative sample of the rules that inductively define its typing judgment, \(\jxpq{\Delta}{P}{A}\), and we refer to \cref{sec:full-definition-pqx} for the full listing.

\begingroup\small
\begin{gatherrules}
  \judgmentbox{\jxpq{\Delta}{P}{A}}{Term or circuit \(P\) approximately has type \(A\)}\\
  \getrule*{x/f/var}
  \qquad
  \getrule*{x/f/pair}
  \\
  \getrule*{x/f/susp}
  \quad
  \getrule*{x/q/gate}
\end{gatherrules}
\begin{gatherrules}
  \judgmentbox{\jxpq{\Delta}{p}{\pat{A}{B}}}{Pattern \(p\) eliminates terms approximately of type \(A\) to produce terms approximately of type \(B\)}
  \\
  \getrule*{x/p/fpu}
  \quad
  \getrule*{x/p/fpp}
\end{gatherrules}
\endgroup

There is a structure-preserving encoding of \pqa typing derivations in \pqx.
It is defined by recursion on the derivation, mapping each rule to its approximate counter-part, and weakening premisses as needed:

\begin{proposition}
  \label{prop:apq:1}
  If \(\japq{\Delta}{P}{A}\), then \(\jxpq{\Delta}{P}{A}\).
\end{proposition}

\Cref{conj:apq:4,prop:apq:1} imply that the approximation of our substructural system enjoys subject reduction.
However, our logical predicates are defined on intrinsically approximately typed terms, and they require that \pqx as a whole enjoys subject reduction:

\begin{proposition}[Subject Reduction]
  \label{conj:apq:1}
  If \(\jxpq{\ctp{\Psi}}{P}{A}\) and \(\step{\cneu{\Psi}}{P}{P'}\), then \(\jxpq{\ctp{\Psi}}{P'}{A}\).
\end{proposition}

\subsection{The Logical Normalization Predicate}
\label{sec:logic-norm-pred}

Our normalization predicate is a type-indexed predicate \( \LR{A} \subseteq \Set{ {\tminctx{\Psi}{P}} \given {\jxpq{\ctp{\Psi}}{P}{A}} } \) on programs-in-context.
It is immediate from the definition that the predicate only contains functionally closed programs (\(\Psi\) only types circuit variables), but that programs can contain free circuit variables.
To handle these open programs, we use a \textit{Kripke} logical predicate.
Kripke logical predicates must be closed under weakening, and more generally, under \defin{neutral substitutions}:
\begin{gatherrules}
  \judgmentbox{\jnctx{\sigma}{\Psi}{\Phi}}{\(\sigma\) is a neutral substitution from \(\Psi\) to \(\Phi\)}
  \\
  \getrule*{ns/empty}
  \qquad
  \getrule*{ns/neu}
\end{gatherrules}

The key feature of neutral substitutions compared to arbitrary substitutions is that they preserve the dynamics of our language:

\begin{theoremEnd}{lemma}[Preservation of dynamics by neutral substitution]
  \label{lemma:apq:6}
  \leavevmode\\
  For all \(\jnctx{\sigma}{\Psi}{\Phi}\),
  \begin{enumerate}
  \item if \(\jneu[\cneu{\Phi}]{R}\), \(\jcan[\cneu{\Phi}]{K}\), or \(\jnorm[\cneu{\Phi}]{V}\), then \(\jneu[\cneu{\Psi}]{\apprs{\sigma}{R}}\), \(\jcan[\cneu{\Psi}]{\apprs{\sigma}{K}}\), or \(\jnorm[\cneu{\Psi}]{\apprs{\sigma}{V}}\), respectively;
  \item if \(\step{\cneu{\Phi}}{P}{P'}\), then \(\step{\cneu{\Psi}}{\apprs{\sigma}{P}}{\apprs{\sigma}{P'}}\).
  \end{enumerate}
\end{theoremEnd}

\begin{proofEnd}
  The first part is by induction on the derivation of \(\jneu[\cneu{\Phi}]{R}\), \(\jcan[\cneu{\Phi}]{K}\), or \(\jnorm[\cneu{\Phi}]{V}\).
  The second part is by induction on the derivation \(\step{\cneu{\Phi}}{P}{P'}\), relying on the first part whenever a rule depends on a normality or neutrality premise.
\end{proofEnd}

\begin{theoremEnd}{corollary}[Closure under composition]
  \label{cor:apq:4}
  If \(\jnctx{\sigma}{\Sigma}{\Phi}\) and \(\jnctx{\rho}{\Phi}{\Psi}\), then \(\jnctx{\apprs{\sigma}{\rho}}{\Sigma}{\Psi}\) (where \(\apprs{\sigma}{\rho}\) applies the neutral substitution \(\sigma\) point-wise to \(\rho\)).
\end{theoremEnd}

\begin{proofEnd}
  By induction on \(\jnctx{\rho}{\Phi}{\Psi}\) using \cref{lemma:apq:6}.
\end{proofEnd}

A key feature of logical relations is that they are defined by induction on the types indexing them, such that they define a desirable property, \eg, normalization, at higher-type in terms of that property at lower type.
Generally, a logical predicate for normalization at a given type is defined by closing its introduction forms (for positive types like \(\PTu\), \(\PTot{A}{B}\) and \(\FTds[\ML][\MU]{A}\)) and elimination forms (for negative types like \(\PTlolly{A}{B}\) and \(\PTus[][]{A}\)) under head expansion.
Normal neutral matches mean that it is no longer sufficient to consider only introduction forms at positive types.
We generalize the above recipe using \cref{lemma:apq:2}'s characterization of normal forms as canonical or neutral forms closed under neutral matching.

We start with the usual definition of a Kripke logical termination predicate.
In the case of positive types, we include all canonical forms by including constructors applied to normalizing terms (\getrn{llr/triv}, \getrn{llr/pair}, and \getrn{llr/ds}).
In the case of negative types, we include terms closed under the corresponding elimination forms (\getrn{llr/fun} and \getrn{llr/us}).
Neutral forms are already normal, and rule \getrn{llr/neu} closes each predicate under neutral forms to ensure that, \eg, \(\llr{\qtp{x}{\QTot{\QTq}{\QTq}}}{x}{\QTot{\QTq}{\QTq}}\).
The rule \getrn{llr/steps} closes each predicate under expansion.
Though this closure is often baked into the predicate for each type, factoring it out simplifies proofs.
To avoid nested inductions on sequences of steps in proofs, \getrn{llr/steps} only extends the relation by one step at a time.
Finally, we close each predicate under neutral matches using \getrn{llr/m}.
For convenience, we write ``\(\qtp{\pi}{S}\)'' for the empty context ``\(\cdot\)'' when \(\qtp{\pi}{S}\) is \(\qtp{\qttriv}{\QTu}\), and for the context ``\(\jneu{x_1}, \qtp{x_1}{S_1}, \jneu{x_2}, \qtp{x_2}{S_2}\)'' when \(\qtp{\pi}{S}\) is \(\qtp{\qtpair{x_1}{x_2}}{\QTot{S_1}{S_2}}\).

\begingroup\small
\begin{gatherrules}
  \judgmentbox{\llr{\Psi}{P}{A}}{Term-in-context \(\tminctx{\Psi}{P}\) satisfies the Kripke logical normalization predicate for type \(A\)}
  \\
  \getrule*{llr/triv}
  \quad
  \getrule*{llr/pair}
  \\
  \getrule*{llr/fun}
  \\
  \getrule*{llr/ds}
  \quad
  \getrule*{llr/us}
  \\
  \getrule*{llr/neu}
  \\
  \getrule*{llr/steps}
  \\
  \getrule*{llr/m}
\end{gatherrules}
\endgroup

\Cref{lemma:apq:3} shows that our logical predicate captures its intended semantic property: normalization.
Its proof is by induction on the derivation of \(\llr{\Psi}{P}{A}\), and it relies on the following lemma in the case of \getrn{llr/us}:

\begin{theoremEnd}{lemma}
  \label{lemma:apq:9}
  If \(\jhalts{\Pi}{(\ptforce{\blue{M}})}\), then \(\jhalts{\Pi}{\blue{M}}\).
\end{theoremEnd}

\begin{proofEnd}
  By inversion on \(\jhalts{\Pi}{(\ptforce{\blue{M}})}\), we get \(\steps{\Pi}{\ptforce{\blue{M}}}{V}\) such that \(\jnorm[\Pi]{V}\).
  We proceed by induction on \(\steps{\Pi}{\ptforce{\blue{M}}}{V}\) to show that \(\steps{\Pi}{P}{W}\) for some \(\jnorm[\Pi]{W}\).
  \begin{proofcases}
  \item[\getrn{steps/refl}] This case is impossible because \(\ptforce{\blue{M}}\) is never normal.
  \item[\getrn{steps/step}] We deduce:
    \begin{enumerate}
    \item  \(\steps{\Pi}{\ptforce{\blue{M}}}{V}\)\hfill rule conclusion
    \item \(\step{\Pi}{\ptforce{\blue{M}}}{Q}\) \hfill rule premise\label{item:apq:338}
    \item \(\steps{\Pi}{Q}{V}\) \hfill rule premise\label{item:apq:339}
    \end{enumerate}
    We proceed by case analysis on the rule used in the derivation of \cref{item:apq:338}.
    The only possibilities are:
    \begin{proofcases}
    \item[\getrn{fstep/force}] Then \(P = \ftsusp{M}\) is normal by \getrn{norm/can} and \getrn{can/susp}, and the result is immediate by reflexivity.
    \item[\getrn{fstep/force/1}] We deduce:
      \begin{enumerate}[label=(\alph*)]
      \item \(\fstep{\Pi}{\ftforce{M}}{\ftforce{N}}\)\hfill rule conclusion
      \item \(\fstep{\Pi}{\blue{M}}{\blue{N}}\) \hfill rule premise\label{item:apq:18}
      \item \(\fsteps{\Pi}{\ftforce{N}}{\blue{V}}\)\hfill \cref{item:apq:339}\label{item:apq:22}
      \item \(\fsteps{\Pi}{\blue{N}}{\blue{W}}\) for some \(\jnorm[\Pi]{\blue{W}}\)\hfill IH for \cref{item:apq:22}\label{item:apq:23}
      \item \(\fsteps{\Pi}{\blue{M}}{\blue{W}}\) for some \(\jnorm[\Pi]{\blue{W}}\)\hfill transitivity, \cref{item:apq:18,item:apq:23}
      \end{enumerate}
    \item[\getrn{fstep/force/cc}] Then \(P\) is normal by the rule premise, so we are done by reflexivity.
    \item[\getrn{cstep/force}] Analogous to \getrn{fstep/force}.
    \item[\getrn{cstep/force/1}] Analogous to \getrn{fstep/force/1}.
    \item[\getrn{cstep/force/cc}] Analogous to \getrn{fstep/force/cc}.
      \qedhere
    \end{proofcases}
  \end{proofcases}
\end{proofEnd}

\begin{theoremEnd}{proposition}
  \label{lemma:apq:3}
  If \(\llr{\Psi}{P}{A}\), then \(\jhalts{\cneu{\Psi}}{P}\).
\end{theoremEnd}

\begin{proofEnd}
  By induction on the derivation of \(\llr{\Psi}{P}{A}\) and \(\llr{\Psi}{\qpat{\pi}{P}}{\qpat{A}{B}}\).
  We implicitly perform inversion on \(\jhalts{\cneu{\Psi}}{P}\) to extract the reduction sequence and normality proof.
  \begin{proofcases}
  \item[\getrn{llr/triv}] Immediate by \getrn{norm/can} and \getrn{can/triv}.
  \item[\getrn{llr/pair}] By induction on the induction hypotheses, using \getrn{fstep/pair/1} and \getrn{fstep/pair/2} in the functional case, and \getrn{cstep/pair/1} and \getrn{cstep/pair/2} in the circuit case, to generate the reduction sequence.
    Normality is given by the induction hypotheses, \getrn{norm/can} and \getrn{can/pair}.
  \item[\getrn{llr/fun}] Immediate by the rule premise.
  \item[\getrn{llr/ds}] By induction on the induction hypothesis, using \getrn{fstep/down} to generate the reduction sequence, and \getrn{norm/can} and \getrn{can/down} to show normality.
  \item[\getrn{llr/us}] By induction on the induction hypothesis and \cref{lemma:apq:9}.
  \item[\getrn{llr/neu}] Immediate by reflexivity and \getrn{norm/neu}.
  \item[\getrn{llr/m}] By induction on the induction hypothesis, using \getrn{fstep/m/q/r} or \getrn{cstep/m/r} as appropriate to generate the reduction sequence, and \getrn{norm/match} to show normality.
  \item[\getrn{llr/steps}] Immediate by definition of transitive closure and the induction hypothesis.\qedhere
  \end{proofcases}
\end{proofEnd}

\subsection{Required Closure Properties and Kripke-hood}

For the logical predicate \(\LR{A}\) to be usable, we must ensure that it is closed under expansion and reduction, and for it to be a \textit{bona fide} Kripke logical predicate, it must be closed under neutral substitutions.
We show each property in turn.
Closure under expansion follows easily from \getrn{llr/steps}:

\begin{theoremEnd}{lemma}[Closure under expansion]
  \label{lemma:apq:16}
  \leavevmode\\
  If \(\steps{\cneu{\Psi}}{P}{Q}\) and \(\llr{\Psi}{Q}{A}\), then \(\llr{\Psi}{P}{A}\).
\end{theoremEnd}

\begin{proofEnd}
  By induction on the derivation of \(\steps{\cneu{\Psi}}{P}{Q}\) using \getrn{llr/steps} in the inductive case.
\end{proofEnd}

Closure under reduction, below, is also straightforward.
It involves an induction on \(\llr{\Psi}{P}{A}\), and the only interesting case is \getrn{llr/steps}.
In this case, we rely on determinism (\cref{prop:apq:2}) to show that \(P\) cannot make a different reduction step that would take it out of the logical relation.

\begin{theoremEnd}{lemma}[Closure under reduction]
  \label{lemma:apq:4}
  \leavevmode\\
  If \(\llr{\Psi}{P}{A}\) and \(\steps{\cneu{\Psi}}{P}{Q}\), then \(\llr{\Psi}{Q}{A}\).
\end{theoremEnd}

\begin{proofEnd}
  By induction on the derivation \(\steps{\cneu{\Psi}}{P}{Q}\).
  The base cases \getrn{steps/refl} are immediate.
  Assume next that \(\steps{\cneu{\Psi}}{P}{Q}\) because \(\step{\cneu{\Psi}}{P}{P'}\) and \(\steps{\cneu{\Psi}}{P'}{Q}\).
  We proceed by induction on the derivation of \(\llr{\Psi}{P}{A}\) to show that \(\llr{\Psi}{P'}{A}\).
  The result will then follow by the outer induction hypothesis for \(\steps{\cneu{\Psi}}{P'}{Q}\).
  \begin{proofcases}
  \item[\getrn{llr/triv}] By finality (\cref{conj:apq:9}) with \getrn{norm/can} and \getrn{can/triv}, this case is impossible.
  \item[\getrn{llr/pair}] By inversion, \(\step{\cneu{\Psi}}{P}{P'}\) must have been by one of the rules \getrn{fstep/pair/1}, \getrn{fstep/pair/2}, \getrn{cstep/pair/1} and \getrn{cstep/pair/2}.
    The result is immediate by the inner induction hypothesis and \getrn{llr/pair}.
  \item[\getrn{llr/fun}] By determinism (\cref{prop:apq:2}), the preservation of reduction by neutral substitutions (\cref{lemma:apq:6}), and the induction hypothesis.
  \item[\getrn{llr/ds}] By inversion, \(\step{\cneu{\Psi}}{P}{P'}\) must have been by \getrn{fstep/down}.
    The result is immediate by the inner induction hypothesis and \getrn{llr/pair}.
  \item[\getrn{llr/us}] By the assumption that \(\step{\cneu{\Psi}}{P}{P'}\) and \getrn{fstep/force/1} or \getrn{cstep/force/1} (as appropriate), we get \(\step{\cneu{\Psi}}{\ptforce{P}}{\ptforce{P'}}\).
    By the induction hypothesis on \(\llr{\Psi}{\ptforce{P}}{A}\), we get \(\llr{\Psi}{\ptforce{P'}}{A}\).
    We deduce \(\llr{\Psi}{P'}{\FTus[][]{A}}\) as desired.
  \item[\getrn{llr/neu}] By finality (\cref{conj:apq:9}) with \getrn{norm/neu}, this case is impossible.
  \item[\getrn{llr/m}] By finality (\cref{conj:apq:9}),  the neutral term \(\red{R}\) being eliminated cannot step, so by inversion, the step must have been by \getrn{fstep/m/q/r} or \getrn{cstep/m/r}.
    The result then follows by the induction hypothesis and \getrn{llr/m}.
  \item[\getrn{llr/steps}] We deduce:
    \begin{enumerate}
    \item \(\llr{\Psi}{P}{A}\)\hfill conclusion
    \item \(\step{\cneu{\Psi}}{P}{Q}\)\hfill premise\label{item:apq:169}
    \item \(\llr{\Psi}{Q}{A}\)\hfill premise\label{item:apq:170}
    \item \(\step{\cneu{\Psi}}{P}{P'}\)\hfill outer induction case\label{item:apq:342}
    \item \(Q = P'\)\hfill \cref{prop:apq:2,item:apq:169,item:apq:342}\label{item:apq:343}
    \item \(\llr{\Psi}{P'}{A}\)\hfill \cref{item:apq:343,item:apq:170}
    \end{enumerate}
  \end{proofcases}
\end{proofEnd}

The fundamental theorem of logical relations states that not only do all well-typed terms satisfy the logical normalization predicate, but also that the relation is closed under substitution by terms in the relation.
Because our language is call-by-value, variables range over functionally closed normal forms instead of arbitrary terms, and we need only consider substitutions of functionally closed normal forms for variables.
\defin{Functionally closed substitutions} \(\jssub{\sigma}{\Psi}{\Delta}\) are inductively defined by:
\begin{gatherrules}
  \judgmentbox{\jssub{\sigma}{\Psi}{\Delta}}{\(\sigma\) is a functionally closed substitution from \(\Psi\) to \(\Delta\)}
  \\
  \getrule*{x/cs/empty}
  \qquad
  \getrule*{x/cs/tm}
\end{gatherrules}

We lift our logical normalization predicate pointwise from terms to substitutions.
The context-indexed family \( \LR{\Delta} \subseteq \Set{ \tminctx{\Psi}{\sigma} \given \jssub{\sigma}{\Psi}{\Delta} } \) of \defin{reducible substitutions} is inductively defined by the following rules.
We implicitly assume that the conclusions are well formed, \ie, that \(\jnorm[\cneu{\Psi}]{V}\) in \getrn{llr/rs/tm}:
\begin{gatherrules}
  \judgmentbox{\llr{\Psi}{\sigma}{\Delta}}{\(\sigma\) is a reducible substitution from \(\Psi\) to \(\Delta\)}
  \\
  \getrule*{llr/rs/nil}
  \quad
  \getrule*{llr/rs/tm}
\end{gatherrules}

Our logical predicate is monotone, so a \textit{bona fide} Kripke logical predicate:

\begin{theoremEnd}{lemma}[Monotonicity]
  \label{lemma:apq:5}\leavevmode
  \begin{enumerate}
  \item If \(\llr{\Phi}{P}{A}\) and \(\jnctx{\sigma}{\Psi}{\Phi}\), then \(\llr{\Psi}{\apprs{\sigma}{P}}{A}\).
  \item If \(\llr{\Phi}{\rho}{\Delta}\) and \(\jnctx{\sigma}{\Psi}{\Phi}\), then \(\llr{\Psi}{\apprs{\sigma}{\rho}}{\Delta}\).
  \end{enumerate}
\end{theoremEnd}

\begin{proofEnd}
  We prove the first point by induction on induction on the derivation of \(\llr{\Phi}{P}{A}\).
  \begin{proofcases}
  \item[\getrn{llr/triv}] Immediate by \getrn{llr/triv}.
  \item[\getrn{llr/pair}] Immediate by the induction hypotheses and \getrn{llr/pair}.
  \item[\getrn{llr/fun}] We show that  \(\llr{\Psi}{\apprs{\psi}{P}}{\PTlolly{A}{B}}\) for an arbitrary \(\jnctx{\psi}{\Psi}{\Phi}\):
    \begin{enumerate}
    \item \(\llr{\Phi}{P}{\PTlolly{A}{B}}\)\hfill rule conclusion
    \item \(\forall \jnctx{\psi}{\Psi}{\Phi} \mathrel{.} \forall \llr{\Psi}{Q}{A} \mathrel{.} \llr{\Psi}{\ptapp{(\apprs{\psi}{P})}{N}}{B}\)\hfill rule premisses\label{item:apq:144}
    \item \(\jhalts{\cneu{\Phi}}{P}\)\hfill rule premise\label{item:apq:344}
    \item \(\jnctx{\psi}{\Psi}{\Phi}\)\hfill lemma assumption\label{item:apq:9}
    \item \(\jhalts{\cneu{\Psi}}{\apprs{\psi}{P}}\)\hfill induction on \cref{item:apq:344}'s premisses using \cref{lemma:apq:6}\label{item:apq:345}
    \item \(\jnctx{\sigma}{\Sigma}{\Psi}\)\hfill assumption for rule's universal quantification\label{item:apq:10}
    \item \(\llr{\Sigma}{Q}{A}\)\hfill assumption for rule's universal quantification\label{item:apq:13}
    \item \(\jnctx{\apprs{\sigma}{\psi}}{\Sigma}{\Phi}\)\hfill\cref{item:apq:9,item:apq:10,cor:apq:4}\label{item:apq:12}
    \item \(\llr{\Sigma}{\ptapp{(\apprs{\apprs{\sigma}{\psi}}{P})}{N}}{B}\)\hfill\cref{item:apq:144,item:apq:12,item:apq:13}\label{item:apq:14}
    \item \(\apprs{\apprs{\sigma}{\psi}}{P} = \apprs{\sigma}{(\apprs{\psi}{P})}\)\hfill substitution properties\label{item:apq:15}
    \item \(\llr{\Sigma}{\ptapp{(\apprs{\sigma}{(\apprs{\psi}{P})}}){Q}}{B}\)\hfill\cref{item:apq:14,item:apq:15}\label{item:apq:145}
    \item \(\llr{\Psi}{\apprs{\psi}{P}}{\PTlolly{A}{B}}\)\hfill\getrn{llr/fun}, \cref{item:apq:10,item:apq:13,item:apq:145,item:apq:345}
    \end{enumerate}
  \item[\getrn{llr/ds}] Immediate by the induction hypothesis and \getrn{llr/ds}.
  \item[\getrn{llr/us}] Immediate by the induction hypothesis and \getrn{llr/us}.
  \item[\getrn{llr/steps}] This case is immediate by the induction hypothesis, \cref{lemma:apq:6}, and \getrn{llr/steps}.
  \item[\getrn{llr/neu}] Immediate by \cref{lemma:apq:6} and \getrn{llr/neu}.
  \item[\getrn{llr/m}] Immediate by \cref{lemma:apq:6}, the induction hypothesis, and \getrn{llr/m}.
  \item[\getrn{llr/rs/nil}] Immediate.
  \item[\getrn{llr/rs/tm}] Immediate by the induction hypotheses.
    \qedhere
  \end{proofcases}
\end{proofEnd}

\subsection{Fundamental Theorem of Logical Relations}
\label{sec:fund-theor-logic}

The fundamental theorem of logical relations (\cref{conj:apq:5}) states if \(\jxpq{\Delta}{P}{A}\) and \(\llr{\Phi}{\sigma}{\Delta}\), then \(\llr{\Phi}{\apprs{\sigma}{P}}{\Delta}\), and it is proven by induction on the derivation of \(\jxpq{\Delta}{P}{A}\).
As usual, the proof follows easily from the induction hypotheses in the case of constructors.
Elimination forms are more complex, especially in call-by-value languages.
Consider, for example, a typical functional languages with positive products, where \(\Gamma \vdash \ms{match}\ M\ \ms{with}\ \ppp{x}{y}{N} : B\) whenever \(\Gamma \vdash M : A_1 \otimes A_2\) and \(\Gamma, x : A_1, y : A_2 \vdash N : B\).
By the induction hypothesis \(\apprs{\sigma}{M}\) is in the relation for \(A_1 \otimes A_2\) and so normalizes to some \(\ptpair{V_1}{V_2}\), which is in the relation by closure under reduction.
We can thus extend \(\sigma\) to \((\sigma, V_1, V_2)\) and appeal to the induction hypothesis for \(N\) to get that \(\apprs{\sigma, V_1, V_2}{N}\) is in the relation for \(B\).
Since \(\apprs{\sigma}{(\ms{match}\ M\ \ms{with}\ \ppp{x}{y}{N})}\) reduces to \(\apprs{\sigma, V_1, V_2}{N}\), we can finally appeal to closure under expansion to conclude the result for this case.
Elimination cases are even more complex for \pqx, because normal forms can involve neutral matches, and elimination forms must commute past these matches to reach the canonical form used to extend the reducible substitution.
We factor out these cases into \cref{lemma:apq:1,lemma:apq:8,lemma:apq:14}---one per eliminator---that state that if \(P\) satisfies the logical predicate at type \(A\) and \(f\) eliminates terms of type \(A\) to produce terms of type \(B\), then \(f(P)\) satisfies the logical predicate at type \(B\).
These lemmas all have similar proofs: they proceed by induction on the derivation of the proof that \(P\) satisfies the logical predicate at type \(A\).
The cases when \(P\) is a constructor for canonical values (\eg, \getrn{llr/triv} or \getrn{llr/pair}) or neutrals (rule \getrn{llr/neu}) are usual.
When considering neutral matches (rule \getrn{llr/m}), we must perform commuting conversions and appeal to the fact that Kripke logical predicates are monotone (so closed under weakening by the variables bound by the pattern), along with closure under reduction and expansion.
The \getrn{llr/steps} cases follow straightforwardly from induction hypotheses.

\begin{theoremEnd}{lemma}[Closure under application]
  \label{lemma:apq:8}
  If \(\llr{\Phi}{P}{\PTlolly{A}{B}}\), then \(\llr{\Psi}{\ptapp{(\apprs{\sigma}{P})}{Q}}{B}\) for all \(\llr{\Psi}{Q}{A}\) and \(\jnctx{\sigma}{\Psi}{\Phi}\).
  In particular, if \(\llr{\Phi}{P}{\PTlolly{A}{B}}\), then \(\llr{\Phi}{\ptapp{P}{Q}}{B}\) for all \(\llr{\Phi}{Q}{A}\).
\end{theoremEnd}

\begin{proofEnd}
  We begin by proving a special case of the result: \(\llr{\Psi}{\qtapp{g}{V}}{\red{S}}\) for all \(\red{g} : \QTlolly{U}{S}\) and \(\llr{\Psi}{\red{V}}{\red{U}}\) such that \(\jnorm[\cneu{\Psi}]{\red{V}}\).
  Assume that \(\llr{\Psi}{\red{V}}{\red{U}}\) and \(\jnorm[\cneu{\Psi}]{\red{V}}\).
  We proceed by induction on the derivation of \(\jnorm[\cneu{\Psi}]{\red{V}}\).
  \begin{proofcases}
  \item[\getrn{norm/can}] The result is immediate by \getrn{neu/app/can} and \getrn{llr/neu}.
  \item[\getrn{norm/neu}] The result is immediate by \getrn{neu/app/neu} and \getrn{llr/neu}.
  \item[\getrn{norm/match}] We deduce:
    \begin{enumerate}
    \item \(\jnorm[\cneu{\Psi}]{\qtm{\red{R}}{\qpat{\pi}{P}}}\)\hfill rule conclusion
    \item \(\jneu[\cneu{\Psi}]{R}\) \hfill rule premise\label{item:apq:351}
    \item \(\jnorm[\cneu{\Psi},\jneu{\pi}]{\red{P}}\) \hfill rule premise\label{item:apq:352}
    \item \(\llr{\Psi}{\qtm{\red{R}}{\qpat{\pi}{P}}}{\red{U}}\)\hfill assumption\label{item:apq:353}
    \item \(\llr{\Psi, \qtp{\pi}{S'}}{\red{P}}{\red{U}}\)\hfill inversion on \cref{item:apq:353} via \getrn{llr/m}\label{item:apq:354}
    \item \(\llr{\Psi, \qtp{\pi}{S'}}{\qtapp{g}{P}}{\red{U}}\)\hfill IH for \cref{item:apq:352} with \cref{item:apq:354}\label{item:apq:357}
    \item \(\llr{\Psi}{\qtm{R}{\qpat{\pi}{\qtapp{g}{P}}}}{\red{U}}\)\hfill \getrn{llr/m}, \cref{item:apq:357}\label{item:apq:358}
    \item \(\cstep{\Psi}{(\qtapp{g}{(\qtm{\red{R}}{\qpat{\pi}{P}})})}{(\qtm{R}{\qpat{\pi}{\qtapp{g}{P}}})}\)\hfill \getrn{cstep/app/cc/2}\label{item:apq:359}
    \item \(\llr{\Psi}{\qtapp{g}{(\qtm{\red{R}}{\qpat{\pi}{P}})}}{\red{S}}\) \hfill \getrn{llr/steps}, \cref{item:apq:358,item:apq:359}
    \end{enumerate}
    In \cref{item:apq:354}, the only possible possible rule is \getrn{llr/m}: the term is a match (ruling out \getrn{llr/neu}, \getrn{llr/triv}, and \getrn{llr/pair}), of simple type (ruling out \getrn{llr/us}, \getrn{llr/ds} and \getrn{llr/fun}), and normal (ruling out \getrn{llr/steps} by \cref{prop:apq:3}).
  \end{proofcases}
  This completes the proof of the special case.

  We proceed by induction on the derivation of \(\llr{\Phi}{P}{\PTlolly{A}{B}}\) to show the first claim.
  \begin{proofcases}
  \item[\getrn{llr/fun}] Immediate by the premisses.
  \item[\getrn{llr/neu}] We deduce:
    \begin{enumerate}
    \item \(\llr{\Phi}{\red{R}}{\QTlolly{U}{S}}\)\hfill rule conclusion\label{item:apq:29}
    \item \(\jneu[\cneu{\Phi}]{R}\)\hfill rule premise\label{item:apq:49}
    \item \(\jxpq{\ctp{\Phi}}{R}{\QTlolly{U}{S}}\)\hfill rule premise\label{item:apq:30}
    \item \(\red{R} = \red{g}\) for some \(\red{g}\)\hfill \cref{item:apq:49,item:apq:30,lemma:apq:2}\label{item:apq:31}
    \item \(\llr{\Psi}{\red{Q}}{\red{U}}\)\hfill assumption\label{item:apq:50}
    \item \(\jnctx{\sigma}{\Psi}{\Phi}\)\hfill assumption\label{item:apq:362}
    \item \(\csteps{\cneu{\Psi}}{\red{Q}}{\red{V}}\) with \(\jnorm[\cneu{\Psi}]{\red{V}}\) \hfill \cref{lemma:apq:3,item:apq:50}\label{item:apq:27}
    \item \(\llr{\Psi}{\red{V}}{\red{U}}\)\hfill induction on \cref{item:apq:27} with \cref{lemma:apq:4}\label{item:apq:361}
    \item \(\llr{\Psi}{\qtapp{g}{V}}{\red{U}}\)\hfill the special case, \cref{item:apq:27,item:apq:361}\label{item:apq:360}
    \item \(\csteps{\cneu{\Psi}}{\qtapp{g}{Q}}{\qtapp{g}{V}}\)\hfill induction on \cref{item:apq:361} with \getrn{cstep/app/2}\label{item:apq:363}
    \item \(\apprs{\sigma}{\red{g}} = \red{g}\)\hfill definition\label{item:apq:364}
    \item \(\csteps{\cneu{\Psi}}{\qtapp{(\apprs{\sigma}{g})}{Q}}{\qtapp{g}{V}}\)\hfill\cref{item:apq:363,item:apq:364}\label{item:apq:365}
    \item \(\llr{\Psi}{\qtapp{(\apprs{\sigma}{g})}{Q}}{\red{S}}\)\hfill \cref{lemma:apq:16,item:apq:365,item:apq:360}
    \end{enumerate}
  \item[\getrn{llr/m}] This case involves closure under reduction and expansion because commuting conversions only occur when applying an elimination form to a normal term, but the induction hypothesis does not necessarily give us a normal term.
    We deduce:
    \begin{enumerate}
    \item \(\llr{\Phi}{\ptm{\red{R}}{\qpat{\pi}{P}}}{\PTlolly{A}{B}}\)\hfill rule conclusion
    \item \(\jneu[\cneu{\Phi}]{R}\)\hfill rule premise\label{item:apq:102}
    \item \(\jxpq{\ctp{\Phi}}{\red{R}}{\red{U}}\)\hfill rule premise\label{item:apq:104}
    \item \(\llr{\Phi, \qtp{\pi}{U}}{P}{\PTlolly{A}{B}}\)\hfill rule premise\label{item:apq:103}
    \item \(\llr{\Psi}{Q}{A}\)\hfill assumption\label{item:apq:105}
    \item \(\jnctx{\sigma}{\Psi}{\Phi}\)\hfill assumption\label{item:apq:366}

    \item \(\jhalts{\cneu{\Phi}, \jneu{\pi}}{P}\)\hfill\cref{lemma:apq:4,item:apq:103}\label{item:apq:372}
    \item \(\jhalts{\cneu{\Psi}}{Q}\)\hfill \cref{lemma:apq:4,item:apq:105}\label{item:apq:373}

    \item \(\steps{\cneu{\Phi}, \jneu{\pi}}{P}{V}\) and \(\jnorm[\cneu{\Phi}, \jneu{\pi}]{V}\)\hfill inversion on \cref{item:apq:372}\label{item:apq:374}
    \item \(\steps{\cneu{\Psi}}{Q}{W}\) and \(\jnorm[\cneu{\Psi}]{W}\)\hfill inversion on \cref{item:apq:373}\label{item:apq:375}

    \item \(\jnctx{\wk{\sigma}}{\Psi, \qtp{\pi}{U}}{\Phi}\)\hfill\cref{lemma:apq:5}\label{item:apq:368}
    \item \(\jnctx{(\wk{\sigma}, \red{\pi})}{(\Psi, \qtp{\pi}{U})}{(\Phi, \qtp{\pi}{U})}\)\\
      \mbox{}\hfill extend \cref{item:apq:368} with variables bound by \(\red{\pi}\) via \getrn{llr/rs/tm} and \getrn{llr/neu}\label{item:apq:369}

    \item \(\steps{\cneu{\Psi}, \jneu{\pi}}{\apprs{\wk{\sigma}, \red{\pi}}{P}}{\apprs{\wk{\sigma}, \red{\pi}}{V}}\) and \(\jnorm[\cneu{\Phi}, \jneu{\pi}]{\apprs{\wk{\sigma}, \red{\pi}}{V}}\)\\
      \mbox{}\hfill\cref{lemma:apq:6,item:apq:374,item:apq:369}\label{item:apq:376}
    \item \(\steps{\cneu{\Psi}, \jneu{\pi}}{\wk{Q}}{\wk{W}}\) and \(\jnorm[\cneu{\Psi}, \jneu{\pi}]{\wk{W}}\)\\
      \mbox{}\hfill \cref{lemma:apq:6,item:apq:375} and neutrality of weakening\label{item:apq:377}

    \item \(\llr{\Psi, \qtp{\pi}{U}}{\wk{Q}}{A}\)\hfill\cref{lemma:apq:5,item:apq:105}\label{item:apq:370}
    \item \(\llr{\Psi, \qtp{\pi}{U}}{\ptapp{(\apprs{\wk{\sigma}, \red{\pi}}{P})}{(\wk{Q})}}{\red{B}}\)\hfill IH for \cref{item:apq:103} with \cref{item:apq:369,item:apq:370}\label{item:apq:371}
    \item \(\steps{\cneu{\Psi}, \jneu{\pi}}{\ptapp{(\apprs{\wk{\sigma}, \red{\pi}}{P})}{(\wk{Q})}}{\ptapp{(\apprs{\wk{\sigma}, \red{\pi}}{V})}{(\wk{W})}}\)\\
      \mbox{}\hfill induction on \cref{item:apq:376} with \getrn{fstep/app/1} or \getrn{cstep/app/1}, and\\
      \mbox{}\hfill on \cref{item:apq:377} with \getrn{fstep/app/2} or \getrn{cstep/app/2} and \cref{item:apq:376}
      \label{item:apq:378}
    \item \(\llr{\Psi, \qtp{\pi}{U}}{\ptapp{(\apprs{\wk{\sigma}, \red{\pi}}{V})}{(\wk{W})}}{\red{B}}\)\hfill \cref{lemma:apq:4,item:apq:378,item:apq:371}\label{item:apq:379}
    \item \(\jneu[\cneu{\Psi}]{\apprs{\sigma}{R}}\)\hfill \cref{lemma:apq:6,item:apq:102,item:apq:366}\label{item:apq:382}
    \item \(\jxpq{\ctp{\Psi}}{\apprs{\sigma}{R}}{\red{U}}\)\hfill substitution property, \cref{item:apq:104,item:apq:366}\label{item:apq:381}
    \item \(\llr{\Psi}{\ptm{\apprs{\sigma}{\red{R}}}{\qpat{\pi}{\ptapp{(\apprs{\wk{\sigma}, \red{\pi}}{V})}{(\wk{W})}}}}{\red{B}}\)\hfill \getrn{llr/m}, \cref{item:apq:379,item:apq:382,item:apq:381}\label{item:apq:380}

    \item \(\step{\cneu{\Psi}}{\ptapp{(\apprs{\sigma}{(\ptm{\red{R}}{\qpat{\pi}{V}})})}{W}}{(\ptm{\apprs{\sigma}{\red{R}}}{\qpat{\pi}{\ptapp{(\apprs{\wk{\sigma}, \red{\pi}}{V})}{(\wk{W})}}})}\)\\
      \mbox{}\hfill \getrn{fstep/app/cc} or \getrn{cstep/app/cc/1}\label{item:apq:383}
    \item \(\llr{\Psi}{\ptapp{(\apprs{\sigma}{(\ptm{\red{R}}{\qpat{\pi}{V}})})}{W}}{B}\) \hfill \getrn{llr/steps}, \cref{item:apq:383}\label{item:apq:384}
    \item \(\steps{\cneu{\Psi}}{\apprs{\sigma}{(\ptm{\red{R}}{\qpat{\pi}{P}})}}{\apprs{\sigma}{(\ptm{\red{R}}{\qpat{\pi}{V}})}}\)\\
      \mbox{}\hfill \cref{lemma:apq:6} applied to an induction on \cref{item:apq:374} using \getrn{fstep/m/q/r} or \getrn{cstep/m/r}\label{item:apq:385}
    \item \(\jnorm[\cneu{\Psi}]{\apprs{\sigma}{(\ptm{\red{R}}{\qpat{\pi}{V}})}}\)\hfill \cref{lemma:apq:6}, \getrn{norm/match}, \cref{item:apq:102,item:apq:374}\label{item:apq:386}
    \item \(\steps{\cneu{\Psi}}{\ptapp{(\apprs{\sigma}{(\ptm{\red{R}}{\qpat{\pi}{P}})})}{Q}}{\ptapp{(\apprs{\sigma}{(\ptm{\red{R}}{\qpat{\pi}{V}})})}{W}}\)\\
      \mbox{}\hfill induction on \cref{item:apq:385} using \getrn{fstep/app/1} or \getrn{cstep/app/2}, and\\
      \mbox{}\hfill induction on \cref{item:apq:375} using \getrn{fstep/app/2} or \getrn{fstep/app/2} with \cref{item:apq:386}
      \label{item:apq:387}

    \item \(\llr{\Phi}{\ptapp{(\ptm{\red{R}}{\qpat{\pi}{P}})}{Q}}{B}\)\hfill \getrn{llr/steps}, \cref{item:apq:387,item:apq:384}
    \end{enumerate}
  \item[\getrn{llr/steps}] We deduce:
    \begin{enumerate}
    \item \(\llr{\Phi}{P}{\PTlolly{A}{B}}\)\hfill conclusion
    \item \(\step{\cneu{\Phi}}{P}{P'}\)\hfill premise\label{item:apq:188}
    \item \(\llr{\Phi}{P'}{\PTlolly{A}{B}}\)\hfill premise\label{item:apq:189}
    \item \(\llr{\Psi}{Q}{A}\)\hfill assumption\label{item:apq:190}
    \item \(\jnctx{\sigma}{\Psi}{\Phi}\)\hfill assumption\label{item:apq:388}
    \item \(\llr{\Psi}{\ptapp{(\apprs{\sigma}{P'})}{Q}}{B}\)\hfill IH for \cref{item:apq:189,item:apq:190,item:apq:388}\label{item:apq:191}
    \item \(\step{\cneu{\Psi}}{\ptapp{\apprs{\sigma}{P}}{Q}}{\ptapp{(\apprs{\sigma}{P'})}{Q}}\)\hfill \getrn{fstep/app/1} or \getrn{cstep/app/1} with \cref{lemma:apq:6,item:apq:188}\label{item:apq:192}
    \item \(\llr{\Psi}{\ptapp{(\apprs{\sigma}{P})}{Q}}{B}\)\hfill \getrn{llr/steps}, \cref{item:apq:191,item:apq:192}
    \end{enumerate}
  \end{proofcases}
  The second claim follows immediately from the first and the fact that the identity substitution is neutral.
\end{proofEnd}

\begin{theoremEnd}{lemma}[Closure under forcing]
  \label{lemma:apq:1}
  \leavevmode\\
  If \(\llr{\Phi}{\blue{M}}{\FTus[][]{A}}\), then \(\llr{\Phi}{\ptforce{\blue{M}}}{A}\).
\end{theoremEnd}

\begin{proofEnd}
  By induction on the derivation of \(\llr{\Phi}{\blue{M}}{\FTus[][]{A}}\).
  \begin{proofcases}
  \item[\getrn{llr/us}] Immediate by the rule premise.
  \item[\getrn{llr/neu}] This case is impossible: all neutrals are circuit terms, while \(\blue{M}\) is a functional term.
  \item[\getrn{llr/m}] Similarly to the \getrn{llr/m} case in the proof of \cref{lemma:apq:8}, we must use closure under reduction and expansion to show the result.
    Once again, this is because commuting conversions only occur when applying an elimination form to a normal term, but the induction hypothesis does not necessarily give us a normal term.
    \begin{enumerate}
    \item \(\llr{\Phi}{\ftm{\red{R}}{\qpat{\pi}{M}}}{\FTus[][]{A}}\)\hfill rule conclusion
    \item \(\jxpq{\ctp{\Phi}}{\red{R}}{\red{B}}\)\hfill rule premise\label{item:apq:315}
    \item \(\jneu[\cneu{\Phi}]{R}\)\hfill rule premise\label{item:apq:330}
    \item \(\llr{\Phi, \qtp{\pi}{B}}{M}{\FTus[][]{A}}\)\hfill rule premise\label{item:apq:332}
    \item \(\llr{\Phi, \qtp{\pi}{B}}{\ptforce{M}}{A}\)\hfill IH for \cref{item:apq:332}\label{item:apq:331}
    \item \(\jhalts{\cneu{\Phi}, \jneu{\pi}}{M}\)\hfill \cref{lemma:apq:3,item:apq:332}\label{item:apq:389}
    \item \(\steps{\cneu{\Phi}, \jneu{\pi}}{M}{V}\)\hfill inversion on \cref{item:apq:389}\label{item:apq:390}
    \item \(\jnorm{\cneu{\Phi}, \jneu{\pi}}{V}\)\hfill inversion on \cref{item:apq:389}\label{item:apq:391}
    \item \(\steps{\cneu{\Phi}, \jneu{\pi}}{\ptforce{M}}{\ptforce{V}}\)\\
      \mbox{}\hfill induction on \cref{item:apq:390} with \getrn{fstep/force/1} or \getrn{cstep/force/1}\label{item:apq:393}
    \item \(\llr{\Phi, \qtp{\pi}{B}}{\ptforce{V}}{\FTus[][]{A}}\)\hfill \cref{lemma:apq:4,item:apq:393,item:apq:332}\label{item:apq:392}
    \item \(\llr{\Phi}{\ptm{\red{R}}{\qpat{\pi}{\ptforce{V}}}}{A}\)\hfill \getrn{llr/m}, \cref{item:apq:392,item:apq:330,item:apq:315}\label{item:apq:399}
    \item \(\jnorm[\cneu{\Phi}]{\ftm{\red{R}}{\qpat{\pi}{V}}}\)\hfill \getrn{norm/match}, \cref{item:apq:330,item:apq:391}\label{item:apq:394}
    \item \(\step{\cneu{\Phi}}{(\ptforce{(\ftm{\red{R}}{\qpat{\pi}{V}})})}{(\ftm{\red{R}}{\qpat{\pi}{\ptforce{V}}})}\)\\
      \mbox{}\hfill \getrn{fstep/force/cc} or \getrn{cstep/force/cc}, \cref{item:apq:394}\label{item:apq:395}
    \item \(\steps{\cneu{\Phi}}{(\ftm{\red{R}}{\qpat{\pi}{M}})}{(\ftm{\red{R}}{\qpat{\pi}{V}})}\)\hfill induction on \cref{item:apq:390} with \getrn{fstep/m/q/r}, \cref{item:apq:330}\label{item:apq:396}
    \item \(\steps{\cneu{\Phi}}{\ptforce(\ftm{\red{R}}{\qpat{\pi}{M}})}{\ptforce(\ftm{\red{R}}{\qpat{\pi}{V}})}\)\\
      \mbox{}\hfill induction on \cref{item:apq:396} with \getrn{fstep/force/1} or \getrn{cstep/force/1}\label{item:apq:397}
    \item \(\steps{\cneu{\Phi}}{\ptforce(\ftm{\red{R}}{\qpat{\pi}{M}})}{(\ftm{\red{R}}{\qpat{\pi}{\ptforce{V}}})}\)\\
      \mbox{}\hfill transitivity, \cref{item:apq:395,item:apq:397}\label{item:apq:398}
    \item \(\llr{\Phi}{\ptforce(\ftm{\red{R}}{\qpat{\pi}{M}})}{A}\)\hfill \cref{lemma:apq:4,item:apq:398,item:apq:399}
    \end{enumerate}
  \item[\getrn{llr/steps}] We deduce:
    \begin{enumerate}
    \item \(\llr{\Phi}{P}{\FTus[][]{A}}\)\hfill rule conclusion
    \item \(\step{\cneu{\Phi}}{P}{P'}\)\hfill rule premise\label{item:apq:401}
    \item \(\llr{\Phi}{P'}{\FTus[][]{A}}\)\hfill rule premise\label{item:apq:400}
    \item \(\llr{\Phi}{\ftforce{P'}}{A}\)\hfill IH for \cref{item:apq:400}
    \item \(\step{\cneu{\Phi}}{\ftforce{P}}{\ftforce{P'}}\)\hfill \getrn{fstep/force/1} or \getrn{cstep/force/1} and \cref{item:apq:401}\label{item:apq:402}
    \item \(\llr{\Phi}{\ftforce{P}}{A}\)\hfill \getrn{llr/steps}, \cref{item:apq:400,item:apq:402}
    \end{enumerate}
  \end{proofcases}
\end{proofEnd}

We remark the parallels between the hypotheses of \cref{lemma:apq:14} and the premisses \getrn{llr/fun}.

\begin{theoremEnd}{lemma}[Closure under matching]
  \label{lemma:apq:14}
  \leavevmode\\
  If \(\llr{\Psi}{P}{A}\), then \(\llr{\Psi}{\ptm{P}{p}}{B}\) for all patterns \(\jxpq{\Psi}{p}{(\pat{A}{B})}\) such that \(\llr{\Phi}{\ecanop{K}{\apprs{\sigma}{p}}}{B}\) for all \(\jnctx{\sigma}{\Phi}{\Psi}\) and \(\llr{\Phi}{K}{A}\) with \(\jcan[\cneu{\Phi}]{K}\).
\end{theoremEnd}

\begin{proofEnd}
  By induction on the derivation of \(\llr{\Psi}{P}{A}\).
  By definition of the judgment \(\jxpq{\Psi}{p}{(\pat{A}{B})}\), \(A\) must be one of \(\PTu\), \(\PTot{A_1}{A_2}\) or \(\FTds[\ML][\MU]{A_\MU}\).
  \begin{proofcases}
  \item[\getrn{llr/triv}, \getrn{llr/pair} or \getrn{llr/ds}] We deduce:
    \begin{enumerate}
    \item \(\llr{\Psi}{P}{A}\)\hfill rule conclusion\label{item:apq:19}
    \item \(\jhalts{\cneu{\Psi}}{P}\)\hfill \cref{lemma:apq:3,item:apq:19}\label{item:apq:414}
    \item \(\jxpq{\Psi}{p}{(\pat{A}{B})}\)\hfill assumption\label{item:apq:419}
    \item \(\llr{\Phi}{\ecanop{K}{\apprs{\sigma}{p}}}{B}\) for all \(\jnctx{\sigma}{\Phi}{\Psi}\), \(\llr{\Phi}{K}{A}\) with \(\jcan[\cneu{\Psi}]{K}\)\\
      \mbox{}\hfill assumption\label{item:apq:420}
    \item \(\steps{\cneu{\Psi}}{P}{V}\)\hfill inversion on \cref{item:apq:414}\label{item:apq:415}
    \item \(\jnorm[\cneu{\Psi}]{V}\)\hfill inversion on \cref{item:apq:414}\label{item:apq:416}
    \item \(\jcan[\cneu{\Psi}]{V}\)\hfill induction on \cref{item:apq:415} using the shape of \(P\) in \cref{item:apq:19},\\
      \mbox{}\hfill and inversion on \cref{item:apq:416}\label{item:apq:417}
    \item \(\llr{\Psi}{V}{A}\)\hfill \cref{lemma:apq:4,item:apq:19,item:apq:415}\label{item:apq:421}
    \item \(\llr{\Psi}{\ecanop{V}{p}}{B}\)\hfill \cref{item:apq:420,item:apq:421,item:apq:417}, \(\jnctx{\ms{id}}{\Psi}{\Psi}\)\label{item:apq:425}
    \item \(\step{\cneu{\Psi}}{\ptm{V}{p}}{\ecanop{V}{p}}\)\hfill \getrn{fstep/m/k} or \getrn{cstep/m/k}, \cref{item:apq:417}\label{item:apq:426}
    \item \(\steps{\cneu{\Psi}}{\ptm{P}{p}}{\ptm{V}{p}}\)\\
      \mbox{}\hfill induction on \cref{item:apq:415} with \getrn{fstep/m/f}, \getrn{fstep/m/q}, or \getrn{cstep/m}\label{item:apq:427}
    \item \(\steps{\cneu{\Psi}}{\ptm{P}{p}}{\ecanop{V}{p}}\)\hfill transitivity, \cref{item:apq:426,item:apq:427}\label{item:apq:428}
    \item \(\llr{\Psi}{\ptm{P}{p}}{B}\)\hfill\cref{lemma:apq:16,item:apq:428,item:apq:425}
    \end{enumerate}
  \item[\getrn{llr/neu}] We deduce:
    \begin{enumerate}
    \item \(\llr{\Psi}{\red{R}}{\red{A}}\)\hfill rule conclusion\label{item:apq:429}
    \item \(\jxpq{\Psi}{\red{R}}{\red{A}}\)\hfill rule premise\label{item:apq:430}
    \item \(\jneu[\cneu{\Psi}]{\red{R}}\)\hfill rule premise\label{item:apq:431}
    \item \(\jxpq{\Psi}{p}{(\pat{\red{A}}{B})}\)\hfill assumption\label{item:apq:432}
    \item \(\llr{\Phi}{\ecanop{K}{\apprs{\sigma}{p}}}{B}\) for all \(\jnctx{\sigma}{\Phi}{\Psi}\), \(\llr{\Phi}{K}{A}\) with \(\jcan[\cneu{\Psi}]{K}\)\\
      \mbox{}\hfill assumption\label{item:apq:433}
    \item \(\red{A} = \QTu\) or \(\red{A} = \QTot{A_1}{A_2}\)\hfill inversion on \cref{item:apq:432}\label{item:apq:435}
    \end{enumerate}\newcounter{tmpenumi}\setcounter{tmpenumi}{\value{enumi}}
    We proceed by case analysis on \cref{item:apq:435}.

    If \(\red{A} = \QTu\), then \(p = (\qpu{Q})\):
    \begin{enumerate}[resume]
    \item \(\jcan[\cneu{\Psi}]{\qttriv}\)\hfill \getrn{can/triv}\label{item:apq:436}
    \item \(\llr{\Psi}{\qttriv}{\QTu}\)\hfill \getrn{llr/triv}\label{item:apq:437}
    \item \(\llr{\Psi}{\ecanop{\qttriv}{p}}{B}\)\hfill \cref{item:apq:433,item:apq:436,item:apq:437}, \(\jnctx{\ms{id}}{\Psi}{\Psi}\)\label{item:apq:438}
    \item \(\llr{\Psi}{Q}{B}\)\hfill\cref{item:apq:438}, \getrn{ecan/triv}\label{item:apq:439}
    \item \(\llr{\Psi}{\ptm{\red{R}}{p}}{B}\)\hfill \getrn{llr/m}, \cref{item:apq:439,item:apq:430,item:apq:431}
    \end{enumerate}

    If \(\red{A} = \QTot{A_1}{A_2}\), then \(p = \qpp{x}{y}{Q}\).
    Set \(\Phi = (\Psi, \qtp{x}{A_1}, \jneu{x}, \qtp{y}{A_2}, \jneu{y})\).
    \begin{enumerate}
      \setcounter{enumi}{\value{tmpenumi}}
    \item \(\jcan[\cneu{\Phi}]{\qtpair{x}{y}}\)\hfill \getrn{can/pair}, \getrn{norm/neu}, \getrn{neu/var}\label{item:apq:440}
    \item \(\llr{\Phi}{\qtpair{x}{y}}{\red{A}}\)\hfill \getrn{llr/pair}, \getrn{llr/neu}\label{item:apq:441}
    \item \(\llr{\Phi}{\ecanop{\qtpair{x}{y}}{\wk{p}}}{\red{B}}\)\hfill \cref{item:apq:433,item:apq:440,item:apq:441}\label{item:apq:442}
    \item \(\llr{\Psi, \qtp{x}{A_1}, \jneu{x}, \qtp{y}{A_2}, \jneu{y}}{Q}{\red{B}}\)\hfill \cref{item:apq:442}, \getrn{ecan/pair}\label{item:apq:443}
    \item \(\llr{\Psi}{\ptm{\red{R}}{p}}{\red{B}}\)\hfill \getrn{llr/m}, \cref{item:apq:443,item:apq:430,item:apq:431}
    \end{enumerate}
  \item[\getrn{llr/m}]
    Similarly to the \getrn{llr/m} cases in the proofs of \cref{lemma:apq:8} and \cref{lemma:apq:1}, we must use closure under reduction and expansion to show the result.
    Once again, this is because commuting conversions only occur when applying an elimination form to a normal term, but the induction hypothesis does not necessarily give us a normal term.
    We deduce:
    \begin{enumerate}
    \item \(\llr{\Psi}{\ptm{\red{R}}{\qpat{\pi}{P}}}{A}\)\hfill rule conclusion
    \item \(\jxpq{\Psi}{\red{R}}{\red{A_0}}\)\hfill rule premise\label{item:apq:447}
    \item \(\jneu[\cneu{\Psi}]{R}\)\hfill rule premise\label{item:apq:449}
    \item \(\llr{\Psi, \qtp{\pi}{A_0}}{P}{A}\)\hfill rule premise\label{item:apq:448}
    \item \(\jxpq{\Psi}{p}{\pat{A}{B}}\)\hfill assumption\label{item:apq:450}
    \item \(\llr{\Phi}{\ecanop{K}{\apprs{\sigma}{p}}}{B}\) for all \(\jnctx{\sigma}{\Phi}{\Psi}\), \(\llr{\Phi}{K}{A}\) with \(\jcan[\Phi]{K}\)\\
      \mbox{}\hfill assumption\label{item:apq:451}
    \item \(\jxpq{\Psi, \qtp{\pi}{A_0}}{\wk{p}}{\pat{A}{B}}\)\hfill weakening on \cref{item:apq:450}\label{item:apq:452}
    \item \(\llr{\Phi}{\ecanop{K}{\apprs{\sigma}{\wk{p}}}}{B}\) for all \(\jnctx{\sigma}{\Phi}{(\Psi, \qtp{\pi}{A_0})}\) and\\
      \(\llr{\Phi}{K}{A}\) with \(\jcan[\Phi]{K}\) \hfill \cref{cor:apq:4,item:apq:451}\label{item:apq:453}
    \item \(\llr{\Psi, \qtp{\pi}{A_0}}{\ptm{P}{\wk{p}}}{B}\)\hfill IH for \cref{item:apq:448} with \cref{item:apq:452,item:apq:453}\label{item:apq:454}
    \item \(\llr{\Psi}{\ptm{\red{R}}{\qpat{\pi}{\ptm{P}{\wk{p}}}}}{B}\)\hfill \getrn{llr/m}, \cref{item:apq:447,item:apq:449,item:apq:454}\label{item:apq:463}
    \item \(\jhalts{\cneu{\Psi}, \jneu{\pi}}{P}\)\hfill \cref{lemma:apq:3,item:apq:448}\label{item:apq:455}
    \item \(\steps{\cneu{\Psi}, \jneu{\pi}}{P}{V}\)\hfill inversion on \cref{item:apq:455}\label{item:apq:456}
    \item \(\jnorm[\cneu{\Psi}, \jneu{\pi}]{V}\)\hfill inversion on \cref{item:apq:455}\label{item:apq:457}
    \item \(\steps{\cneu{\Psi}, \jneu{\pi}}{\ptm{P}{\wk{p}}}{\ptm{V}{\wk{p}}}\)\\
      \mbox{}\hfill induction on \cref{item:apq:456} with \getrn{fstep/m/f}, \getrn{fstep/m/q} or \getrn{cstep/m}\label{item:apq:461}
    \item \(\steps{\cneu{\Psi}}{(\ptm{\red{R}}{\qpat{\pi}{\ptm{P}{\wk{p}}}})}{(\ptm{\red{R}}{\qpat{\pi}{\ptm{V}{\wk{p}}}})}\)\\
      \mbox{}\hfill induction on \cref{item:apq:461} with \getrn{fstep/m/q/r} or \getrn{cstep/m/r}\label{item:apq:462}
    \item \(\llr{\Psi}{\ptm{\red{R}}{\qpat{\pi}{\ptm{V}{\wk{p}}}}}{B}\)\hfill \cref{lemma:apq:4,item:apq:463,item:apq:462}\label{item:apq:465}
    \item \(\steps{\cneu{\Psi}}{(\ptm{\red{R}}{\qpat{\pi}{P}})}{(\ptm{\red{R}}{\qpat{\pi}{V}})}\)\\
      \mbox{}\hfill induction on \cref{item:apq:456} with \getrn{fstep/m/q/r} or \getrn{cstep/m/r}\label{item:apq:458}
    \item \(\steps{\cneu{\Psi}}{\ptm{(\ptm{\red{R}}{\qpat{\pi}{P}})}{p}}{\ptm{(\ptm{\red{R}}{\qpat{\pi}{V}})}{p}}\)\\
      \mbox{}\hfill induction on \cref{item:apq:458} with \getrn{fstep/m/q} or \getrn{cstep/m}\label{item:apq:459}
    \item \(\jnorm[\cneu{\Psi}]{\ptm{\red{R}}{\qpat{\pi}{V}}}\)\hfill \getrn{norm/match}, \cref{item:apq:457,item:apq:449}\label{item:apq:460}
    \item \(\step{\cneu{\Psi}}{\ptm{(\ptm{\red{R}}{\qpat{\pi}{V}})}{p}}{\ptm{\red{R}}{\qpat{\pi}{\ptm{V}{\wk{p}}}}}\)\\
      \mbox{}\hfill \getrn{fstep/m/f/cc}, \getrn{fstep/m/q/cc} or \getrn{cstep/m/cc} with \cref{item:apq:460}
    \item \(\steps{\cneu{\Psi}}{\ptm{(\ptm{\red{R}}{\qpat{\pi}{P}})}{p}}{\ptm{\red{R}}{\qpat{\pi}{\ptm{V}{\wk{p}}}}}\)\\
      \mbox{}\hfill transitivity, \cref{item:apq:459,item:apq:460}\label{item:apq:464}
    \item \(\llr{\Psi}{\ptm{(\ptm{\red{R}}{\qpat{\pi}{P}})}{p}}{B}\) \hfill \cref{lemma:apq:16,item:apq:464,item:apq:465}
    \end{enumerate}
  \item[\getrn{llr/steps}] We deduce:
    \begin{enumerate}
    \item \(\llr{\Psi}{P}{A}\)\hfill rule conclusion\label{item:apq:444}
    \item \(\step{\cneu{\Psi}}{P}{P'}\)\hfill rule premise\label{item:apq:446}
    \item \(\llr{\Psi}{P'}{A}\)\hfill rule premise\label{item:apq:445}
    \item \(\llr{\Psi}{\ptm{P'}{p}}{B}\)\hfill IH for \cref{item:apq:445}
    \item \(\step{\cneu{\Psi}}{\ptm{P}{p}}{\ptm{P'}{p}}\)\hfill \getrn{fstep/m/f}, \getrn{fstep/m/q} or \getrn{cstep/m}, \cref{item:apq:446}
    \item \(\llr{\Psi}{\ptm{P}{p}}{B}\)\hfill \getrn{llr/steps}, \cref{item:apq:445,item:apq:446}
    \end{enumerate}
  \end{proofcases}
\end{proofEnd}

Our statement of the fundamental theorem is two parted because our program typing judgment is mutually defined with a pattern typing judgment.
The part for programs is standard, while the part for patterns gives exactly the hypotheses needed to appeal to \cref{lemma:apq:14}:

\begin{theoremEnd}{theorem}[Fundamental Theorem of Logical Relations]
  \label{conj:apq:5}
  \leavevmode
  \begin{enumerate}
  \item If \(\jxpq{\Delta}{P}{A}\), then \(\llr{\Phi}{\apprs{\sigma}{P}}{A}\) for all reducible substitutions \(\llr{\Phi}{\sigma}{\Delta}\).
  \item If \(\jxpq{\Delta}{p}{(\pat{A}{B})}\), then \(\llr{\Psi}{\ecanop{K}{\apprs{\psi}{\apprs{\phi}{p}}}}{B}\) for all reducible substitutions \(\llr{\Phi}{\phi}{\Delta}\), neutral substitutions \({\jnctx{\psi}{\Psi}{\Phi}}\), and \(\llr{\Psi}{K}{A}\) with \(\jnorm[\cneu{\Psi}]{K}\).
  \end{enumerate}
\end{theoremEnd}

\begin{proofEnd}
  We proceed by induction on the derivation of \(\jxpq{\Delta}{P}{A}\) and \(\jxpq{\Delta}{(\pat{\pi}{P})}{(\pat{A}{B})}\).

  \begin{proofcases}
  \item[\getrn{x/f/var}] Immediate by the definition of substitution and the hypothesis that \(\llr{\Phi}{\sigma}{\Delta}\).
  \item[\getrn{x/f/lam}] We deduce:
    \begin{enumerate}
    \item \(\jxpq{\Delta}{\ftlam{x}{M}}{\FTlolly{A}{B}}\)\hfill rule conclusion\label{item:apq:111}
    \item \(\jxpq{\Delta, \ftp{x}{A}}{\blue{M}}{\blue{B}}\)\hfill rule premise\label{item:apq:112}
    \item \(\llr{\Sigma}{\sigma}{\Delta}\)\hfill assumption\label{item:apq:32}
    \item \(\llr{\Sigma}{\apprs{\sigma}{\blue{M}}}{\blue{B}}\) for all \(\llr{\Sigma}{\sigma}{\Delta, \ftp{x}{A}}\)\hfill IH for \cref{item:apq:112}\label{item:apq:53}
    \item \(\jnctx{\phi}{\Phi}{\Sigma}\)\hfill assumption for universal quantification\label{item:apq:35}
    \item \(\llr{\Phi}{\blue{N}}{\blue{A}}\)\hfill assumption for universal quantification\label{item:apq:42}
    \item \(\steps{\cneu{\Phi}}{\blue{N}}{\blue{V}}\) with \(\jnorm[\cneu{\Phi}]{\blue{V}}\)\hfill\cref{item:apq:42,lemma:apq:3}\label{item:apq:77}
    \item \(\llr{\Phi}{\blue{V}}{\blue{A}}\)\hfill \cref{lemma:apq:4,item:apq:42,item:apq:77}
    \item \(\llr{\Phi}{\apprs{\phi}{\sigma}}{\Delta}\)\hfill\cref{lemma:apq:5,item:apq:32,item:apq:35}\label{item:apq:75}
    \item \(\llr{\Phi}{\apprs{\phi}{\sigma}, \blue{V}}{\Delta, \ftp{x}{A}}\)\hfill\getrn{llr/rs/tm}, \getrn{x/cs/tm}, and \cref{item:apq:77,item:apq:75}\label{item:apq:78}
    \item \(\llr{\Phi}{\apprs{\apprs{\phi}{\sigma}, \blue{V}}{\blue{M}}}{\blue{B}}\)\hfill\cref{item:apq:53,item:apq:78}\label{item:apq:83}
    \item \(\step{\cneu{\Phi}}{\ftapp{(\apprs{\apprs{\phi}{\sigma}}{\ftlam{x}{M}})}{V}}{\apprs{\apprs{\phi}{\sigma}, \blue{V}}{\blue{M}}}\)\\
      \mbox{}\hfill \getrn{fstep/app/beta}, properties of substitution\label{item:apq:175}
    \item \(\steps{\cneu{\Phi}}{\ftapp{(\apprs{\apprs{\phi}{\sigma}}{\ftlam{x}{M}})}{N}}{\ftapp{(\apprs{\apprs{\phi}{\sigma}}{\ftlam{x}{M}})}{V}}\)\\
      \mbox{}\hfill induction on \cref{item:apq:77} with \getrn{fstep/app/2}\label{item:apq:176}
    \item \(\ftapp{(\apprs{\phi}{\apprs{\sigma}{\ftlam{x}{M}}})}{N} = \ftapp{(\apprs{\apprs{\phi}{\sigma}}{\ftlam{x}{M}})}{N}\)\hfill properties of substitution\label{item:apq:177}
    \item \(\steps{\cneu{\Phi}}{\ftapp{(\apprs{\phi}{\apprs{\sigma}{\ftlam{x}{M}}})}{N}}{\apprs{\apprs{\phi}{\sigma}, \blue{V}}{\blue{M}}}\)\hfill transitivity, \cref{item:apq:175,item:apq:176,item:apq:177}\label{item:apq:178}
    \item \(\llr{\Phi}{\ftapp{(\apprs{\phi}{\apprs{\sigma}{\ftlam{x}{M}}})}{N}}{\blue{B}}\)\hfill\getrn{llr/steps}, \cref{item:apq:178,item:apq:83}\label{item:apq:202}
    \item \(\llr{\Phi}{\ftapp{(\apprs{\phi}{\apprs{\sigma}{\ftlam{x}{M}}})}{N}}{\blue{B}}\) for all \(\jnctx{\phi}{\Phi}{\Sigma}\) and \(\llr{\Phi}{\blue{N}}{\blue{A}}\)\\
      \mbox{} \hfill universal quantification, \cref{item:apq:35,item:apq:42,item:apq:202}\label{item:apq:203}
    \item \(\jhalts{\cneu{\Phi}}{\apprs{\apprs{\phi}{\sigma}}{\ftlam{x}{M}}}\)\\
      \mbox{}\hfill \getrn{halts}, \getrn{norm/can}, \getrn{can/lam/f}, \getrn{steps/refl}\label{item:apq:403}
    \item \(\llr{\Phi}{\apprs{\sigma}{(\ftlam{x}{M})}}{\FTlolly{A}{B}}\)\hfill\getrn{llr/fun}, \cref{item:apq:203,item:apq:403}
    \end{enumerate}
  \item[\getrn{x/f/triv}] Immediate by \(\apprs{\sigma}{\fttriv} = \fttriv\) and \getrn{llr/triv}.
  \item[\getrn{x/f/pair}] We deduce:
    \begin{enumerate}
    \item \(\jxpq{\Delta}{\ftpair{M}{N}}{\FTot{A}{B}}\)\hfill rule conclusion
    \item \(\jxpq{\Delta}{\blue{M}}{\blue{A}}\)\hfill rule premise\label{item:apq:109}
    \item \(\jxpq{\Delta}{\blue{N}}{\blue{B}}\)\hfill rule premise\label{item:apq:110}
    \item \(\llr{\Phi}{\sigma}{\Delta}\) \hfill assumption\label{item:apq:167}
    \item \(\llr{\Phi}{\apprs{\sigma}{\blue{M}}}{\blue{A}}\)\hfill IH for \cref{item:apq:109,item:apq:167}\label{item:apq:113}
    \item \(\llr{\Phi}{\apprs{\sigma}{\blue{N}}}{\blue{B}}\)\hfill IH for \cref{item:apq:110,item:apq:167}\label{item:apq:114}
    \item \(\llr{\Phi}{\ftpair{\apprs{\sigma}{M}}{\apprs{\sigma}{N}}}{\FTot{A}{B}}\)\hfill \getrn{llr/pair}, \cref{item:apq:113,item:apq:114}\label{item:apq:168}
    \item \(\ftpair{\apprs{\sigma}{M}}{\apprs{\sigma}{N}} = \apprs{\sigma}{\ftpair{M}{N}}\)\hfill definition of substitution\label{item:apq:120}
    \item \(\llr{\Phi}{\apprs{\sigma}{\ftpair{M}{N}}}{\FTot{A}{B}}\)\hfill\cref{item:apq:168,item:apq:120}
    \end{enumerate}
  \item[\getrn{x/f/susp}] We deduce:
    \begin{enumerate}
    \item \(\jxpq{\Delta}{\ftsusp{P}}{\FTus[][]{A}}\)\hfill rule conclusion
    \item \(\jxpq{\Delta}{P}{A}\)\hfill rule hypothesis\label{item:apq:1}
    \item \(\llr{\Phi}{\sigma}{\Delta}\)\hfill assumption\label{item:apq:335}
    \item \(\llr{\Phi}{\apprs{\sigma}{P}}{A}\)\hfill IH for \cref{item:apq:1} with \cref{item:apq:335}\label{item:apq:146}
    \item \(\steps{\cneu{\Phi}}{\apprs{\sigma}{(\ptforce{\ftsusp{P}})}}{\apprs{\sigma}{P}}\)\hfill \getrn{fstep/force}, definition of substitution\label{item:apq:404}
    \item \(\llr{\Phi}{\apprs{\sigma}{(\ptforce{\ftsusp{P}})}}{A}\)\hfill \getrn{llr/steps}, \cref{item:apq:146,item:apq:404}\label{item:apq:405}
    \item \(\llr{\Phi}{\apprs{\sigma}{\ftsusp{P}}}{\FTus[][]{A}}\)\hfill \getrn{llr/us}, \cref{item:apq:405}, definition of substitution
    \end{enumerate}
  \item[\getrn{x/f/down}] We deduce:
    \begin{enumerate}
    \item \(\jxpq{\Delta}{\ftdown{M}}{\FTds[\ML][\MU]{A}}\)\hfill rule conclusion
    \item \(\jxpq{\Delta}{\blue{M}}{\blue{A}}\)\hfill rule premise\label{item:apq:158}
    \item \(\llr{\Phi}{\sigma}{\Delta}\)\hfill assumption\label{item:apq:180}
    \item \(\llr{\Phi}{\apprs{\sigma}{\blue{M}}}{\blue{A}}\)\hfill induction hypothesis for \cref{item:apq:158} with \cref{item:apq:180}\label{item:apq:159}
    \item \(\apprs{\sigma}{(\ftdown{M})} = \ftdown{\apprs{\sigma}{M}}\)\hfill definition substitution\label{item:apq:164}
    \item \(\llr{\Phi}{\apprs{\sigma}{(\ftdown{M})}}{\FTds[\ML][\MU]{A}}\)\hfill \getrn{llr/ds}, \cref{item:apq:159,item:apq:164}
    \end{enumerate}
  \item[\getrn{x/f/app}] We deduce:
    \begin{enumerate}
    \item \(\jxpq{\Psi}{\ftapp{M}{N}}{\blue{B}}\)\hfill rule conclusion
    \item \(\jxpq{\Psi}{\blue{M}}{\FTlolly{A}{B}}\)\hfill rule premise\label{item:apq:61}
    \item \(\jxpq{\Psi}{\blue{N}}{\blue{A}}\)\hfill rule premise\label{item:apq:64}
    \item \(\llr{\Phi}{\apprs{\sigma}{M}}{\FTlolly{A}{B}}\)\hfill IH for \cref{item:apq:61}\label{item:apq:65}
    \item \(\llr{\Phi}{\apprs{\sigma}{N}}{\blue{A}}\)\hfill IH for \cref{item:apq:64}\label{item:apq:67}
    \item \(\llr{\Phi}{\ftapp{(\apprs{\sigma}{M})}{(\apprs{\sigma}{N})}}{\blue{B}}\)\hfill \cref{lemma:apq:8,item:apq:65,item:apq:67}\label{item:apq:72}
    \item \(\ftapp{(\apprs{\sigma}{M})}{(\apprs{\sigma}{N})} = \apprs{\sigma}{(\ftapp{M}{N})}\)\hfill definition\label{item:apq:71}
    \item \(\llr{\Phi}{\apprs{\sigma}{(\ftapp{M}{N})}}{\blue{B}}\)\hfill \cref{item:apq:72,item:apq:71}
    \end{enumerate}
  \item[\getrn{x/f/force}] We deduce:
    \begin{enumerate}
    \item \(\jxpq{\Delta}{\ftforce{M}}{\blue{A_\ML}}\)\hfill rule conclusion
    \item \(\jxpq{\Delta}{\blue{M}}{\FTus[\ML][\MU]{\blue{A_\ML}}}\)\hfill rule hypothesis\label{item:apq:4}
    \item \(\llr{\Phi}{\sigma}{\Delta}\)\hfill assumption\label{item:apq:166}
    \item \(\llr{\Phi}{\apprs{\sigma}{\blue{M}}}{\FTus[\ML][\MU]{\blue{A_\ML}}}\)\hfill IH for \cref{item:apq:4}\label{item:apq:5}
    \item \(\llr{\Phi}{\ftforce{\apprs{\sigma}{M}}}{\blue{A_\ML}}\)\hfill \cref{lemma:apq:1,item:apq:5}\label{item:apq:6}
    \item \(\apprs{\sigma}{(\ftforce{M})} = \ftforce{\apprs{\sigma}{M}}\)\hfill definition\label{item:apq:198}
    \item \(\llr{\Phi}{\apprs{\sigma}{(\ftforce{M})}}{\blue{A_\ML}}\)\hfill \cref{item:apq:6,item:apq:198}
    \end{enumerate}
  \item[\getrn{x/f/match}] We deduce:
    \begin{enumerate}
    \item \(\jxpq{\Delta}{\ftm{P}{p}}{\blue{B_\MF}}\)\hfill rule conclusion
    \item \(\jxpq{\Delta}{\blue{P}}{A}\)\hfill rule premise\label{item:apq:199}
    \item \(\jxpq{\Delta}{\blue{p}}{\pat{A}{\blue{B_\MF}}}\)\hfill rule premise\label{item:apq:200}
    \item \(\llr{\Phi}{\sigma}{\Delta}\)\hfill assumption\label{item:apq:466}
    \item \(\llr{\Phi}{\apprs{\sigma}{P}}{A}\)\hfill IH for \cref{item:apq:199} with \cref{item:apq:466}\label{item:apq:467}
    \item \(\llr{\Psi}{\ecanop{K}{\apprs{\psi}{\apprs{\phi}{p}}}}{\blue{B_\MF}}\) for all \({\jnctx{\psi}{\Psi}{\Phi}}\) and\\
      \(\llr{\Psi}{K}{A}\) with \(\jnorm[\cneu{\Psi}]{K}\)
      \hfill IH for \cref{item:apq:200} with \cref{item:apq:466}
    \item \(\llr{\Psi}{\ftm{\apprs{\sigma}{P}}{\apprs{\sigma}{p}}}{\blue{B_\MF}}\)\hfill \cref{lemma:apq:14,item:apq:467,item:apq:466}\label{item:apq:468}
    \item \(\llr{\Psi}{\apprs{\sigma}{(\ftm{P}{p})}}{\blue{B_\MF}}\)\hfill definition of substitution, \cref{item:apq:468}
    \end{enumerate}

  \item[\getrn{x/q/var}] Immediate by the definition of substitution and the hypothesis that \(\llr{\Phi}{\sigma}{\Delta}\).
  \item[\getrn{x/q/lam}] We deduce:
    \begin{enumerate}
    \item \(\jxpq{\Delta}{\qtlam{x}{C}}{\QTlolly{S}{U}}\)\hfill rule conclusion\label{item:apq:211}
    \item \(\jxpq{\Delta, \qtp{x}{A}}{\red{C}}{\red{U}}\)\hfill rule premise\label{item:apq:212}
    \item \(\llr{\Sigma}{\sigma}{\Delta}\)\hfill assumption\label{item:apq:232}
    \item \(\llr{\Sigma}{\apprs{\sigma}{\red{C}}}{\red{U}}\) for all \(\llr{\Sigma}{\sigma}{\Delta, \qtp{x}{S}}\)\hfill IH for \cref{item:apq:212}\label{item:apq:253}
    \item \(\jnctx{\phi}{\Phi}{\Sigma}\)\hfill assumption for universal quantification\label{item:apq:235}
    \item \(\llr{\Phi}{\red{D}}{\red{S}}\)\hfill assumption for universal quantification\label{item:apq:242}
    \item \(\llr{\Phi}{\apprs{\phi}{\sigma}}{\Delta}\)\hfill\cref{lemma:apq:5,item:apq:232,item:apq:235}\label{item:apq:275}
    \item \(\llr{\Phi}{\apprs{\phi}{\sigma}, \red{D}}{\Delta, \qtp{x}{S}}\)\hfill\getrn{llr/rs/tm}, \cref{item:apq:242,item:apq:275}\label{item:apq:278}
    \item \(\llr{\Phi}{\apprs{\apprs{\phi}{\sigma}, \red{D}}{\red{C}}}{\red{U}}\)\hfill\cref{item:apq:253,item:apq:278}\label{item:apq:283}
    \item \(\step{\cneu{\Phi}}{\qtapp{(\apprs{\apprs{\phi}{\sigma}}{\qtlam{x}{C}})}{D}}{\apprs{\apprs{\phi}{\sigma}, \red{D}}{\red{C}}}\)\hfill\getrn{cstep/app/beta}\label{item:apq:80}
    \item \(\qtapp{(\apprs{\phi}{\apprs{\sigma}{\qtlam{x}{C}}})}{D} = \qtapp{(\apprs{\apprs{\phi}{\sigma}}{\qtlam{x}{C}})}{D}\)\hfill properties of substitution\label{item:apq:207}
    \item \(\llr{\Phi}{\qtapp{(\apprs{\phi}{\apprs{\sigma}{\qtlam{x}{C}}})}{D}}{\red{U}}\)\hfill\getrn{llr/steps}, \cref{item:apq:283,item:apq:80,item:apq:207}\label{item:apq:205}
    \item \(\llr{\Phi}{\qtapp{(\apprs{\phi}{\apprs{\sigma}{\qtlam{x}{C}}})}{D}}{\red{U}}\) for all \(\jnctx{\phi}{\Phi}{\Sigma}\) and \(\llr{\Phi}{\red{D}}{\red{S}}\)\\
      \mbox{} \hfill universal quantification, \cref{item:apq:235,item:apq:242,item:apq:205}\label{item:apq:206}
    \item \(\jnctx{\wk{\sigma}}{\Sigma, \qtp{x}{A}, \jneu{x}}{\Delta}\)\hfill \cref{cor:apq:4,item:apq:232}\label{item:apq:406}
    \item \(\jnctx{\wk{\sigma}, \red{x}}{\Sigma, \qtp{x}{A}, \jneu{x}}{\Delta, \qtp{x}{A}}\)\\
      \mbox{}\hfill \cref{item:apq:406}, \getrn{llr/rs/tm}, \getrn{neu/var}, \getrn{llr/neu}\label{item:apq:407}
    \item \(\llr{\Sigma, \qtp{x}{A}, \jneu{x}}{\wk{\sigma}, \red{x}}{\Delta, \qtp{x}{A}}\)\\
      \mbox{}\hfill induction on \cref{item:apq:407} with \getrn{llr/neu}\label{item:apq:408}
    \item \(\llr{\Sigma, \qtp{x}{A}, \jneu{x}}{\apprs{\wk{\sigma}, \red{x}}{\red{C}}}{U}\)\hfill \cref{item:apq:253,item:apq:408}\label{item:apq:409}
    \item \(\steps{\cneu{\Sigma}, \jneu{x}}{\apprs{\wk{\sigma}, \red{x}}{\red{C}}}{\red{V}}\) with \(\jnorm[\cneu{\Sigma}, \jneu{x}]{\red{V}}\)\\
      \mbox{}\hfill \cref{lemma:apq:3,item:apq:409}\label{item:apq:410}
    \item \(\steps{\cneu{\Sigma}}{\qtlam{x}{\apprs{\wk{\sigma}, \red{x}}{\red{C}}}}{\qtlam{x}{V}}\) with \(\jnorm[\cneu{\Sigma}]{\qtlam{x}{V}}\)\\
      \mbox{}\hfill induction on \cref{item:apq:410} with \getrn{cstep/lam}; \getrn{norm/can}, \getrn{can/lam/c}, \cref{item:apq:410}\label{item:apq:412}
    \item \(\qtlam{x}{\apprs{\wk{\sigma}, \red{x}}{\red{C}}} = \apprs{\sigma}{\qtlam{x}{C}}\)\hfill definition of substitution\label{item:apq:411}
    \item \(\jhalts{\cneu{\Sigma}}{\qtlam{x}{C}}\)\hfill \getrn{halts}, \cref{item:apq:412,item:apq:411}\label{item:apq:413}
    \item \(\llr{\Sigma}{\apprs{\sigma}{(\qtlam{x}{C})}}{\QTlolly{S}{U}}\)\hfill\getrn{llr/fun}, \cref{item:apq:206,item:apq:413}
    \end{enumerate}
  \item[\getrn{x/q/triv}] Immediate by \(\apprs{\sigma}{\qttriv} = \qttriv\) and \getrn{llr/triv}.
  \item[\getrn{x/q/pair}] Analogous to \getrn{x/f/pair}.
  \item[\getrn{x/q/gate}] Immediate by \getrn{llr/neu}, \getrn{neu/gate}, and \(\apprs{\sigma}{\red{g}} = \red{g}\).
  \item[\getrn{x/q/force}] Analogous to \getrn{x/f/force}, except that the types are \(\red{A_\MQ}\) and \(\FTus[\MQ][\ML]{\red{A_\MQ}}\) instead of \(\blue{A_\ML}\) and \(\FTus[\ML][\MU]{\blue{A_\ML}}\), respectively.
  \item[\getrn{x/q/app}] Analogous to \getrn{x/f/app}.
  \item[\getrn{x/q/match}] Analogous to \getrn{x/f/match}.
  \item[\getrn{x/p/fpu}] We deduce:
    \begin{enumerate}
    \item \(\jxpq{\Delta}{(\fpu{M})}{(\fpat{\FTu[k]}{A_\MF})}\) \hfill rule conclusion
    \item \(\jxpq{\Delta}{\blue{M}}{\blue{A_\MF}}\) \hfill rule premise\label{item:apq:208}
    \item \(\llr{\Phi}{\apprs{\sigma}{\blue{M}}}{\blue{A_\MF}}\) for all \(\llr{\Phi}{\sigma}{\Delta}\) \hfill IH for \cref{item:apq:208}\label{item:apq:475}
    \item \(\llr{\Phi}{\phi}{\Delta}\)\hfill assumption\label{item:apq:469}
    \item \(\jnctx{\psi}{\Psi}{\Phi}\)\hfill assumption\label{item:apq:470}
    \item \(\llr{\Psi}{\blue{K}}{\FTu[k]}\)\hfill assumption\label{item:apq:471}
    \item \(\jcan[\cneu{\Psi}]{\blue{K}}\)\hfill assumption\label{item:apq:472}
    \item \(\jxpq{\Psi}{\blue{K}}{\FTu[k]}\)\hfill definition of \cref{item:apq:471}\label{item:apq:482}
    \item \(\blue{K} = \fttriv\)\hfill inversion, \cref{lemma:apq:2,item:apq:482,item:apq:472}
    \item \(\apprs{\psi}{\apprs{\phi}{\blue{M}}} = \apprs{\apprs{\psi}{\phi}}{\blue{M}}\)\hfill substitution property\label{item:apq:473}
    \item \(\ecan{\fttriv}{(\apprs{\psi}{\apprs{\phi}{\fpu{M}}})}{\apprs{\apprs{\psi}{\phi}}{\blue{M}}}\)\hfill \getrn{ecan/triv}, \cref{item:apq:473}\label{item:apq:490}
    \item \(\llr{\Psi}{\apprs{\psi}{\phi}}{\Delta}\)\hfill \cref{lemma:apq:5,item:apq:470,item:apq:469}\label{item:apq:474}
    \item \(\llr{\Psi}{\apprs{\apprs{\psi}{\phi}}{\blue{M}}}{\blue{A_\MF}}\)\hfill \cref{item:apq:475,item:apq:474}\label{item:apq:489}
    \item \(\llr{\Psi}{\ecanop{K}{\apprs{\psi}{\apprs{\phi}{(\fpu{M})}}}}{\blue{A_\MF}}\) for all \(\llr{\Phi}{\phi}{\Delta}\),\\
      \({\jnctx{\psi}{\Psi}{\Phi}}\), and \(\llr{\Psi}{\blue{K}}{\FTu[k]}\) with \(\jnorm[\cneu{\Psi}]{\blue{K}}\)
      \hfill \cref{item:apq:469,item:apq:470,item:apq:471,item:apq:472,item:apq:489,item:apq:490}
    \end{enumerate}
  \item[\getrn{x/p/fpd}] We deduce:
    \begin{enumerate}
    \item \(\jxpq{\Delta}{(\fpd{x}{M})}{(\fpat{\FTds[\ML][\MU]{A}}{B})}\) \hfill rule conclusion
    \item \(\jxpq{\Delta, \ftp{x}{A}}{\blue{M}}{\blue{B}}\) \hfill rule premise\label{item:apq:476}
    \item \(\llr{\Phi}{\apprs{\sigma}{\blue{M}}}{\blue{B}}\) for all \(\llr{\Phi}{\sigma}{\Delta, \ftp{x}{A}}\) \hfill IH for \cref{item:apq:476}\label{item:apq:477}
    \item \(\llr{\Phi}{\phi}{\Delta}\)\hfill assumption\label{item:apq:478}
    \item \(\jnctx{\psi}{\Psi}{\Phi}\)\hfill assumption\label{item:apq:479}
    \item \(\llr{\Psi}{\blue{K}}{\FTds[\ML][\MU]{A}}\)\hfill assumption\label{item:apq:480}
    \item \(\jcan[\cneu{\Psi}]{\blue{K}}\)\hfill assumption\label{item:apq:481}
    \item \(\jxpq{\Psi}{\blue{K}}{\FTds[\ML][\MU]{A}}\)\hfill definition of \cref{item:apq:480}\label{item:apq:483}
    \item \(\blue{K} = \ftdown{V}\) with \(\jnorm[\cneu{\Psi}]{\blue{V}}\)\hfill inversion, \cref{lemma:apq:2,item:apq:481,item:apq:483}
    \item \(\llr{\Psi}{\blue{V}}{\blue{A}}\)\hfill inversion on \cref{item:apq:480}: \getrn{llr/ds} is the only rule possible\\
      \mbox{}\hfill by \cref{conj:apq:9}, \getrn{norm/can}, and \cref{item:apq:481}\label{item:apq:486}
    \item \(\ecan{\ftdown{V}}{(\apprs{\psi}{\apprs{\phi}{\fpd{x}{M}}})}{\apprs{\apprs{\psi}{\phi}, \blue{V}}{\blue{M}}}\)\\
      \mbox{}\hfill \getrn{ecan/down}, properties of substitution\label{item:apq:484}
    \item \(\llr{\Psi}{\apprs{\psi}{\phi}}{\Delta}\)\hfill \cref{lemma:apq:5,item:apq:478,item:apq:479}\label{item:apq:485}
    \item \(\llr{\Psi}{\apprs{\psi}{\phi}, \blue{V}}{\Delta, \ftp{x}{A}}\)\hfill \cref{item:apq:485}, \getrn{llr/rs/tm}, \cref{item:apq:486}\label{item:apq:487}
    \item \(\llr{\Psi}{\apprs{\apprs{\psi}{\phi}, \blue{V}}{\blue{M}}}{\blue{B}}\)\hfill \cref{item:apq:477,item:apq:487}\label{item:apq:488}
    \item \(\llr{\Psi}{\ecanop{K}{\apprs{\psi}{\apprs{\phi}{(\fpd{x}{M})}}}}{\blue{B}}\) for all \(\llr{\Phi}{\phi}{\Delta}\),\\
      \({\jnctx{\psi}{\Psi}{\Phi}}\), and \(\llr{\Psi}{\blue{K}}{\FTds[\ML][\MU]{A}}\) with \(\jnorm[\cneu{\Psi}]{\blue{K}}\)\\
      \mbox{}\hfill \cref{item:apq:478,item:apq:479,item:apq:480,item:apq:481,item:apq:488,item:apq:484}
    \end{enumerate}
  \item[\getrn{x/p/fpp}] Analogous to \getrn{x/p/fpd} using \getrn{ecan/pair} and \getrn{llr/pair} instead of \getrn{ecan/down} and \getrn{llr/ds}.
  \item[\getrn{x/p/qpu}] Analogous to \getrn{x/p/fpu}.
  \item[\getrn{x/p/qpp}] Analogous to \getrn{x/p/fpd} using \getrn{ecan/pair} and \getrn{llr/pair} instead of \getrn{ecan/down} and \getrn{llr/ds}.
    \qedhere
  \end{proofcases}
\end{proofEnd}

\begin{corollary}[Normalization]
  \label{cor:apq:2}
  If \(\japq{\ctp{\Psi}}{P}{A}\), then \(\jhalts{\cneu{\Psi}}{P}\).
\end{corollary}

\begin{proof}
  If \(\japq{\ctp{\Psi}}{P}{A}\), then  \(\jxpq{\ctp{\Psi}}{P}{A}\) by \cref{prop:apq:1}.
  By \cref{conj:apq:5} with the identity substitution, \(\llr{\Psi}{P}{A}\), so \(\jhalts{\cneu{\Psi}}{P}\) by \cref{lemma:apq:3}.
\end{proof}

Because our language our is deterministic, \cref{cor:apq:2} also implies that well-typed terms are strongly normalizing, \ie, that they have no infinite reduction sequences.

\section{Related Work}
\label{sec:related-work}

\subsection{Quantum Programming Languages}

Our work joins the large family of Proto-Quipper languages that formalize fragments of Quipper~\cite{green_2013:_quipp}.
Quipper is a language embedded in Haskell, and it does not ensure that quantum computation is type safe.
\Textcite{ross_2015:_algeb_logic_method_quant_comput} formalized the first type-safe, stand-alone Quipper fragment, named \pqs because it uses subtyping to deal with the \(!\) modality.
Its type system and small-step semantics are defined in terms of \textit{circuit constructors}, a collection of axiomatically specified set-theoretic functions for appending circuits, naming wires, \etc.

\Citeauthor{rios_selinger_2018:_categ_model_quant}~\cites{rios_selinger_2018:_categ_model_quant,rios_2021:_categ_sound_quant} introduce \pqm.
It uses explicit ``lift'' and ``force'' operators instead of subtyping to box and unbox circuits.
Similar to \pqs, they encode circuits as triples \((\vec l, C, \vec k)\), where \(\vec l\) and \(\vec k\) label input and output wires for a circuit constant \(C\).
Instead of ranging over elements in a set, constants \(C\) now range over morphisms in a symmetric monoidal category, and they give their language a linear/non-linear (LNL) categorical semantics based on an adjunction between \(\mathbf{Set}\) and a family fibration on a category with circuits.
Unfortunately, their choice to label wires with free names \(\vec l, \vec k\) complicates their categorical semantics and forces them to work with an auxiliary monoidal category of \emph{labelled circuits}.
Because we instead directly notate circuits using the internal language of the symmetric monoidal category of circuits (our red circuit description language), we conjecture that \pqa lends itself to a simpler categorical semantics.
Their big-step operational semantics on pairs of circuits and terms relies on fresh name generation and the simplified form of circuit constructors presented in \cref{sec:gentle-intr-quant}.
We avoid these complex operations by using more familiar operations like binding and function composition.
\pqm includes data types not found in \pqa, \eg, sum types and inductively defined types like lists.
Given the clean separation between our functional and quantum layers, it should be straightforward to extend \pqa with these features.

\Textcite{lindenhovius_2018:_enric_linear_non} generalize \pqm to support string diagrams in arbitrary symmetric monoidal categories.
Though their language has the same syntax and operational semantics as \pqm, they give their language an abstract categorical semantics and show how to recover \citeauthor{rios_selinger_2018:_categ_model_quant}'s semantics as a concrete instance of their model.
They also use their semantics to show how to safely extend \pqm with general recursion.
We conjecture that their semantic techniques could be adapted to extend \pqa with recursion.

\pqs and \pqm, like \pqa, only consider static circuit generation.
However, certain quantum algorithms must interleave circuit generation and execution, a functionality called \emph{dynamic lifting}.
This functionality can be found in languages like \pqdyn~\cite{fu_2023:_proto_quipp_with_dynam_liftin} and \pqvar{K}~\cite{colledan_dal_2023:_dynam_liftin_effec}.
The two phases of dynamic lifting---circuit generation and execution---correspond to the two operational semantics of \pqdyn.
Its circuit generation semantics closely resembles its predecessors'.
However, its circuit execution semantics is significantly more complex because measurement is inherently probabilistic.
Extending \pqa to handle these probabilistic phenomena remains an open and interesting problem.

QWire-style languages~\cite{paykin_2017:_qwire,rennela_staton_2020:_class_contr_quant} take a different approach to quantum programming.
They provide a first-order language for circuit programming, embedded in a higher-order host language, and the interaction between their two languages is inspired by a linear/non-linear model.
Though QWire and \pqa both involve circuit programming languages, they serve fundamentally different purposes.
Indeed, programmers use QWire's circuit to directly compose gates on wires, while our red circuit language is only used internally as a syntax for circuits generated by our functional abstractions.

\subsection{Languages with Linear/non-Linear or Adjoint Foundations}

Quantum programming languages must handle both (linear) quantum data and (non-linear) classical data, and they often build on Benton's~\cite{benton_1995:_mixed_linear_non_linear_logic} linear/non-linear (LNL) models to safely do so.
For example, \textcite{selinger_valiron_2008:_linear_linear_model} give an LNL semantics for a call-by-value quantum lambda calculus.
LNL models have also found applications in programming language foundations beyond quantum computation.
For instance, \textcite{paykin_2018:_linear_linear_types} gives formal foundations for embedded, domain-specific languages with linear resources with non-linear host languages.

Adjoint logic~\cite{pruiksma_2018:_adjoin_logic,reed_2009:_judgm_decon_modal_logic,licata_shulman_2016:_adjoin_logic_with_modes} generalizes LNL logic to conservatively combine multiple intuitionistic logics with varying structural properties.
Because the embeddings preserve the proof-theoretic properties of the individual logics, adjoint logic provides a framework for combining different computational interpretations of intuitionistic logics.
It has been used, \eg, to combine different varieties of message-passing communication~\cite{pruiksma_pfenning_2021:_messag_passin_inter_adjoin_logic}, functional programming with session-typed concurrency~\cite{pruiksma_pfenning_2022:_back_to_futur}, and to model fork-join parallelism and concurrent write-once shared-memory semantics~\cite{pruiksma_pfenning_2022:_back_to_futur}.
\pqa's type system uses adjoint-logical foundations to safely integrate its linear/non-linear functional programming language with its linear circuit description language.

\subsection{Linear Logical Relations and Normalization}
\label{sec:logic-relat-norm}

Over the past three decades, a variety of approaches have been
developed to prove meta-theoretic properties such as normalization about substructural
systems. One the earliest works by \textcite{Bierman:PhD94} proves strong
normalization for the linear lambda-calculus only considering
$\beta$-reduction and omitting any commuting conversions using logical
relations. Because he only considers reduction of closed terms, the
complexity that arises of dealing with open terms goes
away. Nevertheless, the overall proof remains complex. %

\Textcite{Accattoli:RTA13} proved strong normalization for linear logic
using proof nets and applying reducibility
candidates. Subsequently, \Textcite{Guerrini:Linearity16}
proves strong normalization for the
linear lambda-calculus only  considering $\beta$-reduction for
functions by fine-tuning the original, syntactic proof of strong normalisation
proposed by Gandy \cite{Gandy:80} for the lambda-calculus. 
The idea is to increase the measure for variable occurrences in a term
by 1. This new measure then can be used to prove strong normalisation of
$\beta$-reduction and of its linear version. Our proof uses logical
relations instead to prove normalization and we leverage
the generic proof for the structural calculus by approximating the
typing of linear constructs. This leads to a clean proof that we
believe is straightforward to mechanize following similar techniques
as in \cite{POPLMarkReloaded:19}. 

Our approach of using approximately typed substructural systems when
defining the logical relation appears to be the first of its kind. We
believe that this perspective could simplify other work that carefully crafts
linear logical relations based on linearly well-scoped and well-typed
terms such as the work by \textcite{Zhao:APLAS10} where the
authors prove parametricity for a polymorphic linear
lambda-calculus. Most recently, \textcite{Aberle:FSCD25}
develop a family of unary logical relations that allow them to prove
consequences of parametricity for a range of substructural type
systems. A key idea is to parameterize the relation by an algebra,
which they exemplify with a monoid and commutative monoid
to interpret ordered and linear type systems, respectively.
Similarly, in modal the semantic model is 
parameterized by a semi-ring that governs the interactions between
different sub-systems. In our work, we do not parameterize the logical
relations predicate but rather we define it using an approximate
notion of well-typedness that disregards any substructural
restrictions. This simplifies the definition of the
logical relation and the actual normalization proof. Our view could
also potentially help simplify the aforementioned works.

\section{Conclusion}
\label{sec:conclusion}

We introduced \pqa, an adjoint-logical rational reconstruction of \pqm.
We avoided the need for complex metatheoretic operations by representing circuits as a fragment of the linear \(\lambda\)-calculus.
This new representation let us recast \pqm's effectful operational semantics as a pure call-by-value reduction semantics.
Our rational reconstruction makes it easy to prove meta-theorems using standard techniques, and we illustrated this fact by proving that our language is normalizing using a logical normalization predicate.
In doing so, we showed that standard logical relations arguments could be used to reason about behavioural properties of substructural systems while minimizing any proof overhead caused by substructurality.

\subsection{Future Work}

\paragraph{Dynamic Lifting}

While \pqa only captures static circuit generation as found in \pqm and its extensions, we believe that it can naturally capture the more complex features offered by other Proto-Quipper languages.
Of those, we are interested in extending our system with \emph{dynamic lifting}~\cite{fu_2023:_proto_quipp_with_dynam_liftin} to capture quantum algorithms that interleave circuit generation and execution.
Recent work by \textcite{sano_2025:_fusin_session_typed} uses a contextual box modality to integrate a pure functional language with an effectful concurrent language, and it provides abstractions for the functional to generate concurrent code and observe its effects.
We conjecture that this approach can be adapted to integrate functional programming with circuit generation and circuit execution.

\paragraph{Circuit Optimization}

Current quantum computers are modest and prone to errors due to ``quantum noise''.
By optimizing quantum circuits, we can reduce the number of gates used and potential errors.
Our explicit circuit syntax sets the stage for Proto-Quipper dialects that build on prior work~\cite{georges_2017:_lincx,sano_2025:_fusin_session_typed} to support pattern-matching on circuit code.
Such an extension could then be used to optimize circuits from within Proto-Quipper.

\paragraph{Mechanization}

An important motivation for our rational reconstruction was to make Proto-Quipper more amenable to mechanization, and its mechanization is ongoing.
We have mechanized the statics, dynamics and subject reduction theorem of \pqx in approximately 260 lines of Beluga~\cite{pientka_dunfield_2010:_belug} code.
We are confident that we can fully mechanize \pqa's meta-theory thanks to \pqa's logically informed design.

\begin{credits}
  \subsubsection{\ackname}
  The authors thank Max Gross for his work on mechanizing \pqs: the challenges he encountered directly inspired this rational reconstruction.
  They also thank him for technical discussions and for his comments on a draft of this work.
  The authors thank Peter Selinger for technical discussions that contributed to the development of \pqa.

\subsubsection{\discintname}
The authors have no competing interests to declare that are relevant to the content of this article.
\end{credits}

\printbibliography

\clearpage
\appendix

\section{Full reduction semantics}
\label{sec:full-reduct-semant}

We include all reduction rules, including those omitted from \cref{sec:reduction}:

\begingroup\small\allowdisplaybreaks
\begin{gatherrules}
  \judgmentbox{\fstep{\Pi}{\blue{M}}{\blue{N}}}{Term \(M\) reduces to \(N\)}
  \\
  \getrule{fstep/app/1}
  \quad
  \getrule{fstep/app/2}
  \\
  \getrule{fstep/app/beta}
  \\
  \getrule{fstep/app/cc}
  \\
  \getrule{fstep/force}
  \quad
  \getrule{fstep/force/1}
  \\
  \getrule{fstep/force/cc}
  \\
  \getrule{fstep/pair/1}
  \quad
  \getrule{fstep/pair/2}
  \\
  \getrule{fstep/down}
  \quad
  \getrule{fstep/m/f}
  \\
  \getrule{fstep/m/k}
  \quad
  \getrule{fstep/m/q}
  \\
  \getrule{fstep/m/q/r}
  \\
  \getrule{fstep/m/f/cc}
  \\
  \getrule{fstep/m/q/cc}
  \\
  \judgmentbox{\cstep{\Pi}{\red{C}}{\red{D}}}{Circuit \(C\) reduces to \(D\)}
  \\
  \getrule*{cstep/app/1}
  \quad
  \getrule*{cstep/app/2}
  \\
  \getrule{cstep/lam}
  \quad
  \getrule*{cstep/app/beta}
  \\
  \getrule*{cstep/app/cc/1}
  \\
  \getrule{cstep/app/cc/2}
  \\
  \getrule*{cstep/force}
  \quad
  \getrule*{cstep/force/1}
  \\
  \getrule*{cstep/force/cc}
  \\
  \getrule*{cstep/pair/1}
  \quad
  \getrule*{cstep/pair/2}
  \\
  \getrule*{cstep/m}
  \quad
  \getrule*{cstep/m/k}
  \\
  \getrule*{cstep/m/r}
  \\
  \getrule*{cstep/m/cc}
\end{gatherrules}
\begin{gatherrules}
  \judgmentbox{\ecan{K}{p}{P}}{Eliminating canonical form \(K\) with pattern \(p\) produces \(P\)}
  \\
  \getrule{ecan/down}
  \qquad
  \getrule{ecan/triv}
  \\
  \getrule{ecan/pair}
\end{gatherrules}
\endgroup

\section{Full definition of \pqx}
\label{sec:full-definition-pqx}

We include all rules defining \pqx, including those omitted from \cref{sec:normalization}:
\begingroup\small
\allowdisplaybreaks
\begin{gatherrules}
  \judgmentbox{\jxpq{\Delta}{P}{A}}{Term or circuit \(P\) approximately has type \(A\)}\\
  \getrule{x/f/var}
  \qquad
  \getrule*{x/f/lam}
  \\
  \getrule*{x/f/triv}
  \qquad
  \getrule{x/f/pair}
  \\
  \getrule{x/f/susp}
  \qquad
  \getrule*{x/f/down}
  \\
  \getrule*{x/f/app}
  \\
  \getrule*{x/f/force}
  \qquad
  \getrule*{x/f/match}
  \\
  \getrule*{x/q/var}
  \qquad
  \getrule*{x/q/lam}
  \\
  \getrule*{x/q/triv}
  \qquad
  \getrule*{x/q/pair}
  \\
  \getrule*{x/q/force}
  \quad
  \getrule*{x/q/app}
  \\
  \getrule*{x/q/match}
  \qquad
  \getrule{x/q/gate}
\end{gatherrules}
\begin{gatherrules}
  \judgmentbox{\jxpq{\Delta}{p}{\pat{A}{B}}}{Pattern \(p\) eliminates terms approximately of type \(A\) to produce terms approximately of type \(B\)}
  \\
  \getrule{x/p/fpu}
  \quad
  \getrule{x/p/fpp}
  \\
  \getrule*{x/p/fpd}
  \quad
  \getrule*{x/p/qpu}
  \\
  \getrule*{x/p/qpp}
\end{gatherrules}
\endgroup

\section{Omitted proofs}

\printProofs

\end{document}